**V.A. Trapeznikov Institute of Control Sciences of Russian Academy of Sciences**

**Dmitry A. Novikov**

# MECHANISM DESIGN and ACTIVE SYSTEMS THEORY:

## 60 Years of Parallel Development

**Abstract.** This survey compares two research traditions within the mathematical theory of designing the rules of interaction (mechanisms) for purposeful participants: mechanism design, together with contract theory and implementation theory, and the theory of active systems created by V.N. Burkov and developed into the theory of control in organizations (TCO) by D.A. Novikov's school; the main results of Yu.B. Germeier's school and of the domestic schools adjacent to active systems theory (AST) are covered as well. The problem statements and principal results of 2006–2026 are presented: for mechanism design (MD) – robust and dynamic mechanism design, multidimensional screening, information design, simplicity of mechanisms and credibility of the designer, market design, algorithmic and learning-based design, mechanisms for AI agents, and robust, relational, dynamic and algorithmic contracts; for AST and TCO – a general typology of control, strategy-proof and incentive-compatible planning mechanisms, decomposition of the agents' game, systems with distributed control and coalitional interaction, reflexive games and informational control, control of staff and structure, dynamic active systems, and project management. A correspondence between the classical results is established (the fair play principle and the revelation principle, the serial resource allocation mechanism and the uniform rule, active expertise mechanisms and generalized median schemes, quasi-compensatory incentive schemes and contracts under observable actions), and priorities are delineated by uniform rules; the areas where the two traditions diverge are identified, as are the dimensions of the problem (staff, structure, institutions, awareness, agents' beliefs) covered only by TCO. The degree to which each class of problems has been studied in the two traditions is presented in expert maps on a four-level (0–3) and a ten-point (0–10) scale. A faceted classification of problems (ten facets) and a typology of results (nine types) of the two schools are proposed. Applications are reviewed, and a development forecast up to 2035 is given, with twenty problems in which the two traditions are expected to converge.

**Keywords:** mechanism design, theory of active systems, theory of control in organizations, contract theory, implementation theory, strategy-proofness, fair play principle, incentives, informational control, reflexive games, coalitional interaction, classification, forecast.

# Contents

# 1. Introduction

This survey is about the mathematical theory of designing the rules (mechanisms) by which participants with their own goals and their own information interact. The field, which originates in L. Hurwicz's work on the informational efficiency of resource allocation processes [293], is called *mechanism design* (the term took hold after Hurwicz's papers of 1972–1973, which posed the problem of designing mechanisms and introduced *incentive compatibility* [294, 295]) – hereafter MD – and includes *implementation theory*, *contract theory*, and the adjacent *auction theory* and *markets design* [495]. The same range of problems has been studied since 1969 within the *theory of active systems* (AST), created by V.N. Burkov (1939–2025) at the Institute of Control Sciences of the Russian Academy of Sciences [117, 121, 87, 297], and, since the early 2000s, within D.A. Novikov's *theory of control in organizations* (TCO), for which AST serves as the historical core [109, 418, 412].

The two research traditions took shape in different scientific communities and, at the early stage, did not cite each other; their substantive origins differ, though they partly overlap. MD grew out of the debate on the comparative efficiency of economic systems (including decentralized planning and intra-firm resource allocation [294]) and out of social choice theory; its original objects are the market, the auction, voting, and the contract, but also the firm and the planning authority. AST grew out of control theory and operations research; its original object is an organization in which a governing body (the *principal*) plans and incentivizes the activity of its subordinates (*active elements*, *agents*). The formal foundation is nevertheless shared: the principal chooses the rules of the game, the agents choose strategies, and the outcome is evaluated at the solution of the resulting game. Representatives of the AST school themselves characterize TCO as a version of mechanism design adapted to the problems of managing organizations [106, 108].

There are two reasons to take up the subject now. First, the last detailed comparison, carried out within the AST school itself – the survey by Burkov, Enaleev and Korgin [102] – is confined to the incentive compatibility and strategy-proofness of planning mechanisms and deliberately leaves incentives and contract theory outside its scope; earlier comparisons date back to the 1990s [104, 121], when surveys of the AST school's results also appeared [125, 99]. Second, over 2006–2026 MD has changed substantially: robust, dynamic and information design have emerged, an algorithmic branch has taken shape, contract theory has acquired robust and computational versions, and since 2023 learning artificial intelligence systems have become both participants in and designers of mechanisms.

The aim of the survey is to give a coherent picture of both research traditions over the past twenty years, to show the exact correspondence of problem statements and results where it exists, to identify the areas where the traditions do not intersect, to propose a common system for classifying problems and results, and, on this basis, to give a development forecast. The genre is a survey for specialists in control theory and game theory: the main results are presented as problem statements and theorems without proofs.

**Structure.** Section 2 gives the general problem statement and a dictionary of correspondences. Section 3 compares the classical cores. Sections 4 and 5 describe the development of MD and contract theory in 2006–2026, and Section 6 the development of AST and TCO. Section 7 contains a comparative analysis, Section 8 the system of classifications, Section 9 applications, and Section 10 the forecast.

The author actively used large language models to collect and systematize information, format and layout text, and generate figures and tables.

## 2. The Two Research Traditions: General Statement, Chronology, Vocabulary

### 2.1. The General Mechanism Synthesis Problem

Both research traditions consider a set of agents $N = \{1, \dots, n\}$ and a distinguished participant who sets the rules: the principal (designer). An *agent* $i$ holds private information – a *type* $\theta_i \in \Theta_i$ (in AST notation, a parameter $r_i \in \Omega_i$) – and preferences $u_i(x, \theta)$ over a set of outcomes $X$. The MD statement starts from the goal: a social choice rule (correspondence) $F: \Theta \rightrightarrows X$, $\Theta = \prod_i \Theta_i$ is given. A mechanism is a game form $\Gamma = \langle M, g \rangle$, where $M = \prod_i M_i$ are the message sets and $g: M \to X$ is the outcome function. Under a chosen solution concept $\mathcal{S}$ (dominant-strategy equilibrium, Nash equilibrium, Bayesian equilibrium, etc.), the mechanism *fully implements* $F$ [304, 359, 480] if

$$g\big(\mathcal{S}(\Gamma, \theta)\big) = F(\theta) \qquad \forall\, \theta \in \Theta, \tag{1}$$

Weak implementation requires only the inclusion $\varnothing \neq g\big(\mathcal{S}(\Gamma, \theta)\big) \subseteq F(\theta)$, and in optimization-type MD statements an even weaker requirement, called *partial implementation*, usually suffices: for every $\theta$ there exists a solution $m^* \in \mathcal{S}(\Gamma, \theta)$ with $g(m^*) \in F(\theta)$; implementation is understood in this sense, for example, in [64]. A direct mechanism ($M_i = \Theta_i$, $g = F$) is dominant-strategy incentive compatible if

$$u_i(F(\theta_i, \theta_{-i}), \theta_i) \geq u_i(F(\theta_i', \theta_{-i}), \theta_i) \qquad \forall\, i, \theta_i, \theta_i', \theta_{-i}, \tag{2}$$

and in the Bayesian setting with a prior type distribution the same inequality is required in expectation over $\theta_{-i}$. The optimization version of the problem: the principal maximizes the expected value of its own criterion (revenue, social welfare) over the set of mechanisms satisfying the incentive compatibility and individual rationality constraints [131, 384].

The AST statement starts from control. The principal chooses a *control* $u \in U$ (a planning procedure, an incentive scheme, constraints, information), the agents choose *actions* $y \in A$, and the *set of solutions of their game* under the given control is denoted by $P(u)$. The *efficiency of control* is defined by the value of the principal's goal function $\Phi(u, y)$ on the set of game solutions, and the texts of the AST school use two indices [120, 424, 412]:

$$K(u) = \max_{y \in P(u)} \Phi(u, y), \qquad K_g(u) = \min_{y \in P(u)} \Phi(u, y), \tag{3}$$

and the synthesis problem consists in maximizing the chosen index over $u \in U$. The first index corresponds to the *benevolence hypothesis* (among the actions of equal value to itself, an agent chooses the one best for the principal) and is the default in AST and TCO. The second, *guaranteed efficiency*, follows the *maximum guaranteed result* principle of the theory of hierarchical games and is used when benevolence cannot be relied upon. In incentive problems, including multi-agent ones, the passage from the first index to the second is secured by arbitrarily small bonuses added to the reward [428] (Section 3.5). In models with information revelation, the control includes a planning procedure $x = \pi(s)$ that assigns plans based on the agents' messages $s \in S$; in the terms of (1) this is an outcome function $g$, and the AST mechanism $\langle S, \pi \rangle$ coincides with the game form [102]. Uncertainty about types is historically described in AST by a set $\Omega$ without a distribution, and the criterion is taken in guaranteed form over the unknown parameters: instead of the expectation over $r$, the minimum over $r \in \Omega$ is computed [98].

The criteria of the two research traditions (MD and AST) are related as follows. Requirement (1) applies to all solutions of the game; the index $K_g$ evaluates a mechanism at the worst solution, the index $K$ at the solution best for the principal. Hence $K$ corresponds to partial implementation, and $K_g$ to a requirement on all equilibria, as in weak and full implementation; in contract theory the analogue of $K_g$ is the requirement of unique implementation, under which the principal evaluates a scheme at the equilibrium of the agents' game worst for itself [266, 479, 528]. Implementation theory starts from a given choice rule and asks whether it is implementable; in AST, a criterion is given and the best control in a given class is sought. The guaranteed treatment of unknown parameters (a minimum over a set instead of an expectation over a distribution) is a separate and more persistent distinctive feature of the AST statement. The general scheme is shown in Fig. 1.

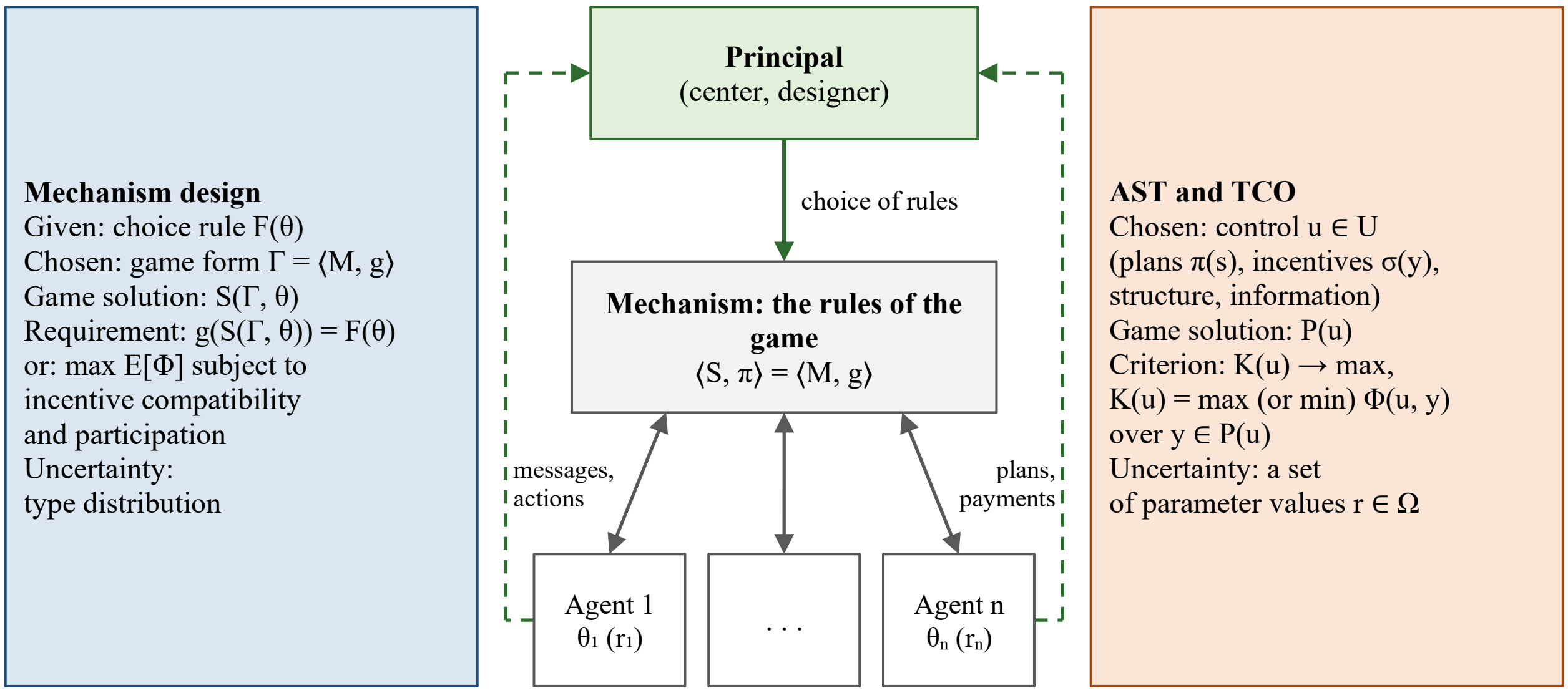

Fig. 1. The general mechanism synthesis problem and its formulation in the two research traditions

## 2.2. Chronology

The MD line is counted from L. Hurwicz's papers of 1960–1973, which introduced the notions of a mechanism as a message-exchange system and of incentive compatibility [294, 293, 295, 495]. The Vickrey auction and the Clarke and Groves mechanisms were obtained in 1961–1973 [150, 257, 517]; the Gibbard–Satterthwaite theorem in 1973–1975 [232, 477]; the *revelation principle* in 1973–1979 [165, 232, 383]; Maskin's theorem on *Nash implementation* in 1977 (published in 1999) [358]; and Myerson's optimal auction in 1981 [384].

Contract theory took shape in parallel: models of *moral hazard* [255, 284], *adverse selection* [50], and incomplete contracts [256, 271]. The Nobel Prizes of 1996 (Mirrlees, Vickrey: “for their fundamental contributions to the economic theory of incentives under asymmetric information”), 2001 (Akerlof, Spence, Stiglitz: “for their analyses of markets with asymmetric information”), 2004 (Kydland, Prescott: “for their contributions to dynamic macroeconomics: the time consistency of economic policy”) and 2005 (Aumann, Schelling: “for having enhanced our understanding of conflict and cooperation through game-theory analysis”) [393, 394, 395, 396] belong to the prehistory and adjacent areas, while the prizes of 2007 (Hurwicz, Maskin, Myerson), 2012 (Roth, Shapley), 2014 (Tirole), 2016 (Hart, Holmström) and 2020 (Milgrom, Wilson) successively recognize the foundations of MD, market design, regulation, contract theory and auction theory [184, 185, 492, 493, 495].

The AST line is counted from 1969, when V.N. Burkov introduced the notions of an *active element* and an *active system* and formulated the *fair play principle* (the open control principle); the first journal publication appeared in 1970 [117], and the first original English-language one (in the proceedings of an international school on differential games) in 1971 [118]; the journal Avtomatika i Telemekhanika is published in English translation as Automation and Remote Control, and the first monograph came out in 1977 [87].

The AST school distinguishes four periods in its history [121]: birth (the 1960s–70s: the fair play principle; at the same time the *theory of hierarchical games* of Yu.B. Germeier was being developed at the Computing Center of the USSR Academy of Sciences [226]); consolidation (the 1980s: incentive-compatible planning, optimality of fair play [101, 98], resource allocation and expertise mechanisms [97], adaptive mechanisms [124], dynamic systems with farsighted elements [481, 482], business games and the first practical implementations); the 1990s (the comparison with contract theory [104], incentives in dynamic and multi-agent systems [406], surveys [125, 99], models with uncertainty, applications to project management [119], the summarizing book [120] and the lecture course [424]); and the twenty-first century (the crystallization of TCO, whose foundational monograph first appeared in 2005 and reached its fourth edition in 2022 [418, 412], incentives in multi-agent systems [428], control mechanisms for dynamic active systems [427, 426], reflexive games [422], strategy-proof mechanisms [110, 328], hierarchy optimization [375, 246], networks [147], complex activity [54]). The anniversary conferences "Theory of Active Systems" were held in 2009, 2014, 2019 and 2024 [121, 498].

The AST school also has an earlier, optimization root. From the mid-1960s V.N. Burkov and his colleagues worked on network planning and scheduling and on resource allocation over networks [116, 86]; this line chronologically precedes the notion of an active system and survives to this day in the method of *network programming* and in work on project management, including that of S.A. Barkalov's Voronezh school [47, 121] (Section 6.9). Other independent directions also took shape within the AST school and are considered below: G.A. Ougolnitsky's theory of sustainable development control for active systems (Southern Federal University), which its authors define as a development of AST [512] (Section 6.7), and M.I. Geraskin's studies of reflexion in oligopoly games (Samara University), going back to joint papers with A.G. Chkhartishvili of 2015–2017, of which [220] appeared in Automation and Remote Control (Section 6.5). Two regional offshoots of AST are considered in Section 6.7: the Kazakh school of A.A. Ashimov, which in the 1980s, jointly with the Institute of Control Sciences, developed and deployed incentive-compatible planning mechanisms at nonferrous metallurgy enterprises [19, 22] and in the 2000s–2010s developed the theory of parametric regulation of macroeconomic systems [18, 23, 24], and the Samara school of coordinated incentives of V.G. Zaskanov and G.M. Grishanov [75, 514, 301].

Yu.B. Germeier's *theory of hierarchical games* developed in parallel with the contract-theory branch of MD. The theorem on the maximum guaranteed result in the game $\Gamma_2$, where the first player's strategy is a function of the second player's choice, was published in 1971 [225], and the monograph in 1976 [226]; the role of information about the partner's goal function and multi-stage conflicts with information exchange were studied by A.F. Kononenko in 1973–1977 [321, 320], with a systematic exposition in [334]. The very setting "the principal announces a function of an observable variable" arose in economics independently and in the same years: the tax schedule in the model of income taxation under private information about abilities [372], payment schedules under observable information in insurance [485], and then the agency models [284, 373, 463]. These papers maximize expected utility and use the forcing contract (punishment for any deviation) only as a remark on the observable-action case [284, 485]. The priority of Germeier's school therefore concerns the general mathematical statement, the theorems on the structure and properties of the game's solution (the first player's optimal strategy), and the guaranteed-result criterion. The game $\Gamma_1$ goes back to Stackelberg [487], but it was Germeier who gave it a rigorous formulation and located it among the other hierarchical games, in particular comparing the first player's payoffs in this game and in the others. The English translation of Germeier's monograph, made by A. Rapoport, appeared in 1986 [227], and the translation of the foundational TCO monograph in 2013 [414]; by 2021, [57, 59, 81, 85, 106, 147, 402, 403, 398, 435] had also been published by international publishers. Nevertheless, the reference lists of the classical MD papers on strategy-proof rules [42, 381] and the subsequent core MD literature contain **no references to these editions** (Section 7.1).

Germeier's closest students at the Computing Center of the USSR Academy of Sciences continued the hierarchical-games line in the early 1970s. F.I. Ereshko and A.F. Kononenko solved the game with the right of the first move under incomplete information about the partner's intentions [201]; I.A. Vatel and F.I. Ereshko published the popular-science brochure “The Mathematics of Conflict and Cooperation” [515] and then an encyclopedic article on games with a hierarchical structure [516]; Germeier and Vatel introduced games with a hierarchical vector of interests [230]; Germeier and Ereshko considered *side payments* in games with a fixed order of moves [228] – a construction that coincides in substance with the compensatory incentive schemes of AST (Section 6.7).

The hierarchical-games line continued at the Computing Center after Germeier's death in 1975: on the basis of his sector, a department of mathematical modeling of conflict situations was created, headed by A.F. Kononenko [333]. The 1980s–1990s saw the monographs by V.A. Gorelik and A.F. Kononenko on game-theoretic models of hierarchical systems [239] and by V.A. Gorelik, M.A. Gorelov and A.F. Kononenko on the analysis of conflict situations in control systems [238], the book by A.F. Kononenko, A.D. Khalezov and V.V. Chumakov on decision making under uncertainty [325], and the Computing Center preprints on exact aggregation in game-theoretic models [11] and on informational aspects of decision making under conflict [242]. F.I. Ereshko published a preprint on hierarchical compromises [196] and two IIASA papers on mathematical methods for the analysis of hierarchical systems [200, 203], and in 1999–2001 introduced the term “reflexive strategies” in his school [202, 198]. These works are considered in Section 6.7.

The AST school itself puts the priority question with restraint (priorities in a given class of problems are a matter of the years in which the main results were obtained). The survey [102] names the 1970 paper as the starting point of the notions of compatibility and truthfulness in AST, traces the beginning of the systematic study of the problem in MD to Hurwicz, and concludes that, starting from different premises, the researchers arrived at a similar technique: analyze the game generated by a mechanism and pass to an equivalent direct mechanism. It also notes that after the 1971 talk, the school's original English-language publications – apart from the journal's cover-to-cover translation (Automation and Remote Control) and talks at IFAC symposia and world congresses – ceased for a long time. In the introduction to the 2005 monograph, D.A. Novikov writes that the theory of active systems, the theory of hierarchical games and mechanism design arose in the late 1960s practically simultaneously and have by now practically merged into a single field whose object is organizational systems, whose subject is control mechanisms, and whose method is mathematical modeling [412]. Chronologically, the fair play principle (1969–1971) precedes the revelation principle in Gibbard's (1973) and Myerson's (1979) formulations. A synchronized presentation of the MD line, of AST and TCO, and of the schools of Yu.B. Germeier and L.A. Petrosyan is given in Fig. 2.

Beyond the two schools named above, the literature contains lines of Russian researchers adjoining the MD core. At the Central Economics and Mathematics Institute of RAS, V.I. Danilov and A.I. Sotskov developed the theory of *social choice mechanisms* in the MD formulation [164]; Danilov's paper [163] gave a necessary and sufficient condition for Nash implementability with three or more participants and belongs to the core of implementation theory (Section 3.4). This line was published in international journals and is cited in core MD terms. The axiomatic theory of preference aggregation is represented by Aleskerov's monograph [7]; it belongs to collective choice theory. Reflexive games go back to the work of V.A. Lefebvre [347] (Section 6.5), and the theory of hierarchical multilevel systems to the book by Mesarovic, Macko and Takahara [367], widely cited in the AST literature. The cooperative dynamic games of L.A. Petrosyan's school (St. Petersburg State University) are shown in Fig. 2 as an adjacent line [530, 531].

The rules by which priority is stated in this survey are uniform across all sections. The date of a result is the year of its first publication, not of its exposition in later monographs. Priorities of the statement, the result, the proof and the criterion are assessed separately. If works of the two research traditions (MD and AST) do not cite each other and are substantively independent, the result is called parallel, and “earlier” refers only to the publication date. A priority claim is accepted only after the statements have been compared

– the class of admissible mechanisms, the principal's criterion, and the participants' awareness; coincidence of names is no such ground. In Sections 7.1 and 11 the priorities are listed without repeating the arguments.

MD work on planning and intra-firm problems. In parallel with AST and independently of it, MD addressed closely related problems. Hurwicz considered informationally decentralized resource allocation [294]; Groves and Loeb studied incentives and truthful reporting by divisions within a firm [259]; Weitzman, Loeb and Magat, and Holmström analyzed the Soviet system of targets and bonuses [286, 355, 524]; and Freixas, Guesnerie and Tirole analyzed planning without commitment by the principal and the ratchet effect [214]. These papers contain no references to Burkov's school, and the 2021 survey of the AST school [102] contains none to them. For the subarea of planning and intra-firm incentives this means parallel development with incomplete mutual citation; the question of either side's priority does not arise here.

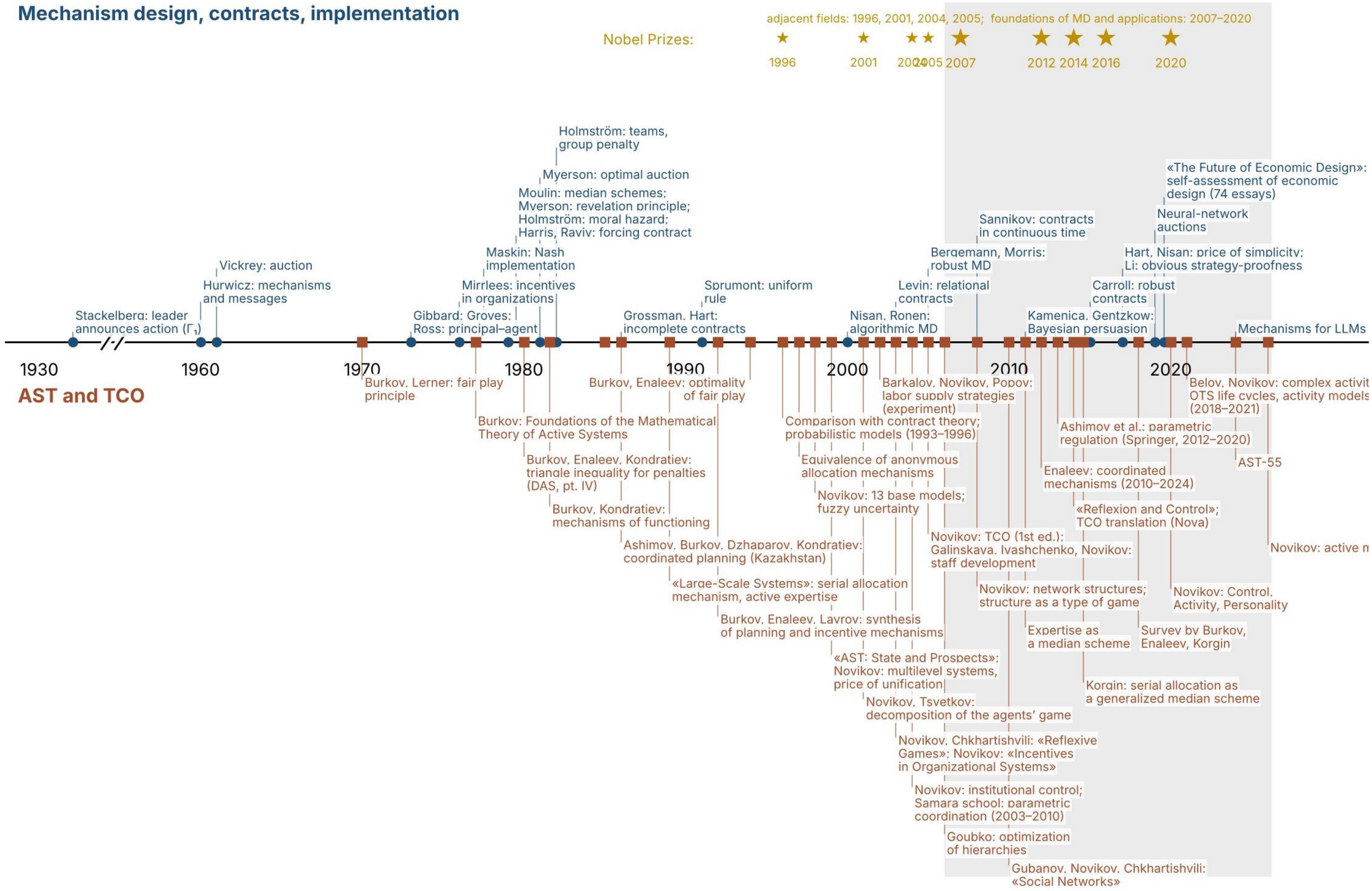
Mechanism design, contracts, implementation
adjacent fields: 1996, 2001, 2004, 2005; foundations of MD and applications: 2007–2020
Nobel Prizes:
1996
2001
2004
2005
2007
2012 2014 2016
2020
Holmström: teams, group penalty
Myerson: optimal auction
Moulin: median schemes; Myerson: revelation principle; Holmström: moral hazard; Harris, Raviv: forcing contract
Maskin: Nash implementation
Mirrlees: incentives in organizations
Gibbard; Groves; Ross: principal–agent
Vickrey: auction
Hurwicz: mechanisms and messages
Stackelberg: leader announces action (Γ1)
Grossman, Hart: incomplete contracts
Sprumont: uniform rule
Nisan, Ronen: algorithmic MD
Levin: relational contracts
Bergemann, Morris: robust MD
Sannikov: contracts in continuous time
Kamenica, Gentzkow: Bayesian persuasion
Carroll: robust contracts
Hart, Nisan: price of simplicity; Li: obvious strategy-proofness
Neural-network auctions
«The Future of Economic Design»: self-assessment of economic design (74 essays)
Mechanisms for LLMs
1930
1960
1970
1980
1990
2000
2010
2020
AST and TCO
Burkov, Lerner: fair play principle
Burkov: Foundations of the Mathematical Theory of Active Systems
Burkov, Enaleev, Kondratiev: triangle inequality for penalties (DAS, pt. IV)
Burkov, Kondratiev: mechanisms of functioning
Burkov, Enaleev: optimality of fair play
Ashimov, Burkov, Dzhaparov, Kondratiev: coordinated planning (Kazakhstan)
«Large-Scale Systems»: serial allocation mechanism, active expertise
Burkov, Enaleev, Lavrov: synthesis of planning and incentive mechanisms
Comparison with contract theory; probabilistic models (1993–1996)
Equivalence of anonymous allocation mechanisms
Novikov: 13 base models; fuzzy uncertainty
«AST: State and Prospects»; Novikov: multilevel systems, price of unification
Novikov, Tsvetkov: decomposition of the agents’ game
Novikov, Chkhartishvili: «Reflexive Games»; Novikov: «Incentives in Organizational Systems»
Novikov: institutional control; Samara school: parametric coordination (2003–2010)
Goubko: optimization of hierarchies
Gubanov, Novikov, Chkhartishvili: «Social Networks»
Barkalov, Novikov, Popov: labor supply strategies (experiment)
Novikov: TCO (1st ed.); Galinskaya, Ivashchenko, Novikov: staff development
Novikov: network structures; structure as a type of game
Expertise as a median scheme
Ashimov et al.: parametric regulation (Springer, 2012–2020)
Enaleev: coordinated mechanisms (2010–2024)
«Reflexion and Control»; TCO translation (Nova)
Korgin: serial allocation as a generalized median scheme
Survey by Burkov, Enaleev, Korgin
Belov, Novikov: complex activit
OTS life cycles, activity models (2018–2021)
AST-55
Novikov: Control, Activity, Personality
Novikov: active m

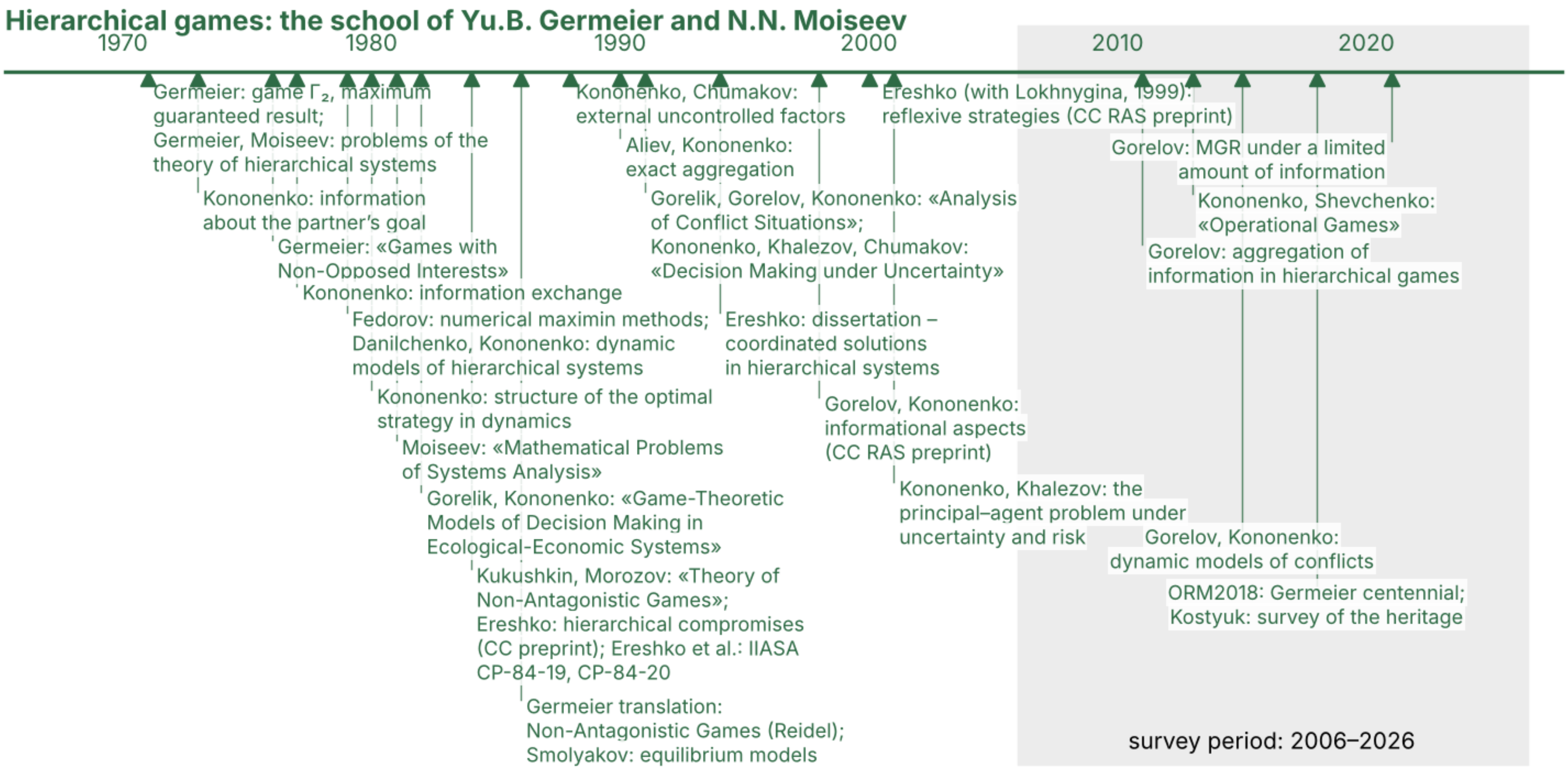
Hierarchical games: the school of Yu.B. Germeier and N.N. Moiseev
1970
1980
1990
2000
2010
2020
Germeier: game Γ₂, maximum guaranteed result;
Germeier, Moiseev: problems of the theory of hierarchical systems
Kononenko: information about the partner’s goal
Germeier: «Games with Non-Opposed Interests»
Kononenko: information exchange
Fedorov: numerical maximin methods;
Danilchenko, Kononenko: dynamic models of hierarchical systems
Kononenko: structure of the optimal strategy in dynamics
Moiseev: «Mathematical Problems of Systems Analysis»
Gorelik, Kononenko: «Game-Theoretic Models of Decision Making in Ecological-Economic Systems»
Kukushkin, Morozov: «Theory of Non-Antagonistic Games»;
Ereshko: hierarchical compromises (CC preprint); Ereshko et al.: IIASA CP-84-19, CP-84-20
Germeier translation: Non-Antagonistic Games (Reidel);
Smolyakov: equilibrium models
Kononenko, Chumakov: external uncontrolled factors
Aliev, Kononenko: exact aggregation
Gorelik, Gorelov, Kononenko: «Analysis of Conflict Situations»;
Kononenko, Khalezov, Chumakov: «Decision Making under Uncertainty»
Ereshko: dissertation – coordinated solutions in hierarchical systems
Gorelov, Kononenko: informational aspects (CC RAS preprint)
Kononenko, Khalezov: the principal–agent problem under uncertainty and risk
Ereshko (with Lokhnygina, 1999): reflexive strategies (CC RAS preprint)
Gorelov: MGR under a limited amount of information
Kononenko, Shevchenko: «Operational Games»
Gorelov: aggregation of information in hierarchical games
Gorelov, Kononenko: dynamic models of conflicts
ORM2018: Germeier centennial;
Kostyuk: survey of the heritage
survey period: 2006–2026

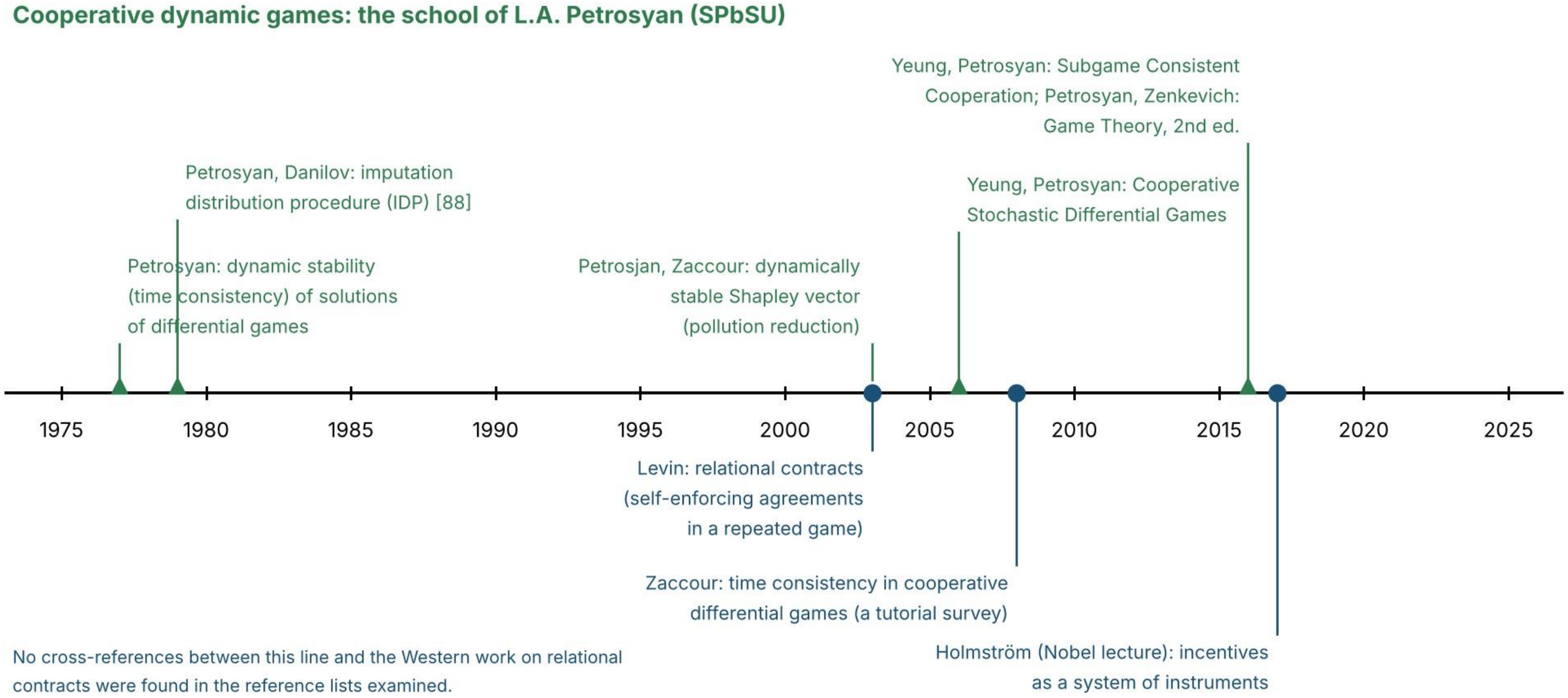

Fig. 2. Chronology of the research traditions, 1934–2026: key results, monographs and Nobel Prizes
(top – the MD line; bottom – AST and TCO; in the other panels below – the schools of Germeier and Petrosyan)

## 2.3. A Dictionary of Correspondences

The terminology of the two research traditions (MD and AST) differs more than their mathematics does. Table 1 fixes the correspondences used below. The Russian rendering of the term revelation principle as «принцип откровенности» was proposed in [102]; earlier works of the AST school used the variant «принцип выявления» [405].

Table 1. A dictionary of correspondences between the basic notions

| AST and TCO | Mechanism design, contract theory | Comment |
|---|---|---|
| Principal; active element (agent) | Principal, designer; agent | In AST the principal is a participant with a goal function of its own |
| Active system, organizational system | Organization | In TCO the model is specified by six components (Section 6.1) |
| Mechanism of functioning, control mechanism | Mechanism, game form | In TCO a control mechanism is a set of procedures for making management decisions [412] |
| Planning procedure (mechanism) $\pi(s)$ | Outcome function $g(m)$, social choice function | The plan $x_i$ is the component of the outcome assigned to an agent |
| Message, bid, counter plan $s_i$ | Message, report, bid $m_i$ | |
| Truthful revelation of information; nonmanipulability | Truth-telling; strategy-proofness, dominant-strategy incentive compatibility (DSIC) | |
| Fair play (open control) principle | Revelation principle; taxation principle | The exact relation is given in Section 3.1 |
| Perfect coordination condition | Characterization of DSIC via option sets | Necessary and sufficient for truth-telling [98] |
| Coordinated mechanism (plans are fulfilled) | Obedience; incentive compatibility in actions | |
| Proper mechanism (coordinated and strategy-proof) | Incentive compatible direct mechanism | |
| Equivalent direct mechanism | Direct revelation mechanism | |
| Incentive scheme $\sigma(y)$ | Contract, wage schedule $w(q)$ | Basic classes: step-type, compensatory, proportional [120] |
| Incentive problem | Moral hazard, hidden action | In the deterministic version, the action is observable |
| Mechanism with information revelation | Adverse selection, screening | |
| Guaranteed efficiency; efficiency under the benevolence hypothesis | Unique (full) implementation; partial implementation | See (1) and (3) |
| Weak influence hypothesis | Price-taking in large economies [456]; justified by strategy-proofness in the large | [112, 33] |
| Dictators (agents who receive their optimal plan) | Positional dictators, unconstrained agents | [102, 328] |
| Serial resource allocation mechanism | Uniform rule, sequential allotment rules | [42, 486] |
| Active expertise mechanism | Generalized median voter scheme | [110, 381] |
| Nontransferable and transferable utility (NTU, TU) | Mechanisms without money; quasi-linear utility | |
| Informational control | Information design, Bayesian persuasion | The formal models differ (Section 6.5) |
| Institutional control | Delegation, rules, restriction of action sets | |

| AST and TCO | Mechanism design, contract theory | Comment |
|---|---|---|
| Structure control | Organization design, hierarchies | |
| Integrated mechanism | Composition of mechanisms | Composition is studied in [114, 489] |
| Unified incentive scheme; price of unification | Simple, anonymous mechanism; price of simplicity | Introduced in [408] (Section 4.4) |
| Dynamic stability of a cooperative solution (Petrosyan's school) | Time consistency; self-enforcing agreement | Outside TCO; see Section 5.3 |
| Distribution of farsightedness; degree of farsightedness; decision horizon | Discounting; myopic and far-sighted agents; planning horizon, limited foresight | Formulas (57), (58); the planning horizon in the theory of simplicity [455] |
| Control regimes: current, rolling (with and without commitment), program-based | Short-term contracts; limited commitment; full commitment | Condition (60) (Section 6.8) |
| History-dependent incentives | History-dependent long-term contract | System (59) |

## 2.4. Methodological Differences

The differences in the underlying premises are collected in Table 2. Three of them are central.

**First** – the description of uncertainty: a prior distribution and an expected criterion in classical Bayesian MD versus a set of possible values and a guaranteed result in AST. In this respect AST is closer to robust MD [64] and related strands than to the classical Bayesian theory. Individual results with a guaranteed criterion have been known in MD since the 1970s: Hurwicz and Shapiro [296] maximize the worst-case (over an unknown cost parameter) ratio of the principal's payoff to the full-information payoff and obtain the optimality of an equal split of output; sets of distributions, beliefs and actions have been used systematically in the theory since the 2000s: prior-free competitive auctions [235], ambiguity among participants [79, 173, 529], robust design [64] and robust contracts [136].

The taxonomy of uncertainty in AST is broader than in MD: the monograph [407] classifies the basic incentive models by the type of uncertainty (internal – about the agent's parameters; external – about the state of nature), its kind (interval, probabilistic, fuzzy) and the awareness pattern (symmetric, asymmetric) – thirteen models M1–M13 in all; contract theory, in the book author's assessment, concentrates on one of them (external probabilistic uncertainty under symmetric awareness, i.e. moral hazard); this assessment refers to the problem of incentivizing an agent's effort, while the screening and optimal-auction models [50, 384] correspond to internal probabilistic uncertainty under asymmetric awareness, so MD as a whole covers at least two models of this taxonomy. Fuzzy uncertainty does not appear in the known MD and contract theory sources (Section 3.5).

**Second** – the status of a mechanism: in MD it is an object of theoretical comparison of institutions or (in market design) an engineering project for a particular market; in AST and TCO it is an element of an organization's control system that must fit together with the other mechanisms.

**Third** – the range of control variables: MD designs the game form for given participants, action sets and information, whereas TCO includes among its controls the set of participants, the structure, the admissible sets and the awareness.

Table 2. Methodological differences between the research traditions

| Attribute | Mechanism design, contract theory, implementation theory | AST and TCO |
|---|---|---|
| Origin | Economic theory, social choice; since 1999 also theoretical computer science | Control theory, operations research, cybernetics |
| Basic object | Market, auction, voting, contract, platform | An organization: a principal and subordinate agents |
| Uncertainty (for the principal) | A prior probability distribution of agents' types (the Bayesian approach); since 2005, also sets of distributions and beliefs | A set of parameter values; interval, probabilistic and fuzzy uncertainty as variants |
| Criterion | Expected revenue or welfare; implementation of a given choice rule | The value of the principal's goal function on the set of solutions of the agents' game – guaranteed or expected, depending on the kind of uncertainty. |
| Solution concept | Dominant strategies, Nash equilibrium and its refinements, Bayesian equilibrium, rationalizability | Dominant strategies and Nash equilibrium; informational and reflexive equilibrium |
| What is controlled | The game form (messages, outcome, payments) – the goal function; in information design, the signal structure | Staff, structure, admissible sets, goal functions, awareness, order of functioning |
| Type of result | Characterizations, impossibility theorems, optimal mechanisms, approximation and complexity bounds | Optimal mechanisms in parametric classes, equivalence of mechanisms, a catalog of standard solutions |
| Validation | Laboratory and field experiments, structural econometrics | Business games, simulation, implementation experience |
| Main publication venues | Econometrica, AER, REStud, JET, TE, GEB; EC, STOC, FOCS | Automation and Remote Control, Control Sciences, Large-Scale Systems Control. Monographs. |

# 3. The Classical Cores of MD and AST and Their Correspondence

## 3.1. The Fair Play Principle and the Revelation Principle

**The revelation principle.** Suppose the mechanism $\Gamma = \langle M, g \rangle$ has an equilibrium $m^*(\theta)$ in dominant strategies (or a Bayesian equilibrium). Then the direct mechanism $f(\theta) = g\big(m^*(\theta)\big)$ satisfies condition (2) (respectively, its Bayesian version), i.e. truthful reporting is an equilibrium of it [165, 232, 383]. The principle reduces the search over all game forms to a search over incentive compatible direct mechanisms and underlies all optimization-type statements of MD. It guarantees the existence of a truthful equilibrium but not its uniqueness; eliminating the other equilibria is the subject of implementation theory (Section 3.4). The result holds if the equilibrium has the form $m^*(\theta) = \big(m_1^*(\theta_1),\ \dots,\ m_i^*(\theta_i),\ \dots,\ m_n^*(\theta_n)\big)$, which is the case after each agent computes the conditional expectation over the other agents' types given its own (i.e. when expected utility is the criterion and/or Bayesian strategies are used). In the latter case the problem splits into *n* single-agent problems, and in a system with a single agent an equivalent strategy-proof mechanism always exists [97, 87]. In the general case, however, $m^*(\theta) = \big(m_1^*(\theta),\ \dots,\ m_i^*(\theta),\ \dots,\ m_n^*(\theta)\big)$ – each agent's equilibrium message depends on the types of all agents – and the equilibrium property of truthful reports requires separate study and justification, i.e. it does not always hold.

**Theorem (Gibbard, Satterthwaite).** *If the outcome set $X$ is finite and contains at least three elements, all strict preferences are admissible, and the choice rule $f$ takes all values in $X$, then $f$ is strategy-proof if and only if it is dictatorial* [232, 359, 477].

All substantive strategy-proof mechanisms therefore exist only on restricted preference domains: single-peaked (Section 3.2), quasi-linear (Section 3.3) and some others [40].

**The fair play principle.** In the AST model the principal does not know the parameters $r_i \in \Omega_i$, the agents report estimates $s_i \in \Omega_i$, the principal assigns plans $x = \pi(s)$, and the agent then chooses a state $y_i$ maximizing $f_i(x_i, y_i, r_i)$. The agent's preference function over the set of plans is $\varphi_i(x_i, r_i) = \max_{y_i} f_i(x_i, y_i, r_i)$. Burkov's idea is to assign each agent the plan that is best from the standpoint of the agent's own preference function evaluated at the estimate it reported; then there is no reason to distort information. The exact formulation uses the perfect coordination condition (PCC): there exist sets $X_i(s_{-i})$, independent of the agent's own message $i$, such that the plan assigned to the agent belongs to this set and is the best element of it:

$$\pi_i(s) \in \operatorname{Arg}\max_{z_i \in X_i(s_{-i})} \varphi_i(z_i, s_i) \qquad \forall\, i \in N, \forall\, s \in \Omega. \tag{4}$$

The fair play law chooses the plan best for the principal among the plans satisfying the PCC:

$$\pi^{\mathrm{FP}}(s) \in \operatorname{Argmax}_{x} \left\{ \Phi(x, s) \colon x_i \in \operatorname{Arg}\max_{z_i \in X_i(s_{-i})} \varphi_i(z_i, s_i),\, i \in N \right\}. \tag{5}$$

**Theorem (Burkov, Enaleev, 1985).** *(a) Truthful reporting is a dominant strategy of every agent for all $r \in \Omega$ if and only if there exist sets $X_i(s_{-i})$ for which* (4) *holds. (b) In the class of mechanisms with a given incentive scheme and the criterion of efficiency guaranteed over $r \in \Omega$, there exists an optimal mechanism whose planning procedure is the fair play law* (5) [98].

In the books of the AST school these two statements are kept separate. Item (a) is given as the fair play principle proper (in the course [424] it is Theorem 5.1; in the proof of necessity, $X_i(s_{-i})$ is taken to be the set of plans reachable by changing the agent's own message), and the uniqueness of the truthful equilibrium is secured by the additional condition of equitable preference functions [120]. Optimality is stated more cautiously than in item (b). In a single-agent system, for any planning mechanism there exists a fair play mechanism of no lower efficiency ([97], Theorem 5.2 in [424]). The same is true in the Bayesian model, where an agent's equilibrium message depends only on its own type and the problem reduces to the single-agent one (Theorem 5.10). In multi-agent systems the result has been proved for particular classes of mechanisms – resource allocation, active expertise, transfer prices – while in the general case, under Nash equilibrium, fair play mechanisms may be suboptimal, since each agent's equilibrium message depends on the others' types; the authors call the relation between optimality and strategy-proofness in the general case an open question [120, 424].

The relation to the MD results is as follows. Item (a) is a characterization of dominant-strategy incentive compatibility via option sets: a direct mechanism is strategy-proof exactly when each agent is effectively offered a menu $X_i(s_{-i})$ independent of its own message and receives the element of the menu best for itself. In the MD literature this statement is known as the *taxation principle* [263, 267, 458]; for large economies the characterization via a menu independent of the agent's own characteristics was given by Hammond [267] (see also [174]); for a finite number of agents and dominant strategies, as in AST, the menu depends on the other agents' messages; it underlies, for example, the modern neural-network architectures for optimal auctions that represent a mechanism as a menu [181]. Item (b) and Theorems 5.2 and 5.10 are the analogue of the revelation principle for the optimization statement: one may restrict attention to strategy-proof direct mechanisms without loss of efficiency; the AST school itself relates the fair play principle to the revelation principle [102]. The differences lie in the criterion (a result guaranteed over the unknown parameters instead of an expectation), in the narrower domain of proved optimality, and in the fact that the AST mechanism includes the agent's action after receiving the plan; a mechanism in which the plan is both fulfilled and the information is reported truthfully is called *proper* by the AST school [120, 405].

For the case where a system-wide control parameter $\lambda(s)$ (a price, a norm) depends weakly on an individual agent's message, AST adopts the *weak influence hypothesis*: agents treat $\lambda$ as independent of their own messages [98, 112]. Under this hypothesis it suffices to coordinate the individual components of the plan: if the perfect coordination condition holds for them, truthful reporting is a dominant strategy [120]; the transfer-price mechanisms (Section 6.1) and a number of other mechanisms with a common control variable become strategy-proof. The conditions for the hypothesis to hold for a large number of agents are attributed in the book [120] to the 1977 monograph; in the 2021 survey the authors assess models with this hypothesis as “rather degenerate” [102]. The MD parallel is price-taking behavior in large economies. Roberts and Postlewaite [456] showed that reporting competitive demand is generally not a best response with finitely many participants, but an individual participant's gain from monopolistic behavior tends to zero as the economy is replicated; Hammond's paper [267] is close in subject. By publication date these works precede [98, 112] by one to nine years, but the results are parallel: there are no mutual references. Much later the notion of *strategy-proofness in the large* was introduced: truthfulness is required only approximately, against any full-support distribution of the others' reports, and in the limit of a large number of participants [33]. The weak influence hypothesis is an assumption about agents' behavior, whereas the results of [33, 456] are provable asymptotic properties of a mechanism. Rates of decay of the gain for AST mechanisms with a common control variable have not yet been obtained (problem 4 in Table 15).

## 3.2. Strategy-Proofness under Single-Peaked Preferences: Active Expertise and Resource Allocation

**Median schemes.** Let the outcome be a point of the interval $[d, D]$, and let agent $i$'s preferences be single-peaked with peak $\tau_i$. Moulin showed that an anonymous efficient rule is strategy-proof if and only if it is the median of the agents' peaks and $n-1$ fixed numbers (“phantom voters”) [381, 526]:

$$f(\tau) = \text{med}(\tau_1, \dots, \tau_n, \alpha_1, \dots, \alpha_{n-1}), \qquad \alpha_k \in [d, D]. \tag{6}$$

Without the efficiency requirement there are $n+1$ phantoms; without anonymity one obtains a minimax representation via numbers $a_S$ assigned to all possible coalitions. The last two statements assume that the rule depends only on the peaks; for strategy-proof rules this is equivalent to the rule's range being an interval, and for efficient rules it holds automatically [526]. The multidimensional generalizations are coordinatewise medians on boxes [78] and generalized median schemes defined by systems of coalitions [41, 40].

**Active expertise.** In AST the same problem arose as a problem of expert assessment [97, 110, 424]. The experts report estimates $s_i \in [d, D]$, and the final decision $x = \pi(s)$ is determined by a procedure that is strictly monotone, continuous and satisfies the unanimity condition $\pi(a, \dots, a) = a$ (the arithmetic mean, for example). Such a procedure is manipulable: an expert whose peak is above the result benefits from inflating its estimate. The AST school constructs from $\pi$ an equivalent direct mechanism. Let $\tau_1 \le \dots \le \tau_n$ and

$$w_k = \pi\left(\underbrace{d,\dots,d}_{k},\underbrace{D,\dots,D}_{n-k}\right),\quad k=0,\dots,n;\qquad x^*(\tau)=\max_{1\le i\le n}\min\{w_{i-1},\tau_i\}. \tag{7}$$

Then $x^*(\tau)$ is the result of the expertise at a Nash equilibrium of the original game, and the direct mechanism $x^*(\cdot)$ is strategy-proof (Theorems 5.4 and 5.5 in [424]). The formula shows that an expert whose peak lies between $w_i$ and $w_{i-1}$ determines the result single-handedly; in the AST school's terms, it is a "*dictator*" on this interval. The right-hand side of (7) equals $\mathrm{med}(\tau_1,\dots,\tau_n,w_1,\dots,w_{n-1})$, i.e. it is scheme (6) with phantoms $\alpha_k = w_k$; for the arithmetic mean, $w_k = (kd + (n-k)D)/n$. In the paper by Burkov, Iskakov and Korgin [110] this construction is carried over to multicriteria expertise: for a procedure on a multidimensional box an equivalent direct mechanism exists if and only if the procedure is separable across criteria, while on a budget set $\{x \ge 0: \sum_j x_j \le C\}$ the coordinatewise median may leave the set.

**Resource allocation.** The principal allocates a resource $R$ among agents with single-peaked preferences over their own shares; the peaks $\tau_i$ are known only to the agents, $\sum_i \tau_i > R$. AST has studied the class of *priority-based mechanisms* [97, 328]:

$$\pi_i(s)=\begin{cases} s_i, & \sum_j s_j \le R,\\ \min\{s_i,\gamma\,\eta_i(s_i)\}, & \sum_j s_j > R,\end{cases}\qquad \gamma\colon \sum_i \pi_i(s) = R, \tag{8}$$

where $\eta_i$ is the priority function: increasing in the bid (straight priorities, in particular proportional allocation $\eta_i = s_i$), decreasing (reverse priorities), or constant (absolute priorities). The classical result of the AST school is stated under four assumptions [424]: the procedure $\pi_i$ is continuous and strictly monotone in the agent's bid; the resource is allocated in full; any agent can receive any amount of the resource smaller than what it has already received; and when the amount to be allocated grows, nobody receives less. Then, at a Nash equilibrium, an agent receiving less than its peak bids the maximum, and an agent whose bid is below the maximum receives exactly its peak. The equilibrium is constructed in at most $n$ steps by a procedure that at every step identifies the agents who receive their peaks ("dictators", or priority consumers), and the corresponding direct mechanism $h(r) = \pi(s^*(r))$ is a fair play mechanism, i.e. strategy-proof and at least as efficient as the original one (Theorem 5.3). The book [120] attributes this result to the cycle of the AST school's works beginning with the monograph [87]; an exposition is given in [97]. All anonymous mechanisms of this class are equivalent to one another and to the serial resource allocation mechanism (SRAM); this result of the book [105] is cited here after the exposition in [328], which also covers the non-anonymous case. In the anonymous case with $\tau_1 \le \dots \le \tau_n$, the SRAM has the form

$$x_i = \min\left\{\tau_i,\ \frac{R-\sum_{j<i} x_j}{n-i+1}\right\},\qquad i=1,\dots,n. \tag{9}$$

In the West the same mechanism is known as the *uniform rule*: when the resource is scarce, $x_i = \min\{\tau_i,\lambda\}$; when it is abundant, $x_i = \max\{\tau_i,\lambda\}$, where $\lambda$ is determined by the balance condition $\sum_i x_i = R$.

**Theorem (Sprumont, 1991; Ching, 1994).** *The uniform rule is the unique division rule that is simultaneously strategy-proof, Pareto efficient and anonymous; anonymity can be weakened to equal treatment of equals* [143, 486].

The non-anonymous case is described by the sequential allotment rules of Barberà, Jackson and Neme: a rule is strategy-proof, efficient and replacement monotonic if and only if it is sequential; when the resource is scarce it has the form $x_i = \min\{\tau_i, b_i(\tau_{-i})\}$ with a consistent system of guaranteed shares [42]. Korgin gave an analytic representation for priority-based mechanisms [328]: the straight-priority mechanism is equivalent to the SRAM

$$x_i = \min\left\{\tau_i, \max_{S \subseteq N:\, i \in S} \frac{\eta_i(R)}{\sum_{j \in S} \eta_j(R)} \left(R - \sum_{j \in N \setminus S} \tau_j\right)\right\}, \tag{10}$$

any straight-priority mechanism is equivalent to a weighted proportional one, while for a reverse-priority mechanism an equivalent straight-priority mechanism does not always exist. The power-law reverse-priority mechanism $\eta_i = B_i s_i^{\alpha}$, $\alpha < 0$, is equivalent to the proportional one with weights $B_i^{1/(1-\alpha)}$; for $\alpha = -1$ and in the absence of dictators one obtains the equilibrium allocation $x_i^* = R\sqrt{B_i}/\sum_j \sqrt{B_j}$; the reverse-priority mechanism itself is already described in [120]. The paper [327] shows that any SRAM can be represented as a generalized median scheme on an extended class of multidimensional preferences; for the anonymous SRAM the phantom agents have peaks $R/n, R/(n-1), \dots, R/2$. The two basic planning mechanisms of AST thereby receive a unified description in terms of generalized median schemes.

*Anonymity and decentralization.* A mechanism is called *anonymous* if it is symmetric under permutations of the agents; anonymity does not presuppose that the agents are identical and requires only the symmetry of the planning procedure [408]. The equivalence of all anonymous mechanisms of class (8) was established in [105]; the monograph [408] derives from it that any anonymous resource allocation mechanism is equivalent to the proportional one (Lemma 2.3.2) and therefore admits arbitrary decentralization: introducing any number of intermediate hierarchy levels and any partition of the agents into subsystems, with bids summed within subsystems, leaves the equilibrium allocation unchanged (Theorem 2.3.1). Reverse-priority mechanisms do not admit arbitrary decentralization, but they do admit perfect aggregation if priority and non-priority agents never end up in the same subsystem (Example 2.3.1 ibid.). Analogous theorems are proved for expertise mechanisms, each of which is equivalent to a linear one (Theorem 2.4.1 and Lemma 2.4.2), for fair play mechanisms with transfer prices under Cobb–Douglas costs (Theorem 2.5.1), and for insurance mechanisms (Theorem 2.6.1). In the axiomatics of the uniform and sequential rules [42, 486] anonymity also serves as an axiom, but the question of whether the rule's properties survive the introduction of intermediate levels is not posed there; a related question – how the rule's recommendation changes when some agents leave with their shares – is posed by the consistency axiom in fair division [499]. The MD analogues have been studied for a different setting: the theory of hierarchical contracting asks when delegating the contracting to a strategic intermediary does not reduce the principal's payoff [364, 363, 377] and how this changes under collusion [37, 339, 501]. The AST theorems on the invariance of the equilibrium allocation under an arbitrary partition of agents into subsystems with aggregated bids concern a different subject – the invariance of a class of rules – and no direct MD analogue of them has been found. The role of an intermediate tier as such was raised in AST earlier as well: the paper [100] is devoted to the price of decentralizing mechanisms of functioning.

The correspondence of results and dates is collected in Table 3. In the expertise problem the MD result was published earlier, and the AST school acknowledges this [102, 110]. The uniform rule as a rationing scheme was introduced by Benassy [130], and Sprumont (1991) owns its characterization via strategy-proofness, Pareto efficiency and anonymity. In the AST school, the equivalent strategy-proof direct mechanism for priority-based resource allocation mechanisms was constructed in the 1989 monograph [97], before Sprumont's publication but after Benassy introduced the rule, and the identity of the anonymous case with the uniform rule was noted in the AST literature later [105, 328]. The AST school's priority thus concerns the construction of the direct mechanism and the proof of its strategy-proofness for the whole class of priority-based mechanisms; the rule itself and its axiomatics were obtained in MD. In his papers of 2009–2014 Korgin explicitly relates his theorems to the results of Sprumont and of Barberà–Jackson–Neme.

### 3.3. Transferable Utility: Groves Mechanisms, Optimal Auctions, Impossibility

Under quasi-linear preferences $u_i = v_i(x, \theta_i) - t_i$, the efficient decision $x^*(\theta) \in \mathrm{Argmax}_x \sum_i v_i(x, \theta_i)$ is implemented in dominant strategies by the Vickrey–Clarke–Groves (VCG) mechanisms [150, 257, 517]:

$$t_i(\theta) = h_i(\theta_{-i}) - \sum_{j \neq i} v_j\left(x^*(\theta), \theta_j\right), \qquad h_i^{\mathrm{Clarke}}(\theta_{-i}) = \max_x \sum_{j \neq i} v_j\left(x, \theta_j\right). \tag{11}$$

Here $t_i$ is the agent's payment. On sufficiently rich domains there are no other efficient strategy-proof mechanisms [252], and by Roberts' theorem [457], for a finite outcome set with at least three elements and an unrestricted domain of quasi-linear preferences, the only onto rules implementable in dominant strategies are affine maximizers $f(\theta) \in \mathrm{Argmax}_x\left(\sum_i k_i\, v_i(x, \theta_i) + C(x)\right)$ with nonnegative weights $k_i$ not all equal to zero; the authors of [102] note that this theorem is rarely mentioned in the classical mechanism design literature, whereas in algorithmic mechanism design it is central: Lavi, Mu'alem and Nisan generalized it to restricted domains, including combinatorial auctions [342], and gave simplified proofs [343]. VCG mechanisms are, generally speaking, not budget-balanced. In the Bayesian setting balance is attainable by the expected externality mechanism [160], constructed independently also by Arrow [16], but, generally, not simultaneously with voluntary participation.

**Theorem (Myerson, Satterthwaite, 1983).** *In the bilateral trade problem with independent private valuations distributed with positive densities on intervals with overlapping interiors, there is no mechanism that is simultaneously Bayesian incentive compatible, individually rational, budget-balanced and ex post efficient* [387].

Revenue optimization was solved by Myerson [384]. For a single object and independent valuations $v_i \sim F_i$ with densities $f_i$, the expected revenue of any incentive compatible mechanism with allocation probabilities $q_i(v)$ equals

$$\mathbb{E}\left[\sum_i \psi_i\,(v_i)\; q_i(v)\right] - \sum_i U_i\left(\underline{v}_i\right), \qquad \psi_i(v_i) = v_i - \frac{1 - F_i(v_i)}{f_i(v_i)}, \tag{12}$$

where $U_i\left(\underline{v}_i\right)$ is the expected payoff of the participant with the lowest valuation, and in the regular case (the virtual valuations $\psi_i$ are increasing) the optimal auction gives the object to the participant with the highest nonnegative virtual valuation. A corollary is the revenue equivalence of the standard auctions with symmetric participants. With correlated valuations the principal can, as a rule, extract the entire surplus [158], which is commonly taken as evidence that Bayesian mechanisms depend excessively on the details of the distribution and as one of the motivations for the robust approach (Section 4.1).

In AST, systems with transferable utility received less attention; the survey [102] counts the VCG mechanisms among them and raises the question of transferring the "positional dictatorship" schemes to them. An original AST application is the Groves–Ledyard-type mechanism [258] for resource allocation [329]: agent $i$ reports a vector $s_i = (s_{1i}, \dots, s_{ni})$, $\sum_j s_{ji} \le R$ – its proposal for allocating the entire resource; the allocation and the penalty for inconsistency equal

$$x_i(s) = \frac{1}{n}\sum_{j\in N} s_{ij}, \qquad p_i(s) = \gamma \sum_{j\in N} (x_j(s) - s_{ji})^2, \quad \gamma > 0, \tag{13}$$

and the transfers are built from the penalties so that they sum to zero. The game has a unique Nash equilibrium, which implements the allocation maximizing $\sum_i u_i\,(x_i)$; budget balance is achieved at the cost of giving up dominant strategies. A laboratory test of this mechanism is described in Section 6.2.

## 3.4. Nash Implementation and Equivalent Direct Mechanisms

If the agents know each other's types but the principal does not, the natural solution concept is Nash equilibrium. A rule $F$ is Maskin monotonic if $x \in F(\theta)$, together with the fact that in passing to $\theta'$ the outcome $x$ does not "fall in the preferences" of any agent relative to any other outcome, implies $x \in F(\theta')$:

$$\begin{aligned} x \in F(\theta), \quad [u_i(x,\theta) \ge u_i(y,\theta) \Rightarrow u_i(x,\theta') \ge u_i(y,\theta')] \ \ \forall\, i, y \\ \Rightarrow \quad x \in F(\theta'). \end{aligned} \tag{14}$$

**Theorem (Maskin).** *If $F$ is fully Nash implementable, then it is monotonic. If $n \ge 3$, $F$ is monotonic and satisfies no veto power, then it is fully Nash implementable* [358, 359].

A complete characterization of Nash implementability for $n \ge 3$ was given by Danilov [163] via the essential monotonicity condition: it is necessary and sufficient, closing the gap between Maskin's necessary and sufficient conditions; with two participants the condition must be supplemented. The monograph by Danilov and Sotskov [164] presents implementation theory from the social choice standpoint and is the main Russian source on implementation theory.

Nonmonotonic rules are implemented under equilibrium refinements: subgame perfect equilibrium in multi-stage mechanisms [379], virtual (approximate) implementation in iteratively undominated strategies [2]; Bayesian implementation requires Bayesian monotonicity [303]. Surveys of the classical theory are given in [304, 359, 480]. Implementation theory has also entered the teaching texts of the AST school: the course [424] contains the definition of Maskin monotonicity, the no-veto-power condition and both parts of Maskin's theorem (Theorems 5.8 and 5.9).

In AST the question is posed in the reverse direction: a mechanism $\pi$ is given, and a direct mechanism equivalent to it in outcomes is sought. For single-peaked preferences and a monotone continuous procedure, the structure of Nash equilibria is simple: if the agent's plan is below its peak it bids the maximum, if above – the minimum, while an interior bid means the agent receives its peak. This yields a general result.

**Theorem (Novikov, 1997).** *For any planning mechanism $\pi(\cdot)$ of the class indicated there exists an equivalent direct mechanism $g(r) = \pi\big(s^*(r)\big)$ in which truthful reporting is a Nash equilibrium; mechanisms* (7) *and* (9) *are special cases of it* [405].

The link between the concepts is given by the result of Dasgupta, Hammond and Maskin [165]: if the preference domain is *rich* (by their definition, the domain of all single-peaked orderings on a finite ordered set of alternatives is rich, in particular), then a single-valued choice rule that is fully Nash implementable is also implementable in dominant strategies by a direct mechanism. Consequently, on such domains the passage from dominant strategies to Nash equilibrium does not enlarge the class of single-valued rules, and the equivalent direct mechanism built from the Nash equilibrium (Novikov's theorem [405]) is strategy-proof. The converse is false in general: in the direct mechanism, a strategy-proof rule (a median scheme, say) may also have untruthful Nash equilibria with a different outcome; the requirement that a mechanism implement the rule simultaneously in dominant strategies and in Nash equilibrium (so-called *secure implementation* [473]) is substantially stronger. For multi-valued rules, non-rich domains (in particular, narrow parametric families of preference functions, where richness must be verified separately) and multiple equilibria the conclusion does not carry over [102]. For fair division problems – under single-peaked preferences – see [381, 500].

## 3.5. Incentives and Contracts

**The deterministic incentive problem.** An agent chooses an action $y \in A$ and bears convex *costs* $c(y)$, $c(0) = 0$; the principal observes the *action*, receives a concave or linear *income* $H(y)$ and pays a nonnegative reward (*incentive*) $\sigma(y)$: $\Phi = H(y) - \sigma(y)$, $f = \sigma(y) - c(y)$. The AST school's solution is the *cost compensation principle*: the principal's minimum cost of implementing action $x$ equals the agent's costs $c(x)$, and the optimal incentive scheme is the quasi-compensatory one (QK-type) [120, 228, 424]

$$\sigma_{QK}(x, y) = \begin{cases} c(x), & y = x, \\ 0, & y \neq x, \end{cases} \qquad x^* \in \underset{x \in A}{\operatorname{Argmax}}[H(x) - c(x)]. \tag{15}$$

In substance, when the plan is fulfilled the agent's costs are compensated (it is rewarded); when it is not, the agent is punished (the reward is minimal – zero), i.e. (15) is a special case of Germeier's theorem (1971, see below). Scheme (15) is optimal under the *benevolence hypothesis*: the agent is indifferent between the plan and the zero action and chooses the plan. If benevolence cannot be relied upon, an arbitrarily small bonus $\delta > 0$ is added to the payment; the plan becomes the agent's unique best choice, and the principal's loss equals $\delta$ [428]. The AST school reserves the term compensatory proper (K-type) for the scheme $\sigma_K(y) = c(y)$; alongside it, the *step-type* (C-type) and *proportional* (L-type) schemes are considered basic, and the general conclusion of the book [120] for models with uncertainty is that the "boundary" C- and K-type schemes are optimal, while proportional ones do not dominate them in the setting considered.

For $n$ agents with nonseparable costs $c_i(y)$ depending on everyone's actions, a stronger result has been obtained.

**Theorem (Novikov, Tsvetkov, 2000).** *The incentive scheme*

$$\sigma_i(y^*, y) = \begin{cases} c_i(y_i^*, y_{-i}) + \delta_i, & y_i = y_i^*, \\ 0, & y_i \neq y_i^*, \end{cases} \tag{16}$$

*with bonuses* $\delta_i \geq 0$ *implements the vector* $y^*$ *as a dominant-strategy equilibrium, unique for* $\delta_i > 0$. *The class of such schemes is* $\delta$*-optimal for* $\delta = \sum_i \delta_i$*: the principal's minimum cost of implementing* $y^*$ *equals* $\sum_i c_i\,(y^*)$*, and the optimal plan maximizes* $H(y) - \sum_i c_i\,(y)$ [429, 428].

The result is called the *principle of decomposition of the agents' game*: by compensating costs whatever the others do, the principal splits the game into independent problems (Theorems 4.2.1a and 4.2.2 in [428]). If an agent's reward depends only on its own action, $\sigma_i = c_i(y^*) + \delta_i$ with $y_i = y_i^*$, the vector $y^*$ is only a Nash equilibrium, and its uniqueness and dominance require additional conditions on the cost functions (Theorem 4.2.1b).

Robustness to coalitions. Goubko [244] tested the incentive schemes of [428] when the agents can collude with side payments: if the incentive depends on the whole action vector, fulfilling the plan is a dominant strategy not only of every agent but of every coalition; if it depends only on the agent's own action, the plan is implemented by every coalition structure only under a system of inequalities on the costs, and the minimum extra payments guaranteeing the plan are found from a linear program whose number of constraints is exponential in the number of agents; if the principal observes only the aggregated result, coalitions can redistribute actions only within its isoquants, which the principal cannot detect. In MD the same question is posed as coalition-proofness of the equilibrium [72]; for incentives in teams, no direct MD analogue of this result has been found.

The same work considers the case where the principal observes only the aggregated result $z = Q(y)$. With separable costs, the scheme that pays each agent its costs under the cheapest way of producing the result $x$ whenever result $z = x$ is achieved, and zero otherwise, makes such cheapest ways Nash equilibria and implements $x$ at minimum cost to the principal (Theorem 4.5.1; because of the equilibrium with zero actions the scheme is in general only $\varepsilon$-optimal); for nonseparable costs an analogous result is obtained under a monotonicity condition (Theorem 4.6.1). If the principal's income depends only on the result and the cost functions are known to it, aggregation does not reduce control efficiency at all under the authors' assumptions (Theorem 4.7.1, called the perfect aggregation theorem).

Outwardly this scheme resembles the group penalty with which Holmström restores efficiency in the team model by abandoning budget balance: everyone receives the stipulated payment if the joint result is no lower than a given level, and nothing otherwise [287]. The difference lies in the equilibrium implemented. The group penalty makes the efficient vector only one of the Nash equilibria of the agents' game (alongside, for example, the equilibrium in which nobody exerts effort), does not decompose the agents' game, and does not implement the required action vector in dominant strategies. Scheme (16), with observable individual actions, implements the vector $y^*$ that minimizes the agents' total cost of achieving it as the unique dominant-strategy equilibrium; the aggregation theorems in [428], when only the joint result is observed, yield – like the group penalty – a Nash equilibrium, but with an explicit choice of the cheapest way of achieving the result.

In the terms of the theory of hierarchical games, the incentive problem is the game $\Gamma_2$, in which the first player (the principal) announces to the second not an action but a strategy-function $\tilde{x}_1(x_2)$ and, at the moment of its own choice, knows the second player's choice [120, 424]; the original result belongs to Germeier [226, 225]. Let $f_1$ and $f_2$ be continuous payoff functions on a product of compact sets $X_1 \times X_2$, $L = \max_{x_2} \min_{x_1} f_2(x_1, x_2)$ the payoff the second player guarantees itself under punishment (the ideology of the optimality of two regimes – reward and punishment – belongs to Germeier), $E = \operatorname{Argmax}_{x_2} \min_{x_1} f_2(x_1, x_2)$ and $D = \{(x_1, x_2): f_2(x_1, x_2) > L\}$.

**Theorem (Germeier, 1971).** *The maximum guaranteed result of the first player in the game with this order of moves equals*

$$\max\{K_0, M\}, \qquad K_0 = \sup_{(x_1,x_2)\in D} f_1(x_1, x_2), \qquad M = \min_{x_2\in E} \max_{x_1\in X_1} f_1(x_1, x_2) \tag{17}$$

*(if the set $D$ is empty, one puts $K_0 = -\infty$). For $K_0 > M$ an $\varepsilon$-optimal strategy exists; for $K_0 \leq M$ the optimal strategy is the one that, by punishing any choice outside $E$, confines the second player to this set* [225].

In the teaching expositions of the AST school the $\varepsilon$-optimal strategy is written out explicitly: choose $x_1^\varepsilon$ in response to $x_2^\varepsilon$, where $(x_1^\varepsilon, x_2^\varepsilon) \in D$ and $f_1(x_1^\varepsilon, x_2^\varepsilon) \geq K_0 - \varepsilon$, and punish the second player for any other choice; the notations $\Gamma_1$ and $\Gamma_2$ also became standard later. Scheme (15) has exactly the structure "reward for choosing the prescribed action, punishment otherwise". The difference between the schools of V.N. Burkov and Yu.B. Germeier lies in the principle of selection from the set of best responses: the theory of hierarchical games takes the guaranteed result as basic, so the optimum is attained up to $\varepsilon$, whereas AST adopts the benevolence hypothesis, under which it is attained exactly.

In MD, (15) is substantively analogous to the contract under an observable action (the forcing contract) [284, 485]: there is no information rent, and the principal captures the entire surplus. The main subject of contract theory is the unobservable-action case considered below.

The closest analogue of the game $\Gamma_2$ is the inverse (reverse) Stackelberg games of control theory [254, 282, 283, 433, 434]: the leader announces a function of the follower's decision. They differ in the criterion (the leader's team-optimal solution, affine incentives, linear-quadratic settings) and contain no theorem on the value of the game analogous to (17); Olsder himself notes that the general approach yields little here and the results are built on examples [433]. Germeier's work is not cited in [254, 282].

**Moral hazard.** The principal observes a result $q$ with distribution $F(q \mid a)$ depending on the action $a$; the agent is risk-averse. The problem

$$\max_{w(\cdot),\, a} \mathbb{E}[q - w(q) \mid a]$$
$$\text{s.t.} \quad a \in \operatorname*{Argmax}_{a'} \mathbb{E}\big[u\big(w(q)\big) \mid a'\big] - c(a'), \qquad \mathbb{E}\big[u\big(w(q)\big) \mid a\big] - c(a) \geq \bar{u} \tag{18}$$

leads – if the first-order approach is applicable (it is justified under the monotone likelihood ratio property and the convexity of the distribution function in the action) – to the optimality condition $1/u'\big(w(q)\big) = \lambda + \mu f_a(q \mid a)/f(q \mid a)$ and to the informativeness principle: if the agent's optimal action is unique and interior, an additional signal should be included in the contract if and only if it carries information about the action beyond the already observed result [255, 284]. In the Holmström–Milgrom model (Brownian noise in continuous time, exponential utility with risk-aversion coefficient $r$, convex costs) the optimal contract is linear in the final result; the problem reduces to a static one with $q = a + \varepsilon$, $\varepsilon \sim N(0, \sigma^2)$, in the class of contracts $w = w_0 + \beta q$, and the optimal share equals [289, 288]

$$\beta = \frac{1}{1 + r\, c''\sigma^2}. \tag{19}$$

In the static model itself, with normal noise and no restriction on the class of contracts, there is no optimum: the first best is approached arbitrarily closely by rare harsh penalties, so linearity is a property of the dynamic setting, not the static one [288]. Close results on the conditions for approximating the first best with observable outcomes and on the failure of the agent's first-order conditions were obtained by Mirrlees [374]; conditions under which the incentive compatibility constraint can be replaced by the first-order condition were obtained by Rogerson [460] and Jewitt [309]. The multitask generalization explains the weakening of incentives when tasks are poorly measurable [290], the team model shows the impossibility of efficiency under a balanced sharing of the joint result [287], and tournament schemes are studied in [345]. The AST school considered probabilistic incentive problems in the 1990s [407] and compared them with contract theory at the same time [104]; in the book [120] the implementability condition for an action in probabilistic schemes is directly identified with the incentive compatibility constraint of contract theory.

A systematic exposition is given in the monograph [407], where the incentive problem is formulated in a general form covering the settings of AST, the theory of hierarchical games, contract theory and implementation theory, and the basic models M1–M13 are classified by the type and kind of uncertainty and by awareness. For the model with external probabilistic uncertainty and symmetric awareness – the basic model of contract theory – under risk neutrality of both parties and a bound on the penalty it is proved that: under a strict monotone likelihood ratio, for any incentive scheme there is a step-type scheme of no lower efficiency (Proposition 3.1.19); for a "simple" agent, whose result does not exceed its action, the compensatory scheme is optimal (Proposition 3.1.15); the cost of incentives under an unobservable action is strictly higher than under an observable one (Proposition 3.1.14 – the analogue of the gap between first best and second best). Under asymmetric awareness, when the state of nature is known to the agent, compensatory or step-type schemes are optimal, and under the benevolence hypothesis the efficiency may exceed the deterministic one (Propositions 3.2.1–3.2.3). The book's general conclusion for all kinds of uncertainty: the "boundary" C- and K-type schemes are optimal, proportional schemes do not dominate them in any of the basic models, and the guaranteed efficiency does not increase as uncertainty grows. The first result of the cycle – the optimality of the step-type scheme under a uniform distribution – dates from 1993 [104]; the first-order approach and the informativeness condition [284] are not used in the book. Related conditions – risk neutrality and limited liability of the agent – are considered in contract theory as well [298].

*Fuzzy uncertainty.* Chapter 4 of the same book [407] is the only systematic study of incentives under fuzzy uncertainty (the journal version dates from 1997). The agent's fuzzy preference relation is reduced, following Orlovsky, to the set of nondominated actions; it is proved that under internal fuzzy uncertainty the compensatory scheme is optimal under the benevolence hypothesis and the step-type one without it (Proposition 4.1.3); under external fuzzy uncertainty step-type schemes suffice (Propositions 4.3.7 and 4.4.2); and the efficiency under the benevolence hypothesis can be no lower than the deterministic one – in contrast to the probabilistic case. MD and contract theory do not consider fuzzy uncertainty, so in facet F2 (Section 8.1) it forms a value filled only by AST.

*The link with labor economics.* In AST the agent's cost function is a primitive of the model, and its link with labor supply was established in two "experimental" books. In [48] four representations of a worker's preferences are compared – the utility function over income and leisure of the standard labor-economics model, the minimum wage rate for which the agent agrees to work a given number of hours, the desired working time at a given rate, and the "reward minus costs" goal function of AST – and the transitions between them are worked out: the cost function is recovered by integrating the minimum rate over hours; the desired time under proportional pay is the inverse of marginal costs; the minimum rate corresponds to the compensatory incentive scheme; and the reverse passage to the utility function is impossible in general, so the representations are equivalent only with respect to externally observable behavior. A questionnaire survey of 1999–2001 (406 valid questionnaires, Moscow and Moscow Region) identified four types of labor supply strategies (hours independent of the rate, monotone growth, monotone decline, growth followed by decline – the backward-bending curve) and showed that the lower branch of the cost function is well approximated by a parabola, while the association between strategy type and social characteristics is statistically insignificant. The identified cost functions were applied to AST problems – crashing a project's duration with minimum extra payments and staff formation – where the passage from personalized rates to a unified one costs an extra 16–30% in the examples (an empirical estimate of the price of unification, Sections 4.4 and 6.6). The 2010 edition [413] pools the surveys of 1999, 2003 and 2009, including the online survey "Labor and Wages" with over 5500 questionnaires, identifies five strategy types and five clusters, introduces aspiration-level indicators and the parameter of a quadratic cost function, and shows that classifiers of strategies by socio-demographic attributes barely beat the naive rule. What matters for the comparison with MD is that this is a rare case, for both research traditions, of empirical identification of the primitives of an incentive model: in contract theory the effort cost function is as a rule postulated, and empirical work tests predictions about the form of contracts; the agent's costs are not recovered there.

**Adverse selection.** The agent knows its cost parameter $\theta \sim F$ on $[\underline{\theta}, \bar{\theta}]$, the costs are $\theta q$, and the value of output to the principal is $S(q)$. For a nondecreasing ratio $F(\theta)/f(\theta)$, the optimal menu of contracts in the standard textbook form is given by the condition

$$S'(q(\theta)) = \theta + \frac{F(\theta)}{f(\theta)}, \qquad U(\theta) = \int_{\theta}^{\bar{\theta}} q\,(z)\, dz, \tag{20}$$

where $U(\theta)$ is the information rent; the output of every type except the most efficient one is distorted downward relative to the first best (the statement follows [340]; the original model with a continuous type is [372, 382], and the monopoly regulation model is [50]). The AST analogue is the theory of coordinated mechanisms with penalties for deviating from the plan, begun in the 1980s [101]; in its version where the agent reports a cost estimate [189, 187] the criterion is guaranteed: the optimal proper mechanism solves the problem $\max_{x(\cdot)}\min_{r\in\Omega}[H(x(r)) - \sigma(x(r))]$ under self-selection constraints; the reducibility to the basic model of contract theory is shown in [189] (Section 6.3).

**Incomplete contracts.** The third classical line is the Grossman–Hart–Moore theory of incomplete contracts, in which the allocation of ownership and control rights determines the incentives for specific investments [256, 271, 492]. There is no direct analogue in AST; the TCO areas closest in subject are structure control and institutional control (Sections 6.1 and 6.6).

Table 3. Correspondence of the classical results of the two research traditions

| Result | MD: year of the first result | AST, TCO, Germeier's school: year of the first result | Comment |
|---|---|---|---|
| Sufficiency of direct truthful mechanisms | 1973 – revelation principle: Gibbard [232]; 1979 – Myerson [383] | 1970 – fair play principle: Burkov, Lerner [117]; 1985 – general case [98] | The criteria differ: implementability or expectation versus the guaranteed result |
| Characterization of strategy-proofness | 1979 – characterization of individually compatible mechanisms via constraint sets independent of one's own characteristics, in large economies: Hammond [267]; 1981 – Guesnerie (after [174]); 1985 – taxation principle: Rochet [458]; 1995 – Guesnerie [263] | 1970 – fair play principle: Burkov, Lerner [117]; 1985 – perfect coordination condition (PCC) (4) [98] | Substantively close: the PCC concerns a finite number of agents and dominant strategies (the menu depends on the others' messages), Hammond's characterization concerns large economies |
| Expertise, one-dimensional choice | 1980 – median schemes: Moulin [381] | 1989 – active expertise mechanism (7) [97] | The identity was shown in 2008 [110] |

| Result | MD: year of the first result | AST, TCO, Germeier's school: year of the first result | Comment |
|---|---|---|---|
| Resource allocation | 1982 – the uniform rule as a rationing scheme: Benassy [130]; 1991 – axiomatic characterization: Sprumont [486]; 1997 – sequential rules [42] | 1977 – priority-based mechanisms [87]; 1989 – the equivalent strategy-proof direct mechanism [97]; 1997 – equivalence of anonymous mechanisms [105]; 2009–2012 – representation (10), the SRAM as a generalized median scheme [327, 328] | The rule – Benassy; the axiomatics – Sprumont; the construction of the equivalent direct mechanism for priority-based mechanisms – the AST school (1977 and 1989) |
| Anonymity and decentralization of planning mechanisms | 1991 – anonymity as an axiom [486]; multilevel procedures are not considered; 1995 – delegation of contracting to a strategic intermediary [363]; 2006 – survey [377] | 1980 – the price of decentralizing mechanisms of functioning [100]; 1999 – arbitrary decentralization of anonymous allocation, expertise and transfer-price mechanisms [408] | The MD analogues concern a different setting (delegation of contracting); no analogue of the theorems on equilibrium invariance under bid aggregation has been found (Section 3.2) |
| Efficiency with transfers | 1961–1973 – VCG [150, 257, 517]; 1977 – the Groves–Ledyard mechanism [258] | 2013 – a Groves–Ledyard-type mechanism (13) [329]; VCG are used as known | In AST, systems with TU are studied less (except for incentive problems) |
| Nash implementation | 1977 – Maskin's theorem (published in 1999) [358]; 1992 – complete characterization for $n \geq 3$: Danilov [163] | 1997 – equivalent direct mechanisms [405]; an exposition of Maskin's theorem in [424] | Different questions: implementability of a rule versus reduction of a mechanism |
| Contract under an observable action | 1971 – Spence, Zeckhauser (the observable-action case: a penalty for any other action) [485]; 1979 – Holmström (the term forcing contract) [284] | 1971 – Germeier's theorem for the game $\Gamma_2$ [225]; 1998–2000 – schemes (15), (16) [407, 428] | They coincide; TCO's priority is dominant strategies for many agents |
| Incentives for a joint result | 1982 – the group penalty with an unbalanced budget: Holmström [287] | 2000 – the aggregation theorems [428] | Similar in construction, different in result: Holmström's group penalty does not decompose the game and does not implement the action vector in dominant strategies (Section 3.5) |
| Hidden action | 1971 – Spence, Zeckhauser (insurance) [485]; 1973 – Ross [463]; 1974–1976 – Mirrlees (the general statement via the distribution of the result) [373]; 1979 – Holmström (the informativeness principle) [284]; 1983 – Grossman, Hart [255] | 1977 – the "simple" active element [87]; 1993 – the probabilistic incentive problem [104]; 1998 – the M1–M13 taxonomy [407] | The statements were compared by the AST school in 1993; the rigorous characterization when the first-order approach fails – [374], [255] |

| Result | MD: year of the first result | AST, TCO, Germeier's school: year of the first result | Comment |
|---|---|---|---|
| Hidden type | 1971 – Mirrlees (optimal income taxation under private information about ability) [372]; 1978 – Mussa, Rosen (nonlinear pricing) [382]; 1979–1982 – Baron, Myerson (monopoly regulation) [50] | 1970 – the fair play principle [117]; 1983 – incentive-compatible planning [101]; 2010 – the proper mechanism under the guaranteed criterion [187] | Bayesian (welfare or profit) versus guaranteed criteria; in Mirrlees the principal is a planner |
| Interval and fuzzy uncertainty | 1978 – the guaranteed ratio to first best under an unknown cost parameter: Hurwicz, Shapiro [296]; 2001 – prior-free auctions [235]; 2005 – sets of distributions, robust design [64]; 2006 – ambiguity among participants [79]; fuzzy – absent | 1985 – the criterion guaranteed over a parameter set [98]; 1997–1998 – fuzzy uncertainty [407] | The first guaranteed criteria in MD date no later than 1978 (in one environment); fuzzy uncertainty – AST only |
| Unification and simplicity of mechanisms | 1993–2009 – losses under restricted menus and simple mechanisms: Wilson [527], Rogerson [461], Hartline, Roughgarden [276]; 2013–2020 – multidimensional screening [35, 274, 353] | 1999 – the price of unification [408] | Different classes of restrictions and loss measures (Section 4.4); AST's priority is in defining the price of unification for incentive schemes and its absolute and relative measures |
| Agents' limited planning horizon (farsightedness) | 1995 – a limited forecast horizon, equilibrium with correct forecasts: Jehiel [306]; 2023 – simplicity standards under a limited horizon: Pycia, Troyan [455] | 1986 – farsighted elements, coincidence of choice with forecast and plan: Shchepkin [481, 482]; 2002 – distributions of farsightedness, decision horizons, the role-exchange effect [427] | Close statements; the horizon is used as a simplicity standard in MD and as an optimization constraint in AST |
| Hierarchical game $\Gamma_1$: the leader announces an action | 1934 – Stackelberg [487] | 1971–1976 – the game $\Gamma_1$ in the general theory of hierarchical games [226, 225] | The general statement and the maximum guaranteed result – Germeier's school |
| Hierarchical game $\Gamma_2$: a strategy-function (the incentive problem) | 1971 – Spence, Zeckhauser [485]; 1971 – Mirrlees (the tax as a function of income) [372]; 1973 – Ross (a function of the result) [463]; 1979 – Holmström [284] | 1971 – Germeier's theorem (17) [225] – the general result; 1973–1977 – information about the partner, information exchange [321, 320] | The statements are simultaneous; Germeier's school's priority is the guaranteed-result criterion and the general theorem on the value of the game; the closest analogue by discipline is inverse Stackelberg games (Section 3.5); the translation [227] is not cited in the contract theory literature |

| Result | MD: year of the first result | AST, TCO, Germeier's school: year of the first result | Comment |
|---|---|---|---|
| Self-enforcing agreements | 1980 – the theory of self-enforcing agreements: Telser [491]; 1987–1989 – implicit contracts as equilibria of repeated games: Bull, MacLeod, Malcomson [84, 356]; 2003 – characterization of optimal relational contracts: Levin [351] | 1977 – dynamic stability of cooperative solutions [446] (Petrosyan's school, outside AST) | Different formal apparatus; no mutual references (Section 5.3) |
| The principal's commitment in a multi-period model | 1977 – time inconsistency of optimal plans: Kydland, Prescott [335]; 1980 – the ratchet effect under a given revision rule: Weitzman [525]; 1985 – the first model of selection without commitment by the principal: Freixas, Guesnerie, Tirole [214]; 1988 – continuous type, pooling: Laffont, Tirole [337]; 2022 – the revelation principle under limited commitment [176] | 2002 – the current, rolling (with and without commitment) and program-based regimes; the adequacy principle (60) [427] | Different sources of the value of commitment: information disclosure versus limited farsightedness |

## 3.6. The Foundational Monographs of the Two Research Traditions

The classical results of both research traditions (MD and AST) are fixed in monographs, and a comparison of their texts reveals differences that journal articles do not show. Table 16 (see the Appendix) lists the books on which this survey relies and indicates the text used as the source for each set of results.

An analysis of the monographs yields seven observations.

*The way of systematizing.* The AST school's monographs are built up from the basic model (one principal, one agent, one period, complete awareness) to its extensions, and from the general control problem to standard mechanisms. The book [120] classifies models along seven dimensions (staff, structure, number of periods, goal functions, admissible sets, order of functioning, awareness) and, for systems with uncertainty, identifies thirteen basic models. MD textbooks are organized by the kind of information asymmetry and the solution concept: hidden type, hidden action and nonverifiability in Laffont and Martimort [340]; screening, Bayesian and dominant-strategy mechanisms in Börgers [131]; and in the algorithmic literature, by the computational model and the approximation measure [10, 275]. The first way fits the task of choosing a mechanism for a specific control problem; the second grew up in a theory that maps the boundaries of the attainable.

*The default criterion.* In the AST school's texts the benevolence hypothesis is the default: the course [424] states this explicitly, and in the monograph [412] control efficiency is first defined as the maximum of the principal's goal function over the agent's rational choice set, and only then is guaranteed efficiency introduced as an alternative. The guaranteed result is applied to the agents' unknown parameters [98], and in incentive problems the benevolence hypothesis is given up at the price of arbitrarily small bonuses [428]. The widespread characterization of AST as a guaranteed-result theory is therefore accurate only with respect to uncertainty: equilibrium selection is governed by the benevolence hypothesis, and it is the theory of hierarchical games that follows the guaranteed approach consistently [226, 225].

*The attitude toward MD.* The book [120] characterizes contract theory and implementation theory briefly, naming their main authors, and refers the detailed comparison to the journal surveys of the 1990s. The course [424] includes an exposition of implementation theory, and the monograph [412] opens with the thesis that AST, the theory of hierarchical games and mechanism design have practically merged.

*The content of the theorems.* Checking the formulations led to refinements incorporated in Sections 3.1–3.5. The fair play principle has two sides – the characterization of strategy-proofness and optimality – and the second is proved not for all classes of systems. The equivalent direct resource allocation mechanism is constructed under four assumptions, and the coincidence with serial allotment is established for anonymous mechanisms. The optimal incentive scheme of the basic model contains no bonus, which is needed only when the benevolence hypothesis is dropped. On the MD side, the linearity of the Holmström–Milgrom contract, Moulin's median schemes and robust implementation require conditions that survey expositions often omit.

*Results overlooked in previous comparisons.* The AST school's books contain statements with close MD analogues that have never been explicitly related to them. These are the guaranteed-efficiency bounds for tournament mechanisms, i.e. approximation-type results: the book [120] and the AST school's textbooks show that the guaranteed efficiency of the simple tournament can be arbitrarily small, while the direct ("knapsack") tournament guarantees at least half of the optimal effect, the question of the maximum guaranteed efficiency of tournament mechanisms being named open. Bounds of this type are akin to the prophet inequality: a simple rule (a threshold on the distribution of the maximum, Samuel-Cahn [474]; the direct tournament in [120]) guarantees a 1/2 share of the optimum. The settings differ: in the prophet inequality the quantities are independent and the benchmark is the expected maximum under known realizations, whereas in AST tournaments the uncertainty is given by a set of possible values and the agents behave strategically; a direct comparison has not yet been carried out. The same applies to the theorems on incentives based on the aggregated result [428], outwardly similar to group penalties in the team model but differing from them in the implemented equilibrium (Section 3.5), to the theorems on the price of unification [408], and to the theorems on the decentralization of planning mechanisms [408], which have later analogues in the MD theory of delegation [377].

*The monographs of 2003–2004.* Between the summarizing AST book [120] and the foundational TCO monograph [412], three editions appeared that cemented the transition from AST to TCO. The monograph on incentives [409] (312 pp.) brings together models of individual and collective incentives, relates them to forms and systems of remuneration, includes the results of the experimental study of labor supply strategies [48], and applies the solutions of incentive problems to optimizing the staff of organizational systems; it is addressed, among others, to labor economists. In [409] the incentive problem of control theory, the contract-theory model with probabilistic uncertainty and the income–leisure model of labor economics are set out in separate sections, and collective incentives are broken down into incentives for individual results, incentives for the results of collective activity, unified, team and rank-order schemes, and pay scales. The book on reflexive games [421] is considered in Section 6.5, and the brochure on institutional control [411] in Section 6.1.

*English-language editions.* The basic AST texts exist in English: Yu.B. Germeier's monograph translated by A. Rapoport [227], the foundational TCO monograph [414], the collective books [85, 106], and the books on reflexion, networks, mobs, cybernetics and complex activity [57, 59, 81, 147, 403, 398]. The book [414] reproduces the structure of the monograph [412] (chapters: control in organizations; incentive, planning, organization and monitoring mechanisms; staff control; structure control; informational and institutional control). **None of these AST editions, nor the survey [99], is cited in the core MD literature** (Section 7.1); the AST school's references to the MD literature, by contrast, have been systematic since the early 1990s – see [104, 125].

## 4. Mechanism Design in 2006–2026

By 2006 the classical Bayesian theory was essentially complete: the revelation principle, optimal single-parameter mechanisms, the main impossibility theorems and the implementability conditions had all been obtained. The development of the following twenty years is conveniently described as a sequence of overlapping waves, each relaxing one assumption of the classical model (Fig. 3): the principal's knowledge of the distribution and of the agents' beliefs (*robust design*); the static character of the model (*dynamic design*); the one-dimensionality of the type (*multidimensional screening*); the exogeneity of information (*information design*); the agents' unbounded rationality and the principal's full commitment (simplicity of mechanisms and credibility of the designer); the presence of money and prices (market design without transfers); unlimited computational resources and knowledge of distributions (algorithmic and learning-based design); and, finally, the nature of the participants themselves (mechanisms for learning algorithms and AI agents). The periodization relies on the chronology of the landmark publications and the assessments of the survey literature [67, 70, 139, 470]. The structure of the field is shown in Fig. 4, and a summary is given in Table 4.

The field's self-assessment is given in the volume "The Future of Economic Design" [494] (2019; a collection of short essays in eleven parts by more than eighty authors), for which the editors Laslier, Moulin, Sanver and Zwicker asked MD researchers to state their entrenched but hard-to-prove convictions about where the discipline came from and where it is going. The parts cover general questions, aggregation and voting, algorithms and complexity, axiomatics, behavioral aspects, fair division, implementation, interpersonal relations, law, matching and markets, and new technologies. The following theses matter for the comparison with AST. Carroll [138] calls for a "cultural shift" away from the cult of exactly solving the optimal mechanism problem toward models with weak parametric structure and worst-case guarantees – that is, toward the criterion adopted in AST since the 1970s; Jehiel [308] proposes robust design when agents' beliefs are unavailable, via analogy-based expectation equilibrium; Nisan [391] and Roughgarden (Ch. 22) put the complexity and simplicity of mechanisms at the center of the theory; Mookherjee [378] states that MD does not explain the allocation of authority, the structure of organizations or the prevention of abuse of power, and that to do so it must incorporate imperfect commitment, collusion and communication costs – a list almost coinciding with the subject of TCO (structure control, multilevel systems, institutional control); Mishra (Ch. 32), Yildiz (Ch. 50) and Dutta (Ch. 43) outline implementation with boundedly rational agents; Sen (Ch. 49) and Peters (Ch. 48) formulate open questions on manipulability and random rules; Núñez (Ch. 47) demands that mechanisms be transparent to participants; Mezzetti (Ch. 46) couples decision choice with information disclosure – an analogue of informational control; Dütting and coauthors [180] present learning-based mechanism synthesis (Section 4.8); Saijo (Ch. 35) describes "future design" – institutions activating the willingness to sacrifice the present for future generations, tested in laboratory, field and municipal experiments; Jackson (Ch. 72) and Pycia (Ch. 74) discuss the role of theory in the age of abundant data and the link between market design and technology. The volume contains neither references to AST nor an essay on planning within organizations, but the questions of Carroll, Mookherjee, Nisan and Núñez – the guaranteed criterion, hierarchies and authority, the simplicity and intelligibility of a mechanism – belong to the areas where this survey records AST's comparative advantages (Section 7.3), which supports the forecast of possible convergence (Section 10).

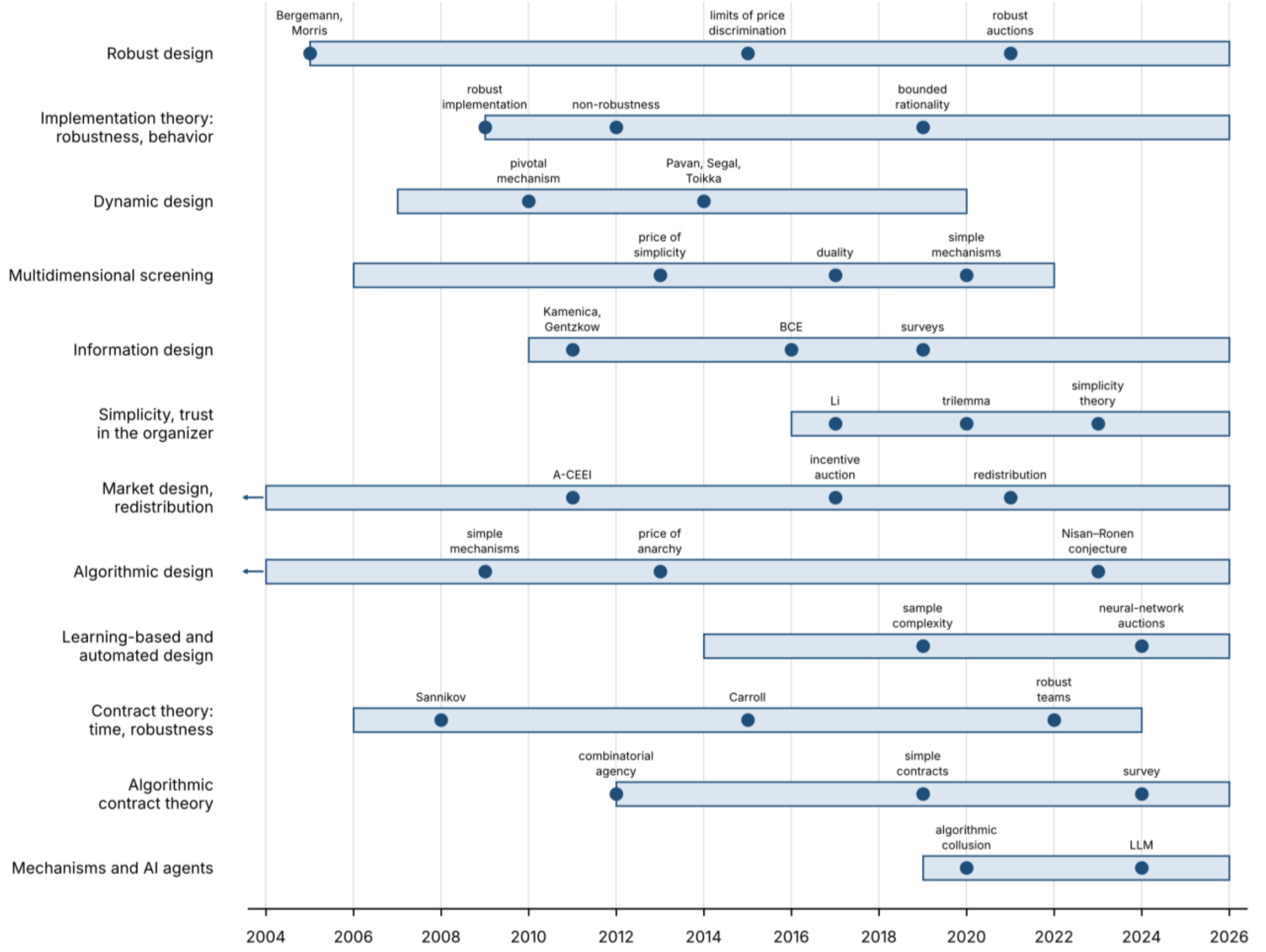

Fig. 3. Waves of development of mechanism design and contract theory in 2006–2026
(a segment marks the period of peak activity; the labels are landmark publications)

## 4.1. Robust Design

The program was formulated as a response to Wilson's critique: a mechanism should not rely on details that are common knowledge only within the model. Bergemann and Morris [64] consider implementation on all type spaces, including the universal one, and show that in separable environments (which include single-valued social choice rules and quasi-linear environments without a budget-balance requirement) a choice rule is partially implementable in Bayesian equilibrium on every type space if and only if it is ex post incentive compatible:

$$u_i(f(\theta), \theta) \geq u_i(f(\theta_i', \theta_{-i}), \theta) \qquad \forall\, i, \theta, \theta_i'. \tag{21}$$

With private values, (21) coincides with (2), so the result justifies dominant strategies through robustness to beliefs; for choice correspondences and under a budget-balance requirement the equivalence generally fails. Another justification was obtained by Chung and Ely [149]: a seller who knows the distribution of valuations but not the buyers' beliefs about one another, and who evaluates a mechanism by the worst-case beliefs, cannot, under a regularity condition, obtain more than the best dominant-strategy mechanism yields. For full implementation this is not enough.

**Theorem (Bergemann, Morris, 2009).** *Let the types be one-dimensional and the agent's preferences depend on the aggregate of types $h_i(\theta)$ and satisfy the strict single-crossing condition. Then the strict condition* (21) *together with the contraction property is sufficient for robust implementability of the choice rule by the direct mechanism, and necessary for its robust implementability by any mechanism if the rule is responsive to types. In the linear model $h_i(\theta) = \theta_i + \gamma \sum_{j \neq i} \theta_j$ the contraction property is equivalent to the condition* $|\gamma| < 1/(n-1)$ [65].

The content of the condition: the interdependence of preferences must be weak enough for the iterative elimination of non-rationalizable messages to converge to the truth. The requirement is substantially stronger than ex post incentive compatibility, for which $\gamma < 1$ suffices in the same linear model.

The second line optimizes the guaranteed value of the principal's criterion over a set of information structures or distributions. Bergemann, Brooks and Morris [63] described all pairs "consumer surplus – monopolist's profit" attainable under some market segmentation: it is the triangle

$$\{(u, \pi): u \geq 0, \quad \pi \geq \pi^*, \quad u + \pi \leq w^*\}, \tag{22}$$

where $\pi^*$ is the profit under a single monopoly price and $w^*$ the maximum total surplus. Brooks and Du constructed auctions maximizing guaranteed revenue under common values [82]. For multidimensional screening, Carroll [137] proved that if the seller knows the marginal distributions $F_k$ of the valuations of each of the $K$ goods but not the joint distribution, then guaranteed revenue is maximized by selling separately:

$$\max_{\text{mechanisms}} \min_{F \in \mathcal{F}(F_1, \ldots, F_K)} \mathrm{Rev}(F) = \sum_{k=1}^{K} \max_{p} p\left(1 - F_k(p)\right). \tag{23}$$

Carroll's survey [139] characterizes the strand as still taking shape: there is no unified theory, and questions remain about the justification of the maximin criterion and the dynamic consistency of non-Bayesian models.

What matters for the comparison is that (23) uses the same guaranteed result over the parameters unknown to the principal as AST does (Section 2.1). The difference is in the object of uncertainty: in robust MD "nature" chooses a distribution or an information structure over a known support, in AST it chooses the parameter value itself. The AST variant is closer to the prior-free setting of theoretical computer science (Section 4.8).

## 4.2. Implementation Theory

The main negative result of the period concerns multi-stage mechanisms. Moore–Repullo mechanisms implement almost any rule in subgame perfect equilibrium under complete information among the agents [379]. Aghion, Fudenberg, Holden, Kunimoto and Tercieux [5] showed that this conclusion does not survive small information perturbations: if the rule is not Maskin monotonic, then under an arbitrarily small departure from common knowledge any implementing mechanism acquires sequential equilibria with undesirable outcomes. Maskin monotonicity thereby regained its role as the central condition; implementation in rationalizable strategies requires its strict version [68].

Other lines of the period relax the rationality assumptions: behavioral implementation, in which agents are described by choice correspondences rather than preferences [168], and implementation under bounded depth of reasoning [169]. A laboratory test of dynamic implementation mechanisms is given in [141]. A separate result links the concepts: with linear utilities and independent one-dimensional private types, for any Bayesian incentive compatible mechanism there exists a dominant-strategy mechanism with the same expected utilities of the agents [231].

The strand's main unsolved problems remain implementation by "nice" (finite, direct) mechanisms under incomplete information and a systematic theory of behavioral implementability.

## 4.3. Dynamic Design

In dynamic problems the agents' types change over time and decisions are made sequentially. Dynamic analogues of the VCG mechanisms have been obtained for the efficient allocation problem. In the dynamic pivot mechanism of Bergemann and Välimäki [69], an agent's payment in every period equals the externality its presence imposes on the others, so the agent's net payoff coincides with its marginal contribution to social welfare, and participation remains voluntary at every moment. Athey and Segal [27] built an efficient mechanism balanced in every period. For revenue maximization, Pavan, Segal and Toikka [443] obtained a dynamic envelope formula. For a single agent with type process $\theta_t$ and utility $\sum_t \delta^t u_t(\theta_t, x_t)$, the derivative of the equilibrium payoff with respect to the initial type equals

$$\frac{\partial V(\theta_0)}{\partial \theta_0} = \mathbb{E}[\sum_{t\geq 0} \delta^t \; I_t(\theta^t) \; \frac{\partial u_t(\theta_t, x_t)}{\partial \theta_t} \mid \theta_0], \tag{24}$$

where $I_t$ is the impulse response of the period-$t$ type to the initial type (for the process $\theta_t = \kappa\theta_{t-1} + \varepsilon_t$ we have $I_t = \kappa^t$). This yields an expression for the dynamic virtual surplus generalizing (12): the allocation distortions in period $t$ are proportional to $I_t$ and die out if the initial information loses relevance. The results rest on the first-order approach; there are no general conditions under which local incentive compatibility implies global, and this remains the strand's main open question [70].

In AST, dynamics is the subject of adaptive mechanisms, in which the principal recomputes norms from observed results and farsighted agents take this into account [124]; the MD analogue is the ratchet effect: the practice of planning from the achieved level, described for Soviet enterprises [71], was formalized by Weitzman [525] for a given target revision rule and by Freixas, Guesnerie and Tirole [214] as an equilibrium without commitment by the planner; later the effect entered regulation theory [185, 338]. In addition, the AST school developed a theory of deterministic dynamic active systems (DAS) with farsighted participants – from A.V. Shchepkin's papers [481, 482] to the survey [406] and the monograph [427] (Section 6.8). In these models the participants are fully informed about the current parameters, and uncertainty concerns only the future; hence no information rent arises, and AST has no systematic analogue of formula (24). The subject of DAS is different: how the participants' farsightedness, their decision horizons and the principal's commitment affect the implementability of plans and the efficiency of control.

## 4.4. Multidimensional Screening

Already for a single buyer with additive valuations of several goods, the optimal mechanism is generally non-deterministic and has no simple structure [459]. The period's results are of two kinds. Exact: Daskalakis, Deckelbaum and Tzamos [166] obtained a strong duality reducing the problem to optimal transport of measures, together with conditions for the optimality of pure bundling. Approximate: for a single buyer and $k$ independent goods, separate selling guarantees at least a share $c/\log^2 k$ of the optimal revenue [274] (the exact order of the guarantee is $1/\log k$ [353]), and the better of two simple mechanisms – separate selling or a single bundle – a constant share [35]:

$$\mathrm{Rev} \le 6 \cdot \max\{\mathrm{SRev}, \mathrm{BRev}\}. \tag{25}$$

Together with (23), these results shaped the view of the "*price of simplicity*" as an object of theory in its own right.

The notion has a predecessor in AST. The monograph [408] introduced the "*price of unification*" – the difference between the efficiency of the optimal personalized incentive scheme and of the optimal unified one, identical for all agents (formula (1.7.8) ibid.); unification is motivated by reducing the informational load on the principal. In an example with three agents and quadratic costs the relative loss is 5/11, and for cost functions of the form $c_i(y_i) = r_i\, \varphi(y_i/r_i)$ it is shown that unified proportional schemes are optimal and the price of unification is zero (Section 1.7 in [408]).

Optimal unified incentive schemes for a finite and an infinite number of agents were studied by Karavaev [313; 314]. Unified schemes are part of the catalog of mechanisms [153] and of the monographs [418, 412]. The correspondence with the "price of simplicity" is exact in the statement – losses from restricting the class of mechanisms (facet F9 in Section 8.1) – and different in the measure: AST measures the absolute difference with the agents' model known to the principal, MD a guaranteed ratio over a set of distributions. Losses from restricting the class of mechanisms have been assessed in MD as well: for a finite number of tariff options – Wilson [527]; for simple menus of contracts – Rogerson [461]; for auctions with reserve prices – Hartline and Roughgarden [276]. In multidimensional screening the bounds were obtained in 2013–2020 [35, 274, 353], after the price of unification for incentive schemes was introduced [408] (1999). These results differ above all in the restricted class: AST measures the loss from a single scheme common to all agents when the agents' model is known, MD a guaranteed share of the optimum over a set of distributions. AST's priority concerns the definition of the price of unification for incentive schemes and its absolute and relative measures.

## 4.5. Information Design

In information design problems the principal controls the players' information while the rules of the game are fixed. In the basic Bayesian persuasion model of Kamenica and Gentzkow [311], the sender chooses a signal about a state $\omega$ with prior distribution $\mu_0$ and commits to disclosing it; the receiver, having updated its belief to $\mu$, chooses an action $a^*(\mu)$. Choosing a signal is equivalent to choosing a distribution of posterior beliefs $\tau$ subject to the single constraint of Bayes plausibility $\mathbb{E}_\tau[\mu] = \mu_0$.

The commitment-based statement belongs to Kamenica and Gentzkow; the technique itself (the concave envelope of the value function, splitting the prior) goes back to Aumann and Maschler's work on repeated games with incomplete information [30].

**Theorem (Kamenica, Gentzkow, 2011).** *The sender's maximum expected payoff equals the value at the point $\mu_0$ of the concave envelope of the function $\hat{v}(\mu) = \mathbb{E}_\mu\, v(a^*(\mu), \omega)$:*

$$V(\mu_0) = \max_{\tau:\, \mathbb{E}_\tau[\mu] = \mu_0} \mathbb{E}_\tau[\hat{v}(\mu)] = (\mathrm{cav}\hat{v})(\mu_0), \tag{26}$$

*and disclosing information benefits the sender if and only if* $V(\mu_0) > \hat{v}(\mu_0)$.

For many-player games, Bergemann and Morris [66] introduced Bayes correlated equilibrium: the set of outcomes attainable under all expansions of a given information structure is described by linear obedience constraints, and the information design problem becomes a linear program over this set. A unified exposition is given in [67]. An adjacent strand sells information as a good in the form of a menu of statistical experiments [62].

The AST analogue in subject matter is informational control in TCO (Section 6.5). The formal models differ: in Bayesian persuasion the principal is bound by commitment and by constraint (26), and the agents are Bayes-rational; in reflexive games the principal shapes an awareness structure that may be inconsistent and false, and the constraint is the stability of the resulting equilibrium.

Over 2015–2025 information design expanded toward settings characteristic of TCO (survey: [310]). Public and private signals (analogues of uniform and non-uniform informational regulation) differ in effect: when agents have a strong taste for coordination, a more precise public signal can reduce welfare [380], and private persuasion of many receivers requires a technique of its own [14]. Design evaluated at the worst equilibrium [360] corresponds to the guaranteed criterion of AST, and competition among senders [219] to the informational confrontation of principals. Models with credulous receivers and lying costs [316], of strategic deception of partners with bounded depth of reasoning [155], and of persuasion with models [478] remove the sender's commitment and come close to reflexive control; the detailed comparison is given in Section 6.5.6 (Fig. 9).

## 4.6. Simplicity, Credibility of the Designer, Limited Commitment

Strategy-proofness does not guarantee that participants will compute and use their dominant strategy. Li [352] introduced the notion of *obvious strategy-proofness* (OSP): a strategy is obviously dominant if, at any point of first possible deviation, the worst outcome from following it is no worse than the best outcome from deviating. The English auction satisfies this requirement, while the sealed-bid second-price auction does not; OSP mechanisms are exactly those the principal can run while holding only bilateral commitments to each agent. A general theory of gradations of simplicity is built in [455]. The property depends on the extensive form of the mechanism and turns out to be demanding: the top trading cycles mechanism with three or more agents is not OSP-implementable [352], and mechanisms selecting a stable matching are in general not obviously strategy-proof for either side [25]. For mechanisms coinciding with the basic AST mechanisms, exact answers have been obtained. Among generalized median schemes only some are OSP-implementable (ordinary median voting is not among them) [36]. By contrast, in the model with indivisible resource units all sequential allotment rules are OSP-implementable [15]: the SRAM can be run as a sequential procedure in which truthful behavior is obviously optimal. AST did not formulate an obvious strategy-proofness standard, but the notion of a limited planning horizon, central to the theory of [455], has been studied in AST since the 1980s (Section 6.8). In [455] an agent plans only part of its future decisions, and simplicity requires the prescribed action to be better under any development beyond the horizon. In DAS [427, 482] a farsighted agent maximizes a weighted sum of payoffs within its horizon, and the principal tailors plans, incentive schemes and commitments to the given horizons. This yields conclusions absent from [455]: an increase in a participant's farsightedness can reduce control efficiency, and an agent more farsighted than the principal can impose future plans on the principal (the *role-exchange effect*).

The second line relaxes the assumption that the principal is bound by the rules. Akbarpour and Li [6] call a mechanism credible if the organizer does not benefit from deviations from the announced rules that the participants cannot detect, and prove a trilemma: an optimal auction cannot simultaneously be static, strategy-proof and credible; among static formats only the first-price auction has this property, and among strategy-proof ones only the ascending auction. For limited commitment in dynamics, Doval and Skreta [176] obtained an analogue of the revelation principle: it suffices to consider mechanisms whose output encodes the principal's posterior belief about the agent's type. In DAS the commitment question is posed for the deterministic model: rolling control regimes with and without the principal's commitment are compared, and condition (60) is found under which commitments for a given number of periods leave efficiency unchanged [427].

## 4.7. Market Design, Mechanisms without Money, Redistribution

Market design turned matching theory into an engineering discipline [184, 464] and the school choice problem into a standard MD problem [1]. The period's theoretical results: approximate competitive equilibrium from equal incomes for combinatorial assignment [83], strategy-proofness in the large as a practical criterion for large markets [33], and allocation under costly verification of messages, where the optimal mechanism has a favored agent and thresholds [60, 388]. Since 2021 distributional goals have been introduced: if the marginal utility of money differs across participants, the optimal mechanism may include rationing and non-price instruments [177]. The auction branch is linked to the 2020 prize: multi-object formats, clock auctions and the radio spectrum reallocation [370, 493].

Mechanisms without money are the area of greatest overlap with AST: resource allocation under single-peaked preferences, expertise, verification of messages and penalties have been studied by the AST school since the 1980s (Sections 3.2 and 6.3), but with no connection to matching problems and without the notions of stability and fairness (envy-freeness) central to market design.

## 4.8. Algorithmic and Learning-Based Design

Algorithmic MD, begun by the paper of Nisan and Ronen [392] and codified in [10, 275], adds computational constraints to the incentive compatibility constraints and replaces exact optimality with a guaranteed worst-case share of the optimum. The main groups of results are as follows.

*The price of truthfulness.* Requiring dominant-strategy truthfulness together with polynomial complexity leads to provable losses: for combinatorial auctions with submodular valuations, truthful mechanisms using a polynomial number of value queries cannot guarantee more than a share of order $m^{-1/2}$ of the optimal welfare, whereas without the truthfulness requirement a constant share is attainable [175]; a similar gap is established for the combinatorial public project problem [442]. In 2023 the Nisan–Ronen conjecture was proved: the best deterministic truthful mechanism for scheduling on $n$ unrelated machines has an approximation ratio of exactly $n$ [148].

*Simple mechanisms.* The VCG auction with monopoly reserve prices guarantees half of the optimal revenue for independent valuations if the distributions have a monotone hazard rate and the environment is downward-closed, or if the distributions are regular and the environment is a matroid [275, 276]. Sequential posted prices rest on the prophet inequality, by which a threshold strategy guarantees half of the expected maximum [275] (a threshold equal to the median of the maximum guarantees at least half of the expected maximum [474]); for multidimensional screening, bound (25) has been obtained. Optimal multidimensional mechanisms are built by reduction to welfare maximization [132]. The research program and open questions are set out in [470].

*The price of anarchy.* Instead of requiring truthfulness, one bounds the quality of the equilibria of simple manipulable mechanisms. For smooth mechanisms the bounds survive composition and extend to equilibria under no-regret learning [469, 489]; for example, simultaneous first-price auctions with fractionally subadditive valuations guarantee at least a share $1 - 1/e$ of the optimal welfare. This is a direct analogue of the guaranteed efficiency $K_g$ of (3), but in relative form.

*Learnability.* If the distribution is known only through a sample, the question of sufficient sample size arises. For single-parameter problems with $n$ participants and regular distributions, the sample complexity of achieving a share $1 - \varepsilon$ of the optimal revenue is $\tilde{\Theta}(n\varepsilon^{-3})$ [264]. In repeated auctions the participants are represented by bidding algorithms with budget and target constraints; for such autobidders the relative price of anarchy of the second-price auction equals 2 [3].

*Automated design.* Conitzer and Sandholm's idea of computing a mechanism for a specific problem [152] has received a scalable implementation. In the “differentiable economics” approach [181] an auction is specified by a neural network with parameters $w$ (an allocation rule $g^w$ and payments $p^w$), and strategy-proofness is replaced by a penalized expected regret:

$$\begin{aligned} &\min_{w} \; -\mathbb{E}_v\left[\sum_i p_i^w(v)\right] \\ &\text{s.t.} \quad \mathrm{rgt}_i(w) = \mathbb{E}_v\left[\max_{v_i'} u_i^w\big(v_i; (v_i', v_{-i})\big) - u_i^w(v_i; v)\right] = 0 \;\; \forall i. \end{aligned} \tag{27}$$

Menu-based architectures ensure exact strategy-proofness by construction (for a single buyer such an architecture was proposed already in [181]); a survey of the strand is given in [159]. Automated design amounts to a numerical solution of the hierarchical principal–agents game without optimality guarantees.

*Information elicitation.* The MD analogue of expertise problems, as far as incentives for truthful reports go, is the theory of proper scoring rules and peer prediction. A rule $S(p, \omega)$ is called strictly proper if the expected score is maximized only by reporting the true distribution; all such rules are generated by strictly convex functions $G$ [234]:

$$S(p, \omega) = G(p) + \langle \nabla G(p), \delta_\omega - p \rangle. \tag{28}$$

When the ground truth is unavailable, reports are scored by their consistency with the reports of other participants [452].

*Mechanisms without money.* Approximate design without transfers was initiated in [454] on the facility location problem, where the median is optimal and strategy-proof.

*Decentralized systems.* For blockchains, Roughgarden [468] formulated the transaction fee mechanism design problem with three requirements (incentive compatibility for users, for the block producer, and robustness to off-protocol collusion) and showed that the EIP-1559 mechanism with a burned base fee satisfies them outside periods of sharply rising demand.

## 4.9. Mechanisms for Language Models and AI Agents

Since 2023 three lines have taken shape. The first is mechanisms whose outcome is generated text. In the token auction [182] the agents are represented by language models specifying distributions $p_i$ over the next token, together with one-dimensional bids $b_i$; the mechanism aggregates the distributions into $q(b,p)$ and assigns payments. The agents' preferences are given only by a partial order of proximity to $p_i$. It is shown that incentive compatibility is equivalent to monotonicity of the aggregation in the bid, and the payment is expressed by an analogue of Myerson's formula (the sign of the integrand is written so that the payment is nonnegative):

$$z_i(b_i) = \frac{1}{2}\int_0^{b_i} [\, \| q(b') - p_i \|_1 - \| q(b_i) - p_i \|_1 ]\, db', \tag{29}$$

and the linear mixture $q(t) = \sum_i b_i\, p_i(t) / \sum_i b_i$ is monotone, while the log-linear one is not.

The second line is *language models as participants*: in the role of pricing agents, without any instructions to collude, they arrive at supracompetitive prices [212], as Q-learning algorithms did before them [133]. Collusion here is generated by the learning dynamics itself, without communication between agents, and is not captured by the classical notions of coalitional strategy-proofness. The third line is social choice and contract theory as a framework for AI alignment: aggregating conflicting human feedback as a social choice problem [151] and goal alignment as an incomplete contract problem [265].

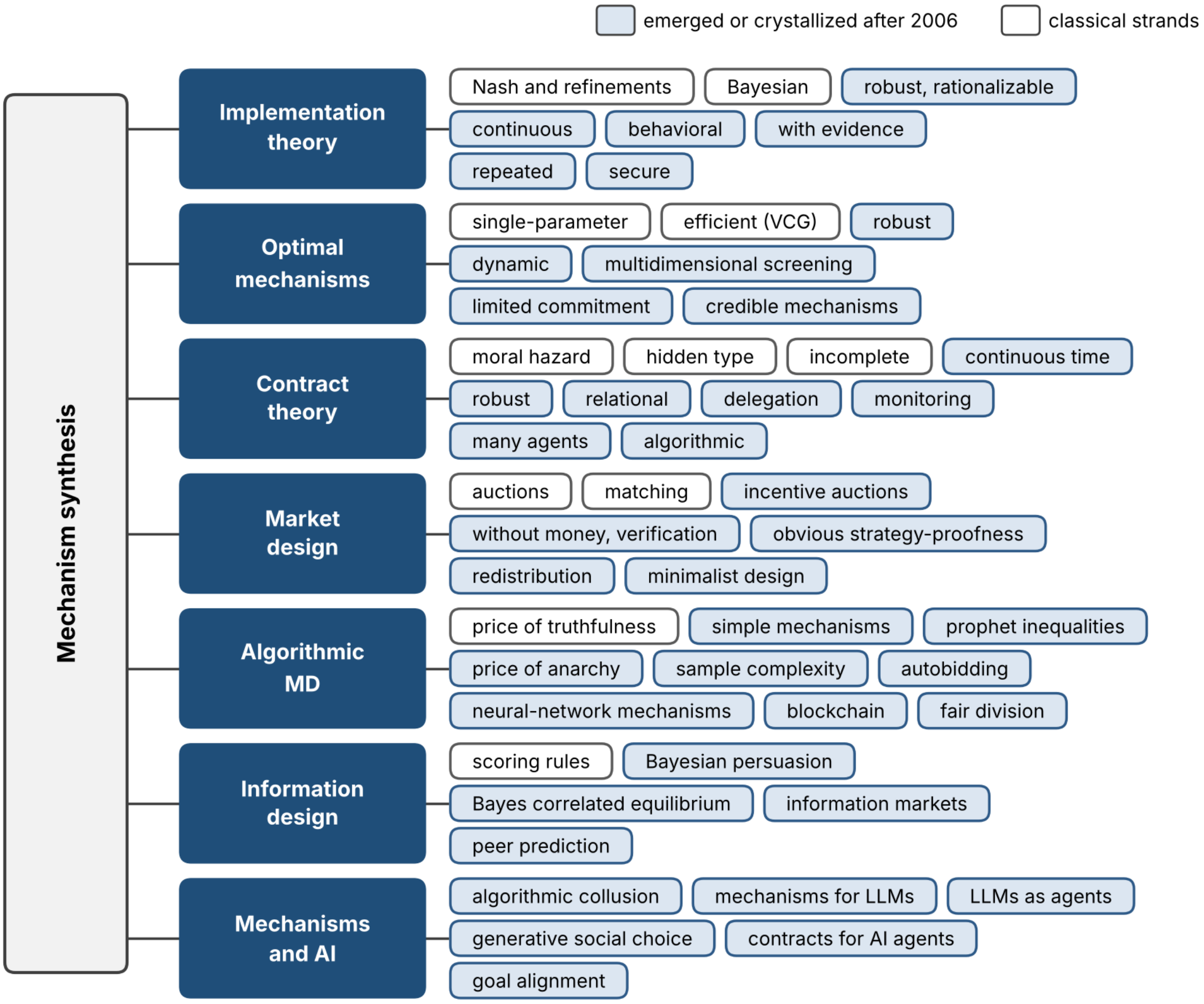

Fig. 4. The structure of mechanism design and adjacent theories by 2026
(strands that emerged or crystallized after 2006 are highlighted)

Table 4. The main lines of development of mechanism design in 2006–2026

| Line | Assumption relaxed; setting | Landmark results | Sources |
|---|---|---|---|
| Robust design | The principal does not know beliefs or the distribution; the maximin criterion | Equivalence of robust and ex post implementability; the contraction condition; the surplus triangle (22); separate selling (23) | [63, 65, 64, 82, 137, 139, 149] |
| Implementation theory | Robustness to information perturbations; finite mechanisms; behavioral agents | Non-robustness of implementation in perfect equilibria; strict monotonicity as the condition for rationalizable implementation; behavioral implementation | [5, 68, 168, 169] |
| Dynamic design | Types change over time | The dynamic pivot mechanism; the balanced efficient mechanism; formula (24) | [27, 70, 69, 443] |
| Multidimensional screening | Multidimensional types | Duality and transport of measures; bound (25); the exact order of the separate-selling guarantee | [35, 166, 274, 353] |
| Information design | The agents' information is an object of choice | The concave envelope (26); Bayes correlated equilibrium; selling information | [62, 66, 67, 311] |
| Simplicity and credibility of the designer | Agents' bounded rationality, the principal's limited commitment | Obvious strategy-proofness; the Akbarpour–Li trilemma; the revelation principle under limited commitment | [6, 25, 176, 352, 455] |
| Market design | No prices, or prices restricted; fairness goals | School choice; approximate competitive equilibrium from equal incomes; verification of messages; redistribution through markets | [1, 60, 83, 177] |
| Algorithmic design | Computational and communication constraints; the worst case | The price of truthfulness; simple mechanisms; the price of anarchy; the Nisan–Ronen conjecture | [148, 175, 276, 392, 489] |
| Learning-based and automated design | The distribution is known from a sample; the mechanism is computed | Sample complexity; problem (27); menu architectures | [159, 181, 264] |
| Mechanisms and AI | The participants are algorithms and language models | The token auction (29); collusion of learning agents; social choice for alignment | [133, 151, 182, 212] |

# 5. Contract Theory in 2006–2026

By the mid-2000s classical contract theory had been set out in the textbooks [77, 340]. Subsequent development followed several lines: continuous time, robustness, self-enforcing (relational) and incomplete contracts, delegation, interaction of many agents, and the computational setting. The correspondence of these lines to the models of AST and TCO is given in Table 5.

## 5.1. Dynamic Contracts in Continuous Time

Sannikov [475] considered an agent continuously choosing effort $A_t$, with the principal observing output $dX_t = A_t\, dt + \sigma\, dZ_t$, where $Z_t$ is a Brownian motion; the agent receives a consumption flow $C_t$ and has instantaneous utility $u(C_t) - h(A_t)$; $r$ is the common discount rate. The key device is the passage to a single state variable, the agent's promised continuation utility $W_t$. An incentive compatible contract specifies its dynamics

$$dW_t = r\big(W_t - u(C_t) + h(A_t)\big)dt + r\,Y_t(dX_t - A_t\,dt), \qquad Y_t = h'(A_t), \tag{30}$$

where the sensitivity $Y_t$ of continuation utility to output is determined by the incentive condition. The principal's profit $F(W)$ as a function of promised utility satisfies the Hamilton–Jacobi–Bellman equation

$$\begin{aligned} rF(W) = \max_{a,\,c} \{&r(a-c) + F'^{(W)}\, r\big(W - u(c) + h(a)\big) \\ &+ 1/2\, F''(W)\, r^2\sigma^2 h'(a)^2\} \end{aligned} \tag{31}$$

with boundary conditions that determine the agent's "retirement" moments. The solution yields qualitative conclusions unavailable in discrete models: how effort and pay depend on history, the conditions for dismissal and retirement, and the role of deferred compensation.

In AST, dynamic active systems have been considered in discrete time [120, 406, 427, 426] (Section 6.8); it has no continuous-time incentive models with private information. Dynamic hierarchical games with inducement (incentives) are studied in G.A. Ougolnitsky's Rostov school [441] (Section 6.7) under complete awareness of the participants.

## 5.2. Robust Contracts

The Bayesian moral hazard model (18) prescribes contracts finely tuned to the distributions, whereas in practice simple linear schemes often prevail. An explanation via robustness was proposed by Carroll [136]. The principal knows some set $\mathcal{A}_0$ of actions available to the agent, each a pair (an output distribution $F$, costs $c$), but allows that the true set $\mathcal{A} \supseteq \mathcal{A}_0$ is larger. Output is nonnegative, and a contract $w(y) \geq 0$ is evaluated by the worst case: $V_P(w) = \inf_{\mathcal{A}\supseteq\mathcal{A}_0} V_P(w \mid \mathcal{A})$.

**Theorem (Carroll, 2015).** *Let the principal and the agent be risk-neutral, let the agent be protected by limited liability and, when indifferent, choose the action best for the principal. If among the actions known to the principal there is one with positive expected surplus, then there exists an optimal contract linear in output:* $w(y) = \alpha\, y$. *The principal's guaranteed profit under such a contract and its largest value equal*

$$\begin{aligned} V_P(\alpha) &= \frac{1-\alpha}{\alpha} \max_{(F,c)\in\mathcal{A}_0} (\alpha\, \mathbb{E}_F[y] - c), \\ \max_\alpha V_P(\alpha) &= \max_{(F,c)\in\mathcal{A}_0} (\sqrt{\mathbb{E}_F[y]} - \sqrt{c})^2. \end{aligned} \tag{32}$$

The meaning of the result: a linear contract ties the principal's payoff to the agent's payoff with coefficient $(1-\alpha)/\alpha$, so any action unknown to the principal that benefits the agent also benefits the principal. In the team problem, robustness singles out schemes linear in the joint result [161]. In the computational setting it is shown that linear contracts are maximin-optimal if the principal knows only the expected results of the actions, and that their worst-case shortfall from the optimal contract grows linearly with the number of actions [183].

Result (32) is the closest MD parallel to the AST methodology: uncertainty is given by a set, the criterion is guaranteed, and the optimum is attained by a simple parametric incentive scheme. Earlier, the optimality of linear sharing under the maximin criterion (the ratio to first best) was obtained by Hurwicz and Shapiro [296] for quadratic costs with an unknown parameter; Carroll [136] cites this work, noting the different criterion and argument. The AST school's conclusion about linear (proportional) schemes is the opposite: by the verdict of the book [120], they do not dominate the compensatory and step-type schemes, and in an example of the course [424], with Cobb–Douglas-type costs, a single rate costs the principal more than cost compensation by a factor equal to the exponent of the cost function. The conclusions do not contradict each other: in Carroll's model the principal does not know the agent's action set, whereas in the AST school's models the cost function is known to it. A justification of linear schemes via uncertainty about the action set is absent from the AST school's work.

## 5.3. Relational and Incomplete Contracts

A *relational contract* is an agreement not enforced by formal obligations (a court) but sustained by the value of the future relationship. The early work on self-enforcing agreements belongs to this class [84, 356, 491]; Levin [351] characterized optimal relational contracts. In Levin's model [351] the optimal contract can be sought among stationary ones, and the discretionary bonus $b(\cdot)$ is bounded by the discounted future "surplus":

$$\sup_{\varphi} b(\varphi) - \inf_{\varphi} b(\varphi) \leq \frac{\delta}{1-\delta}\,(s-\bar{s}), \tag{33}$$

where $s$ is the expected joint surplus per period and $\bar{s}$ the surplus upon termination of the relationship.

In the theory of incomplete contracts, Hart and Moore proposed viewing the contract as a reference point shaping the parties' sense of fairness [272]. Holmström's Nobel lecture [288] states the general conclusion: incentives should be designed as a system of interacting instruments, of which pay for performance is only one. This thesis coincides with TCO's stance on the integration of mechanisms [114].

AST and TCO have not studied self-enforcing agreements: the principal is assumed to abide by the announced rules. In game theory this subject was developed in another school – L.A. Petrosyan's *cooperative theory of dynamic games*. The dynamic stability (time consistency) of a cooperative solution, i.e. its persistence along the optimal trajectory, was introduced in 1977 [446]; in 1979 an imputation distribution procedure was proposed that makes the Shapley value and the core time-consistent [447]; later, strong dynamic stability was introduced and conditions for sustainable cooperation were formulated, combining dynamic stability, strategic stability (support of the cooperative trajectory by an equilibrium in punishment strategies) and protection against irrational behavior [448, 531]. This line entered the differential games literature [445, 532] but is bibliographically unconnected with relational contract theory: the citation lists of [351] and [445] contain no mutual references. A substantive parallel nevertheless exists: both approaches ask under what conditions an agreement not backed by external enforcement is honored by the participants over time; contract theory finds the answer in the equilibrium of a repeated game with incentives (33), Petrosyan's school in redistributing the cooperative payoff along the trajectory.

## 5.4. Delegation and Behavioral Contracts

In the delegation problem transfers are unavailable, and the principal controls by restricting the decision set $D$ from which the informed agent chooses on its own:

$$\max_{D\subseteq Y} \mathbb{E}_{\theta}[u_P(y^*(\theta),\theta)], \qquad y^*(\theta) \in \operatorname{Arg}\max_{y\in D} u_A(y,\theta). \tag{34}$$

The setting goes back to Holmström's paper [285]; information exchange in the absence of transfers was studied by Melumad and Shibano [365]: when interests diverge substantially, communication benefits the decision maker and hurts the informed party. Alonso and Matouschek [13] characterized the optimal sets and the conditions under which interval delegation is optimal. In the TCO classification, problem (34) belongs to *institutional control*, i.e. to choosing the sets of admissible actions (the monograph [411], Section 6.1) [412]; in the single-agent case the set $D$ plays the role of the fair play law's menu (Section 3.1), but the principal's criterion in MD is the expectation rather than the guaranteed result. Deviations of agents' behavior from the rational choice model, and their implications for the form of contracts, are systematized in [336].

## 5.5. Contracts with Many Agents

If the agents' actions are complementary, the agents' game under given contracts may have several equilibria, and a principal who evaluates a scheme by the worst of them must ensure uniqueness. Winter [528] showed that under increasing returns the optimal scheme discriminates even among identical agents: the first receives a reward sufficient to work alone, each next one – sufficient provided the previous ones work. This is a special case of the divide-and-conquer principle in contracting with externalities [479]. Halac, Lipnowski and Rappoport [266] showed that the principal can cut costs by keeping the agents uncertain about their place in this order.

Compare this with the Novikov–Tsvetkov theorem. Scheme (16) achieves a unique dominant-strategy equilibrium without discrimination, at the price of arbitrarily small bonuses, but at the cost of making an agent's reward depend on the observed actions of the others. If the reward depends only on the agent's own action, the same monograph guarantees only a Nash equilibrium, and uniqueness requires additional conditions [428] – the very difficulty that Winter and Segal address. The MD models assume that only the joint result or the agent's own action is observable, and obtain discrimination as the price of uniqueness. A general difference shows up here: **AST more often assumes a better-informed principal and obtains stronger statements**.

## 5.6. Algorithmic Contract Theory

The computational branch was started by the combinatorial agency model [34] and codified by the survey of Dütting, Feldman and Talgam-Cohen [179]. In the combinatorial contracts model [178] the agent chooses a subset $S$ of $m$ actions; the project's success probability is $f(S)$, the costs are $\sum_{j\in S} c_j$. Under a linear contract with share $\alpha$ the agent solves the problem

$$S^*(\alpha) \in \operatorname{Arg}\max_{S\subseteq[m]} \left( \alpha\, f(S) - \sum_{j\in S} c_j \right), \tag{35}$$

which coincides with the demand query of combinatorial auction theory. If $f$ has the gross substitutes property, the optimal contract is computable in polynomial time; for submodular $f$ the problem is NP-hard. The survey [179] also covers contracts for agents with hidden types, learning contracts from data, incentives in classification problems, ambiguous and incomplete contracts, and contracts for social goals; among the open questions it names, in particular, the existence of a strongly polynomial algorithm for computing the optimal contract.

Table 5. Lines of contract theory in 2006–2026 and the closest AST and TCO models

| Line | Setting and main result | Sources | Closest AST and TCO models |
|---|---|---|---|
| Continuous time | Promised utility as the state; equations (30), (31) | [475] | Adaptive mechanisms [124]; dynamic systems in discrete time [120]; adaptation of dynamic contracts under stochastic uncertainty [53, 59] |
| Robustness | An unknown action set; optimality of linear contracts (32) | [136, 161, 183] | The result guaranteed over unknown parameters; interval and fuzzy uncertainty [407]; coordinated mechanisms under interval uncertainty [191] |
| Relational contracts | Self-enforcing agreements; constraint (33) | [351] | No analogue in AST or TCO; dynamic stability of cooperative solutions in Petrosyan's school [446, 445, 531] |
| Incomplete contracts | Control rights; the contract as a reference point | [256, 272, 271] | Structure control; distributed control |
| Delegation | Choice of the set of admissible decisions (34) | [13] | Institutional control [412] |
| Several agents | Unique implementation, discrimination, rank uncertainty | [266, 479, 528] | Decomposition of the agents' game (16) [428]; networked active systems [55] |
| Algorithmic theory | Complexity, approximation, learning; problem (35) | [34, 178, 179] | Optimization of group incentive schemes [93]; complexity of control problems [399] |
| Behavioral theory | Deviations from the agent's rationality | [336] | Bounded rationality and reflexion [417, 404] |

# 6. The Theory of Active Systems and the Theory of Control in Organizations in 2006–2026

## 6.1. TCO as a General Framework

By 2006 the results of AST had been generalized by D.A. Novikov into the *theory of control in organizations*; the foundational monograph appeared in 2005 and has gone through four editions [418, 412], English-language expositions are given in [85, 106] and in the translation of the monograph [414], and surveys in [109, 108].

An *organizational system* (OS) is defined as an association of people jointly pursuing some program or goal and acting on the basis of definite procedures and rules, and a control mechanism as a set of procedures for making management decisions [412]. The OS model is specified by six components: the staff of participants; the organizational structure; the sets of admissible strategies; the preferences (goal functions); the awareness; and the order of functioning. Control is classified by the component the principal changes: staff control, structure control, institutional control (of constraints and norms of activity), motivational control (of preferences and interests), and informational control (of the information the participants hold at the moment of decision making). Control of the order of functioning reduces, in game-theoretic models, to control of structure and/or order of functioning and is not singled out separately. The author calls institutional control the harshest, motivational control softer, and informational control the softest and least studied.

Institutional control is the subject of a separate 2004 monograph [411]. Control of constraints is modeled by the principal's choice of the agent's admissible action set: the agent chooses its best action on this set, the efficiency of control is the value of the principal's goal function, which depends on the action and on the cost of maintaining the constraint, and in a parametric family of nested sets the extreme cases are complete centralization (a singleton) and complete decentralization. A "price of prohibition" is introduced – the principal's minimum cost of a constraint that excludes actions more profitable to the agent than the required one; Proposition 1 states that if the price of prohibiting each action does not exceed the minimum cost of inducing it (the difference between the maximum of the agent's goal function and its value at this action), then the efficiency of institutional control is no lower than that of motivational control, and joint control chooses a balance of prohibitions and incentives. In multi-agent systems global constraints are handled by penalties, by extending strategies, or by coordination, and the principle of decomposition of the agents' game is applied; for boundedly rational agents three types of thresholds and the corresponding relaxations of Nash equilibrium are introduced. Control of norms is defined via a norm as a mapping from the state of nature to an action vector: a norm is coordinated if under common knowledge nobody benefits from deviating from it, and the problem is solved by maximizing efficiency over the intersection of the sets of coordinated and admissible norms. Proposition 2 shows that a payment for following the norm equal to the maximum gain from deviation makes the norm the unique equilibrium and is cost-minimal among motivational controls, so that for single-valued norms flexible motivational control is at least as efficient as institutional control and computationally simpler; unified norms are no more efficient than personalized ones; under incomplete awareness of the agents, the apparatus of reflexive games gives conditions under which a depth-two awareness structure makes any admissible action vector the unique informational equilibrium (the example of lump-sum pay). The author concludes that analytic solutions have been obtained only for simple cases, and that the greatest effect comes from combining institutional control with motivational and informational control. The closest MD analogue is the theory of delegation [13, 285] (Section 5.4), where the principal also chooses the agent's set of admissible decisions, but the criterion is the expectation and the constraint is free; the treatment of norms as equilibria draws on North's institutional economics, and control of norms under inconsistent awareness of the agents has no MD analogue.

The monograph's formal statement covers three types of control with the staff and structure fixed [412]. A control $u = (u_A, u_v, u_I)$ specifies the agent's admissible set, its preferences and its awareness, and thereby the rational choice set $P(u)$; efficiency is defined by the indices (3), by default under the benevolence hypothesis. The general solution scheme is two-step:

$$X(y) = \{u \in U\colon y \in P(u)\},$$
$$G(y) = \max_{u \in X(y)} \Phi(u, y), \qquad y^* \in \operatorname{Arg}\max_{y \in P(U)} G(y), \tag{36}$$

where $P(U)$ is the set of actions implementable by at least one control (written in the notation of (3)): first, for each action, the best control implementing it is found (the principal's minimum "cost" of implementation), then the action is chosen. The properties of coordination in incentive problems and of strategy-proofness in planning problems serve to narrow the class of controls over which the optimum is sought. The same decomposition underlies Grossman and Hart's analysis of moral hazard: first the minimum cost of implementing each action, then the choice of the action [255]. The models are built from hierarchical games: a principal's game of type $\Gamma_1$ or $\Gamma_2$ is superimposed on the agents' game [120, 424]; the game-theoretic machinery is set out in [247]. The *basic model* (one principal, one agent, one-shot decision making, complete awareness) is extended in seven directions: dynamic, multi-agent and multilevel systems, systems with distributed control, with uncertainty, with joint-activity constraints, and with information revelation [412].

The correspondence between the model's components, the types of control and the branches of MD is shown in Fig. 5. It captures the main structural difference: **what in the MD literature constitutes several separate theories with communities of their own (MD, contract theory, delegation, information design, organizational economics) is, in TCO, a set of sections of one scheme with a single problem statement**.

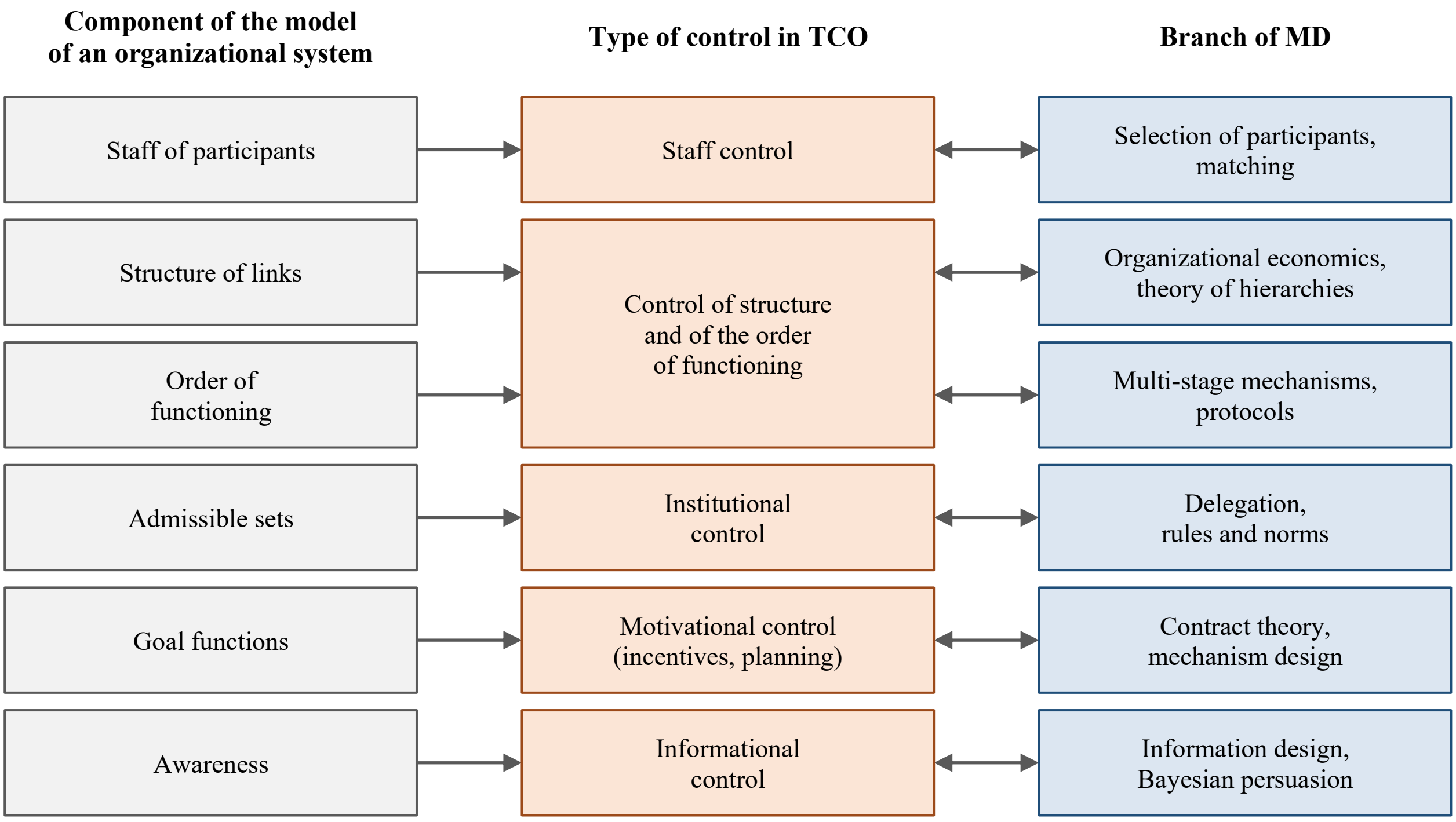

Fig. 5. The components of the organizational system model, the types of control in TCO, and the corresponding branches of MD

The applied layer of the theory is formed by a catalog of standard mechanisms grouped by management function. In the 2005 monograph the table contains 21 mechanisms [412]: planning (resource allocation, active expertise, transfer prices, tournaments, exchange), organization (joint financing, counter-expensive mechanisms, cost–benefit, self-financing, insurance, production cycle optimization, assignment), incentives (for individual results, for the results of collective activity, unified, team pay, incentives in matrix structures), and monitoring (integrated rating, mechanisms of consent, multi-channel mechanisms, supplementary agreements).

The 2011 textbook describes 20 basic and 4 integrated mechanisms [153]; earlier versions of the catalog are given in [119, 120]. A typical example is the transfer-price mechanism [424]. Divisions with Cobb–Douglas-type cost functions report estimates $s_i$ of the parameters $r_i$, and the principal allocates the assignment $R$ and sets a single price:

$$c_i(y_i, r_i) = \frac{1}{\alpha}\, y_i^{\alpha} r_i^{1-\alpha}, \qquad x_i(s) = \frac{s_i}{V} R, \qquad \lambda(s) = \left(\frac{R}{V}\right)^{\alpha-1}, \qquad V = \sum_{j \in N} s_j. \tag{37}$$

The plan $x_i$ maximizes division $\lambda y_i - c_i(y_i, s_i)$'s profit at price $\lambda$, i.e. the perfect coordination condition holds, and under the weak influence hypothesis truthful reporting is a dominant strategy; the total cost of fulfilling the assignment is then minimal. By Theorem 5.6 of the course [424], for any transfer-price mechanism there exists a fair play mechanism of no lower efficiency. The course's theorems have MD predecessors in subject matter: intra-firm transfer prices [281], mechanisms for allocating resources among divisions with truthful reporting [257, 259], and the optimality of transfer-price menus under information asymmetry [268]. The catalog of standard mechanisms as a systematized set organized by management function has no direct MD analogue: the closest genre is the "toolkits" of market design [464].

## 6.2. Strategy-Proof Planning Mechanisms

The main theoretical result of the period in the AST "core" is the incorporation of the school's classical mechanisms into the general theory of strategy-proofness. It was obtained in a cycle of works by N.A. Korgin and coauthors: the identity of the active expertise mechanism (7) with the median scheme and its extension to the multicriteria case [110]; the analytic representation (10) of non-anonymous SRAMs and the equivalence criterion for straight- and reverse-priority mechanisms [328]; and the representation of the SRAM as a generalized median scheme [327]. The pivotal fact of these works is the general form of a strategy-proof allocation mechanism $x_i = \min\{\tau_i, b_i(\tau_{-i})\}$, i.e. condition (4) with sets $X_i(s_{-i}) = [0, b_i(s_{-i})]$.

The summarizing survey [102] offers a unified interpretation: strategy-proofness is achieved by delegating the decision right to the agents. In systems with nontransferable utility, the fair play toolkit reduces to forms of "positional dictatorship" (constant and dictatorial rules, median schemes, serial allotment). In systems with transferable utility, coordination through transfers is added (VCG mechanisms, affine maximizers), but coalitional strategy-proofness is unattainable. The survey also formulates a four-step "active planning" scheme: synthesize the procedure $f$ optimal under truthful information; analyze its manipulability; estimate the error $\Delta_f(\rho) = \max_\omega \parallel f(\omega) - \pi\big(s^*(\omega)\big) \parallel$ introduced by strategic behavior in the mechanism $\rho = \langle S, \pi \rangle$; and synthesize the mechanism minimizing this error. The last setting is close to the MD program of approximate implementation and price-of-anarchy bounds, but uses an absolute rather than a relative loss measure.

The second line is experimental. Mechanism (13) and several alternative resource allocation mechanisms with transferable utility were tested in business games [329, 330, 331]. In the first series (19 games, 769 iterations) an efficient or near-efficient allocation was reached in 8 games, and via cooperation among the participants; the best-response dynamics, for which convergence had been proved, was never observed [329]. The comparison concluded that none of the mechanisms with transfers studied outperformed the transfer-free SRAM in efficiency [102]. This conclusion resonates with the MD work that treats a mechanism's simplicity for participants as a requirement in its own right and tests it in the laboratory [141, 352].

*Dictatorship sets.* The predecessor of Korgin's cycle is S.N. Petrakov's book [444], in which the strategy-proofness conditions for direct planning mechanisms with nontransferable utility and generalized single-peaked separable preferences are written in terms of *dictatorship sets*: the message space is partitioned into regions, in each of which a certain group of agents (the "dictators") receive the plans optimal for them. Sufficient conditions for strategy-proofness, necessary and sufficient conditions for coalitional strategy-proofness, conditions for vector plans and their verification for resource allocation and active expertise mechanisms are obtained; a separate chapter is devoted to the existence of an equivalent direct mechanism for indirect mechanisms (constructive conditions via the Jacobian matrix of the planning procedure; dependence on the set of admissible messages). The same line includes the paper on decentralizing planning mechanisms: anonymous resource allocation, active expertise and transfer-price mechanisms admit arbitrary decomposition into a multilevel structure without loss of efficiency [425]. Relative to MD, dictatorship sets are a constructive form of the same "positional dictatorship" that the survey [102] names the general principle of strategy-proofness without transfers, and that collective choice theory knows as generalized median schemes [41].

*Coalitions of agents in resource allocation mechanisms.* The cooperative extension of this line is given in Goubko's monograph [244] (2003). Coalitional interactions are classified by the transferability of the resource and of utility. With nontransferable resource and utility, in monotone continuous straight-priority mechanisms the strong Nash equilibrium coincides with the ordinary one: neither coalitions of dictators, nor mixed coalitions, nor coalitions of non-dictators can gain (Theorem 7), so robustness to collusion comes without additional conditions, unlike coalitional strategy-proofness in [444]. Under full transferability the coalition's characteristic function is built from the dictatorship sets of the two-coalition game (a coalition against its united complement), and a sufficient condition for the non-emptiness of the core is expressed in terms of the peaks alone: for any balanced cover, the weighted sum of the coalitions' equilibrium resources does not exceed the total resource (Theorem 8); for mechanisms with the "zero-bid property" and linear production functions with a capacity constraint this condition is necessary and sufficient (Corollary 4). The core is nonempty if all peaks are at least the equal share of the resource, or if there is a single non-dictator (Corollaries 5, 6). Two conclusions are substantive: information about the lower bounds of the peaks, useless in the noncooperative model (Theorem 6), becomes decisive in the cooperative one; and the profile with one efficient agent, the worst for noncooperative allocation, yields efficiency 1 in the cooperative model, since a balanced coalition divides the resource the way the principal would under complete information. In the class of continuous monotone mechanisms, balancedness at all profiles is not guaranteed; it is guaranteed by the constant-priority mechanism, which uses no bids (Corollary 7), leading the author to propose an "internal market" for the resource among the agents. The closest MD constructions are different: the coalition-proof equilibria of Bernheim, Peleg and Whinston [72] refine Aumann's strong equilibrium [28], and the core-selecting auctions of Day and Milgrom [167] entrust the choice of the imputation to the mechanism, whereas in Goubko the principal chooses the mechanism so that the agents form the grand coalition and delegates the allocation to it; these works are not cited in the monograph – the comparison is made by the author of this survey.

## 6.3. Incentive-Compatible Mechanisms and the Contract-Theory Problem

The second "core" line continues the 1980s work on incentive-compatible planning [101] and belongs to A.K. Enaleev [189, 187, 190, 191]. The mechanism $\mu = \{\pi(\cdot), f(\cdot,\cdot)\}$ consists of a planning procedure and the agent's goal function $f(x, y)$, which depends on the plan and the action. The set of coordinated plans that the agent finds profitable to fulfill is

$$P = \{x \in X : f(x, y) \le f(x, x) \ \ \forall\, y \in Y\}. \tag{38}$$

If $f(x, y) = h(y) + \sigma(x) - \chi(x, y)$, where $\chi$ is the penalty for deviating from the plan, then a sufficient condition for maximal coordination is the triangle inequality for penalties:

$$\chi(x,y) \leq \chi(x,z) + \chi(z,y) \qquad \forall\, x,y,z. \tag{39}$$

Condition (39) goes back to the fourth part of the cycle "Two-Level Active Systems" [100] (1980; the cycle was started by Burkov and Kondratiev in 1977 [111]), which Enaleev cites as the source of the statement: if the penalty function satisfies the triangle inequality, the mechanism is *maximally coordinated* – the set of plans the agent fulfills coincides with the set of its best responses to the incentive scheme [193]. The condition's significance for coordinated planning and incentives is that it separates the two components of the mechanism: under (39) the class of fulfilled plans is determined by the agent's preferences alone, the penalty merely holding the agent to the plan, so in joint synthesis the plan can be sought on the agent's best-response set while the penalties are taken with the minimal sufficient growth rate. Enaleev [193, 188, 192] introduces the maximal penalty growth index and proves that under (39) the properness of a mechanism (truthful reporting and plan fulfillment) is equivalent to perfect coordination, that for any admissible planning procedure there exists a proper one of no lower efficiency, and that there exists an optimal proper mechanism that is simultaneously maximally and perfectly coordinated; in the two-agent setting [194] the triangle inequality and symmetry are also required of the agents' loss functions from mutual discoordination.

In the model with the agent's cost parameter unknown to the principal, $f(x,y,r) = \sigma(y) - \chi(x,y) - c(y,r)$, $r \in [r_{\min}, r_{\max}]$, the agent reports an estimate of the parameter, receives a plan and chooses an action; admissible rewards are limited by a fund, and the growth rate of penalties by a given function.

**Theorem (Enaleev, 2010–2014).** *Under monotonicity of marginal costs in the parameter (the Spence–Mirrlees condition), there exists a mechanism optimal under the guaranteed criterion that is proper: the agent reports a truthful estimate and fulfills the plan. In this mechanism the penalties have the maximum admissible growth rate, the incentives the minimum growth rate sufficient for self-selection, and the planning procedure is the fair play law* [189, 187].

The optimal incentive scheme consists of a reward for fulfilling the plan and a reward for its tension; the second part is the analogue of the information rent in (20). Without the fund constraint and with an unbounded penalty, the problem reduces to $\max_{x(\cdot)}\min_r\left[H\big(x(r)\big) - \sigma\big(x(r)\big)\right]$ under self-selection constraints, i.e. to the basic adverse selection model with the guaranteed criterion in place of the expected one [189]. The generalizations cover several agents with a common reward fund [190] and network structures, where penalties also arise between adjacent agents and conditions are obtained under which fulfilling the plans is a Nash equilibrium or a dominant strategy [191]. By the setting (an explicit plan, penalties of bounded growth, verification of fulfillment) these models are closest to the MD literature on mechanisms with verification and limited punishments [60, 388], and by the structure "an announced plan plus an asymmetric bonus" to Weitzman's scheme [524], which analyzed the Soviet reform of 1971, and to its sequels [286, 355]. The difference is that Weitzman and a number of his successors analyze a given piecewise-linear scheme under the expected criterion with a random result, whereas AST synthesizes the optimal mechanism under the guaranteed criterion for a deterministic action.

*Joint synthesis of planning and incentives.* In MD, planning under private information (screening and delegation, Sections 3.5 and 5.4) and incentives under hidden action (moral hazard) develop as separate theories with different solution concepts, and their combination in models with a simultaneous hidden type and hidden action remains a specialized topic. In AST a mechanism is from the outset a pair "planning procedure – incentive scheme", and the synthesis problem is posed for the pair: the 1983–1984 cycle [101] obtained necessary and sufficient optimality conditions for proper mechanisms under complete awareness of the principal and, in the third part, under incomplete awareness, synthesizing the planning procedure and the incentive scheme simultaneously; the paper by Burkov, Enaleev and Lavrov [103] is devoted to the synthesis of optimal planning and incentive mechanisms; in [188] the planning procedure and the incentive scheme are synthesized jointly and sufficient optimality conditions are obtained; the survey [102] formulates the principal's choice of the planning mechanism, the reward functions and the penalty functions as a single problem. Jointness does not reduce to solving "first the plan, then the incentives" sequentially: the set of fulfilled plans itself depends on the penalties, and the admissible class of penalties (bounded growth rate, the fund) restricts the implementable plans, so the optimum is sought over the pair. There is no explicit irreducibility theorem in these works, but [194] shows, on two-agent interaction structures, where decomposition is and is not possible: with independent agents the problem splits (first the optimal plan on the sets of coordinated plans, then the penalty with the minimum growth index), while if the principal acts on only one agent, influencing the second through it, the first agent's plan and penalty must be chosen jointly. MD has not addressed joint synthesis problems; the closest setting in structure is the models with simultaneous adverse selection and moral hazard; comparing them with the joint synthesis of AST remains a promising task.

### 6.4. Incentives: Groups, Networks, Multiple Principals

Theorem (16) became the basis for extensions. In the distributed control model [430] an agent is subordinate to several principals (double subordination), each contributing a part of the compensation of its costs. The principals' game has two regimes: when their interests are not too different, the *cooperation mode* is realized, in which the principals jointly compensate the agent's costs and each gains no less than under individual control of the agent; when their interests are nearly opposite, the *competition mode* arises, with an auction solution in which one principal (the dictator) single-handedly compensates the costs and overpays the agent; this is the principle of decomposition of the principals' game. For the general setting with one agent and a compact action set, Karavaev [313, 312] obtained necessary and sufficient conditions for the cooperation mode with outcome $x^*$: each principal's income at $x^*$ is no less than its maximum profit in the individual game with the agent, and the principals' total profit at $x^*$ is no less than the sum of such maxima – that is, the gain from joint control is at least the sum of the gains from separate control. He also showed that the principals' strategies can be restricted to functions with finitely many peaks, that with two principals any Pareto-suboptimal equilibrium is dominated by an optimal one and an optimal equilibrium always exists, and that with two equally strong principals the only equilibria left are the optimal ones, in which the agent captures all the profit. On priority: in Karavaev's monograph the distinction between the regimes is accompanied by a reference to [430]. The common agency problem of Bernheim and Whinston [73], published 15 years earlier, is close in setting; it is mentioned in Karavaev's monograph, but no comparison of the AST school's results with the MD results is made there. The AST results should therefore be considered independent, and comparing them with the MD theory of common agency is a promising task.

*The cooperative version of the principals' game* is considered by Goubko [244] (Ch. 2). For the model of [430] with one agent with convex costs and several ($n$) principals, it is shown that any ε-Nash equilibrium of the principals' game can be replaced by an equilibrium with the same payoffs in which each principal's incentive differs from zero in at most $n$ points (Lemmas 1–3), and the equilibria of two types are characterized: C-type ("cooperation", the agent's payoff is zero) and K-type ("competition", the agent receives the same positive payoff at all threat points);

the conditions are necessary and sufficient (Theorem 1), and the non-emptiness condition for the C-type zone coincides with Karavaev's cooperation-mode condition. For coalitions of principals, four ways of constructing the characteristic function are proposed: the guaranteed result with bluffing allowed and with bluffing forbidden, and the guaranteed and the average result over Pareto-optimal equilibria; for the first three, sufficient conditions for the non-emptiness of the core are proved via properties of the primitives (smooth concave nondecreasing incomes of the principals and convex separable costs of the agent, or collinearity of the coalitions' optimal plans; Theorems 2, 3), while under the average result the game has constant sum and, if essential, is unbalanced – "optimistic" principals do not seek a coalition. Coordination of the principals' coalition with top management is achieved by a flat profit tax and an income tax with rate $\beta_i = \frac{1}{1+\alpha_i/\alpha_0}$, where $\alpha_i$ is the weight of project *i*'s income in the management's criterion and $\alpha_0$ the weight of the costs. Goubko's characteristic function is defined by the Nash equilibrium of a coalition against its united complement, which corresponds to one of the two assumptions of Hart and Kurz's theory of endogenous coalition formation [273] (the other players stay united), whereas Chander's γ-core [140] adopts the other assumption (the rest break up into singletons); the first assumption gives a stricter stability condition. The works [73, 140, 273] are not cited in the monograph; as with Karavaev, no comparison with common agency is made.

In the *networked active systems* of Belov and Novikov [55], an agent's result depends on its actions and on the results of other agents according to a technology network; the problems of incentive-compatible planning and incentives are solved for them, with an application to "extended enterprises". The problem of optimizing group incentive schemes [93] consists in partitioning jobs into groups with a unified pay scheme so as to minimize the wage fund; it reduces to a shortest path problem. Counter-expensive pricing mechanisms, in which the profitability norm depends on product efficiency in such a way that cutting costs benefits even a monopolist, have been in the AST school's catalog since the 1980s [120]. Petrov and Chkhartishvili [449] considered incentives in a network game with linear best responses under divergent beliefs of the agents: for goal functions $f_a(x) = \theta_a x_a - 1/2\, x_a^2 + \beta x_a \sum_b g_{ab}\, x_b$, the informational equilibrium exists, is unique for sufficiently small $\beta$, and has the form

$$x^* = (I - \beta G_I)^{-1}\theta, \qquad (40)$$

where the matrix $G_I$ is built from the agents' awareness structure (under common knowledge, $G_I = G$); the paper studies how informing the agents about the parameters changes the equilibrium. This work draws on the literature on network games: the authors cite Ballester et al. [38]; the models of targeted interventions [217] are also close.

## 6.5. Informational Control and Reflexive Games

### *6.5.1. Point-Type Awareness Structures and Informational Equilibrium*

The term "*reflexive games*" and the idea of reflexive control as shaping the partner's beliefs go back to V.A. Lefebvre's work of 1965–1973 [347] and his later work with an algebraic formalization [348]; the rigorous game-theoretic formulation with the notions of an awareness structure and an informational equilibrium belongs to Novikov and Chkhartishvili and was first systematically set out in the 2003 monograph [421], developed in [422, 404], which note the qualitative character of Lefebvre's work.

The chronology of the two lines shows that they ran in parallel: Lefebvre's "Initial Ideas of the Logic of Reflexive Games" [346] is two years older than Harsanyi's Bayesian games [270], and “The Algebra of Conflict” [349] appeared before Aumann's work on common knowledge [29]. Lefebvre's models were algebraic and qualitative, with no payoff functions or equilibrium; they are almost never cited in economic game theory. The predecessor of the game-theoretic form within AST, besides Germeier's school, was the “active forecast” [420]. The machinery of point-type structures was subsequently carried over to social networks (2010) and mobs (2016), and since 2008 it has been developing toward multiple structures (Section 6.5.2). In Germeier's school the term “reflexive strategies in controlled systems” was used by Ereshko and Lokhnygina in 1999–2001 [202, 198, 199] independently of the TCO line (Section 6.7).

The 2003 monograph [421] consists of four chapters: information in decision making (the rational choice model, games and equilibria, ways of describing awareness); *strategic reflexion* (reasoning about opponents' decision principles under complete awareness, reflexion in bimatrix games, the boundedness of the rational reflexion rank); *informational reflexion* (the awareness structure, informational equilibrium, the graph of a reflexive game, regular structures, the link between reflexion rank and equilibrium, reflexive control); and applied models (implicit control, informational control through the mass media, reflexion in psychology and in works of fiction). The distinction between strategic and informational reflexion, phantom agents, regular awareness structures, the problem of reflexive control as choosing the awareness structure under which the informational equilibrium is most favorable to the principal, and the result on the finiteness of the maximal rational reflexion rank in a number of game classes were all introduced already in that edition; the stability of equilibria, their division into true and false, and reflexive partitions appeared in later work and entered the 2013 monograph [422]. In the 2003 introduction the approach is related to Harsanyi's games, common knowledge in the sense of Lewis and Aumann, Howard's metagames, and the experimental literature on belief hierarchies; the universal type space of Mertens and Zamir is not yet discussed in the 2003 edition. Relative to MD this means that reflexive games arose as a generalization of Nash equilibrium to incomplete and inconsistent awareness; the rapprochement with Bayesian theory came several years later.

The reflexive games of Novikov and Chkhartishvili [422, 404] describe the agents' mutual awareness by a tree of beliefs: $\theta_i$ is agent $i$'s belief about the uncertain parameter, $\theta_{ij}$ its belief about agent $j$'s belief, and so on. The awareness structure $I = (I_1, \dots, I_n)$, $I_i = (\theta_i, \theta_{ij}, \theta_{ijk}, \dots)$, unlike a Bayesian game, consists of point beliefs, need not be consistent, and presupposes no common prior. The authors view their approach as intermediate between two “extremes”: Harsanyi's Bayesian games [270], where the type distribution is common knowledge, and the Mertens–Zamir universal belief space, which covers all hierarchies but is too unwieldy for obtaining solutions [422].

Alongside the real agents, *phantom* agents participate in the game – agents existing in the minds of others. An *informational equilibrium* is a set of actions $x^*_\sigma$ of the real and phantom agents (here $\sigma$ is a sequence of indices), each of which is a best response to the actions that the given agent expects from the others:

$$x^*_{\sigma i} \in \operatorname{Arg} \max_{x_i \in X_i} f_i\big(\theta_{\sigma i}, x^*_{\sigma i 1}, \dots, x^*_{\sigma i, i-1}, x_i, x^*_{\sigma i, i+1}, \dots, x^*_{\sigma i n}\big). \tag{41}$$

The definition additionally requires the awareness structure to have finite complexity and agents with identical structures to choose identical actions. Under common knowledge, equilibrium (41) coincides with Nash equilibrium.

An informational equilibrium is *stable* if the results the agents observe coincide with those they expected; a stable equilibrium may be false, i.e. not an equilibrium under the true awareness.

The informational control problem consists in the principal's choosing the awareness structure from an admissible set and is posed in two variants, under the benevolence hypothesis and under the guaranteed result:

$$\max_{I\in\mathbb{I}} \max_{x\in\mathcal{X}(I)} \Phi(x,I) \qquad \text{or} \qquad \max_{I\in\mathbb{I}} \min_{x\in\mathcal{X}(I)} \Phi(x,I), \tag{42}$$

where $\mathcal{X}(I)$ is the set of equilibrium actions of the real agents, and the maximum is taken over the structures for which it is nonempty; one can additionally require the equilibrium to be stable and account for the principal's cost of forming the structure. The kinds of influence are informational regulation (a message about the state of nature), reflexive control (a message about the opponents' beliefs) and the active forecast [412, 422]. The results on reflexion rank show that deepening reasoning pays only up to a limit: in two-person games with stationary reflexive mappings the maximal rational subjective rank is two, and structures of infinite depth, under a certain condition, make any admissible action an equilibrium (Propositions 2.6.4 and 2.10.1 in [422]). Strategic reflexion (reasoning about the depth of the partners' reasoning) is described by the method of reflexive partitions, whose special cases include the MD models of level-$k$ reasoning and the cognitive hierarchy [404].

The link between point-type structures and the Bayesian formalism was established by Chkhartishvili [144]: point-type structures of infinite depth are described in terms of Bayesian Nash equilibrium. In type-space terms, each class of identical agents $\sigma i$ is assigned a player type $i$ with a degenerate belief at the point $\big(\theta_{\sigma i}, (\sigma ij)_{j\neq i}\big)$; a structure of finite complexity is a finite type space without a common prior, and condition (41) coincides with the condition of a pure-strategy Bayesian equilibrium on this space, with the single proviso that identical agents choose equal actions. Point-type reflexive games thus remain within the Bayesian formalism and single out within it the class of finite type spaces with degenerate and inconsistent beliefs, which yields tractable models and explicit solutions.

### *6.5.2. Multiple and Generalized Multiple Awareness Structures*

A point-type structure does not describe the agent's own lack of confidence: each of its instances is certain of one value at each level. *Multiple awareness structures* (MAS) lift this restriction [146, 145]. Let $A_i$ be the set of instances of agent $i$ (the real one and phantoms), let a possible world $\omega = (\theta, a_1, \dots, a_n)$ specify a state of nature and one instance of each agent, let instance $a_i$ consider the worlds $\mu(a_i)$ possible, and let $\omega^*$ be the real world. Identity (an agent is present in all worlds it considers possible and knows its own beliefs), unity of the world (every world and agent are linked to $\omega^*$ by a chain "world – an agent in it – a world that agent considers possible") and correctness ($\mu(a_i) \neq \varnothing$) are required. An equilibrium is a set of functions $\varphi_i: A_i \to X_i$ in which each instance chooses its action by the guaranteed result over the set of worlds possible for it:

$$\varphi_i(a_i) \in \operatorname{Arg}\max_{x_i\in X_i} \min_{\omega\in\mu(a_i)} f_i\Big(\theta(\omega),\, x_i,\, \varphi_{-i}\big(a_{-i}(\omega)\big)\Big), \qquad a_i \in A_i,\ i \in N. \tag{43}$$

The point-type structure is the special case where each $\mu(a_i)$ consists of a single world: the minimum is over one point, and (43) coincides with (41). Observing the value of the function $w_i(\theta, x)$, an agent's instance removes from $\mu(a_i)$ the worlds where the observation would have differed, after which the worlds and agents that have lost their link to the real world are deleted; the puzzles "three wise men" and "sum and product" are worked through on this transformation scheme [146]. The guaranteed-result criterion makes MAS an analogue of models with Knightian uncertainty rather than of Bayesian ones (Section 6.5.3); the composition of the game in MAS is fixed.

*Generalized multiple structures* (GMAS) abandon the fixed composition as well: each world is assigned its own set of agents, and the number of participants in an imagined world may be larger or smaller than the real one [210]. An interval GMAS is a tuple $\langle \Theta, \Omega, \omega^*, A, \theta(\cdot), \xi(\cdot), \eta(\cdot) \rangle$, where $\Omega$ is the set of worlds, $A$ the set of agents, $\theta(\omega)$ the state of nature in world $\omega$, $\xi(\omega) \subseteq A$ the agents present in world $\omega$, and $\eta(a) \subseteq \Omega$ the worlds agent $a$ considers possible; a probabilistic GMAS additionally specifies distributions $p_a$ on $\eta(a)$ (notation after [209]). The real participants are the agents $\xi(\omega^*)$; the rest are phantom. Besides identity ($\omega \in \eta(a) \Rightarrow a \in \xi(\omega)$), connectedness and correctness, *regularity* is introduced: $a \in \xi(\omega) \Rightarrow \omega \in \eta(a)$, i.e. an agent considers possible every world in which it is present. In a regular structure $\eta(a) = H(a) = \{\omega : a \in \xi(\omega)\}$; without regularity an agent may be present in a world it considers impossible – this is how false beliefs are modeled, including beliefs about the composition (certainty that there are no competitors when there are). The equilibria of the interval and probabilistic GMAS:

$$x_a^* \in \operatorname{Arg}\max_{x_a \in X_a} \min_{\omega \in \eta(a)} f_a\big(\omega, x_a, x^*_{\xi(\omega)\{a\}}\big),$$
$$x_a^* \in \operatorname{Arg}\max_{x_a \in X_a} \sum_{\omega \in \eta(a)} p_a(\omega) f_a\big(\omega, x_a, x^*_{\xi(\omega)\setminus\{a\}}\big), \qquad a \in A. \tag{44}$$

An agent's payoff depends on the world (the state and the composition) and on the actions of the agents present in it, so the goal functions must be defined for profiles of variable length – via agents' roles or an aggregate of actions, for example. The probabilistic variant is built to deliver Bayesian Nash equilibria; the interval one generalizes (43). MAS is the special case of GMAS in which the set of agents is partitioned into $n$ classes and each world contains exactly one representative of each class; the point-type structure is the special case of MAS (Table 6, Fig. 6).

Table 6. Types of awareness structures in AST

| **Structure** | **What is uncertain** | **The agent's belief** | **Choice criterion** | **Set of participants** |
|---|---|---|---|---|
| Point-type [421] | The state of nature and the others' beliefs; the agent itself has no doubt | One point at each level of the hierarchy | Best response | Fixed, common knowledge |
| MAS [145] | The state of nature and the others' beliefs | A set of worlds $\mu(a_i)$ | Guaranteed result | Fixed |
| Interval GMAS [210] | The state, beliefs, composition | A set of worlds $\eta(a)$ | Guaranteed result | Depends on the world |
| Probabilistic GMAS [210] | The state, beliefs, composition | A distribution $p_a$ on $\eta(a)$ | Expected payoff | Depends on the world |

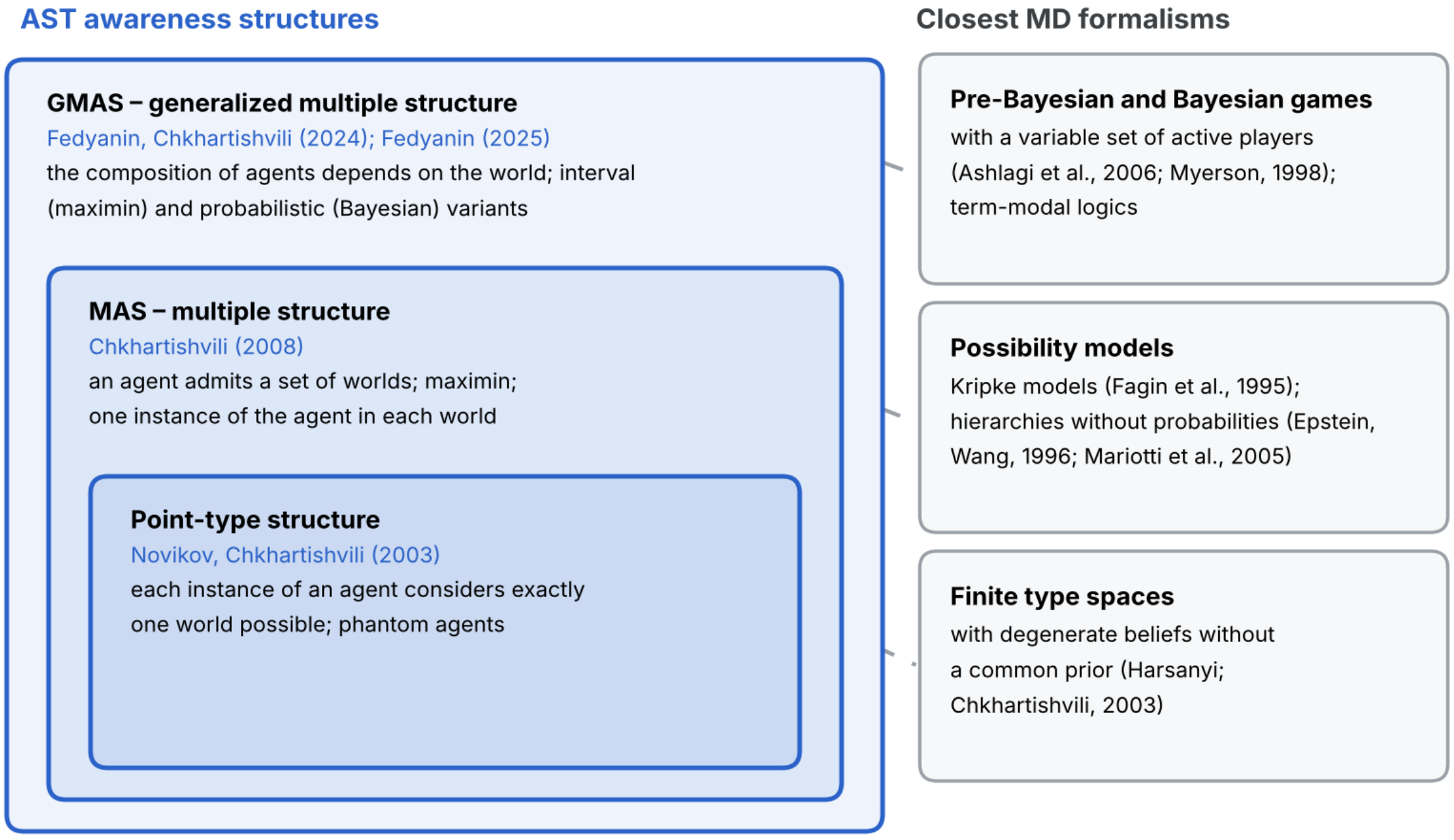

Fig. 6. The nesting of AST awareness structures and the closest formalisms of MD and epistemic logic

*Transformations*. Upon observing the value of its observation function, an agent splits into copies by the distinct values: the copy with value $v$ is present in the worlds of $H(a)$ with this value and considers possible the worlds of $\eta(a)$ with the same value; probabilities are updated by Bayes' rule, and disconnected parts are deleted; the number of agents may grow, the number of worlds may not. A public announcement $g(\omega^*) = 0$ removes from $\Omega$ and from all $\eta(a)$ the worlds with $g(\omega) \neq 0$ and preserves regularity. In the non-regular case an agent may observe something it considered impossible, and its set of possible worlds becomes empty; this case requires separate rules, whose MD counterparts are updating on a zero-probability event and growing awareness [280, 315].

*The example* "the child and the bus" [210] shows composition uncertainty in its pure form (Fig. 7). The worlds: B – only the bus is on the road, C – only the child, B&C – both; the real world is B&C. The driver does not know whether there is a child: $\eta(\text{driver}) = \{B, B\&C\}$; the child does not know whether a bus is coming: $\eta(\text{child}) = \{C, B\&C\}$; each allows that it is playing alone. The horn is the public announcement "there is a bus": world C is removed, the child learns that there are two players in the game, while the driver still does not know whether there is a child. One message has turned two-sided composition uncertainty into one-sided, and this is known to both of them.

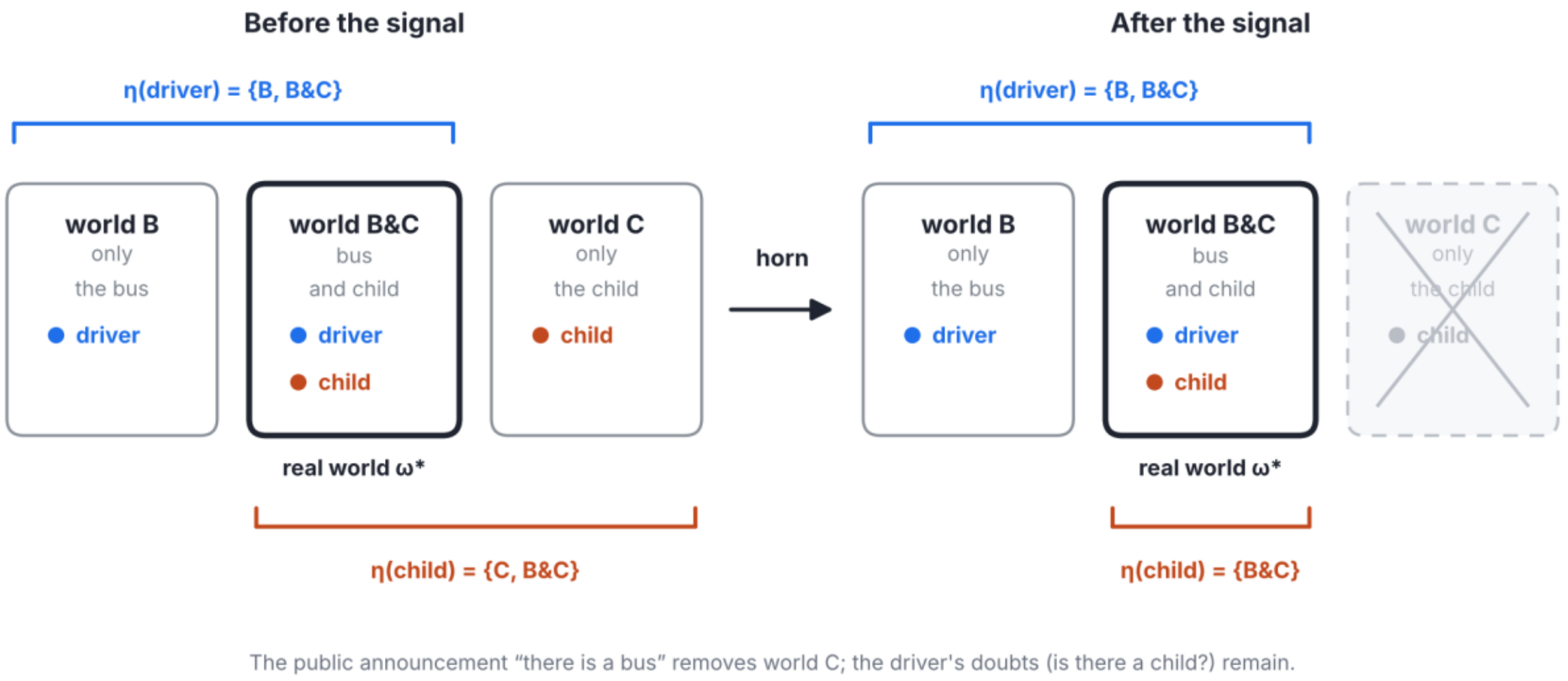

Fig. 7. The "child and bus" GMAS before and after the driver's signal. The world frames show their composition; the braces are the sets of worlds the agents consider possible; the bold frame is the real world

Fedyanin's dissertation [209] proves the equivalence of regular GMAS to pointed Kripke structures of the logic S5, obtains a doubly exponential bound on the complexity of exhaustively enumerating the structures, lower bounds on the number of worlds that must be excluded to make the agents fully informed in one or two steps, and applications (opinion polarization, smart-city sensors, groups of underwater robots, transport safety).

*Control in multiple structures.* The principal's instrument is messages that exclude worlds: a public one removes the same worlds for everyone, a private one changes one agent's beliefs. The principal's task is to choose a sequence of messages so that, after the transformation, equilibrium (44) is favorable to it; an important special case is full awareness, when every real agent considers only the real world possible. What is new compared to the point-type case is *control of beliefs about the composition*: in TCO, staff control (who participates) is traditionally separated from informational control (what the agents know), and GMAS join these types. Without changing the real composition, the principal can announce the existence of a participant (the driver's horn), convince the agents of the existence of a competitor that does not exist, or conceal a participant by leaving worlds without it in the others' beliefs. The MD analogues are decisions on disclosing the number of auction participants, the organizer's shill bids, and the value of not knowing the number of players (Section 6.5.4).

### 6.5.3. The Epistemic Foundations of MD and Formal Correspondences

MD theory describes the same belief hierarchies as reflexive games, but, as a rule, probabilistically and with a common prior. In Harsanyi [270] an agent's type specifies its belief about the state and the others' types, and the Bayesian equilibrium maximizes the expected payoff of each type; the common prior yields Aumann's theorem: if the posterior probabilities of an event are common knowledge, they are equal [29]. Reflexive games have no such restriction: agents may know that they disagree and still keep their opinions. Mertens and Zamir [366], and then Brandenburger and Dekel [80], showed that consistent hierarchies form a *universal type space* $T_i^* \cong \Delta(\Theta \times T_{-i}^*)$ into which any type space embeds; the authors of reflexive games place their approach between this construction and Harsanyi's Bayesian game (Section 6.5.1).

*Possibility sets.* Non-point beliefs without probabilities were developed in epistemic logic and the theory of Knightian uncertainty. Already in Aumann, knowledge is given by partitions of the state space; in multi-agent epistemic logic these are Kripke models with accessibility relations [207], knowledge corresponding to the logic S5 and fallible belief to KD45. The MAS and GMAS conditions read in these terms: correctness is axiom D, regularity axiom T, and identity provides introspection; the equivalence of regular GMAS to S5 structures is proved in [209], while the correspondence of correct non-regular GMAS to the logic KD45 is a conjecture of the author of this survey. For hierarchies without probabilities, analogues of the universal space have been built [195, 357], into which finite MAS embed naturally. The maximin over a set of worlds is a special case of maxmin expected utility [233] when the set of priors contains all distributions on $\eta(a)$; in robust games [4] a player maximizes its worst-case payoff with the others' strategies fixed, which is close to (43), but in MAS different worlds contain different instances of the partners with different actions. A summary of the correspondences is given in Table 9 (Section 7.1).

*The role of higher orders.* MD has shown that higher-order beliefs can be decisive: in the "email game", knowledge of any finite depth yields a different outcome than common knowledge [471]; in global games, small noise in private signals destroys common knowledge and leaves a unique equilibrium [135]; with a sufficiently rich payoff space, any rationalizable action becomes uniquely rationalizable under a small perturbation of higher-order beliefs [523]. The AST results point in the opposite direction: under stationary reflexive mappings, depth beyond two does not enlarge the set of implementable outcomes (Propositions 2.6.4 and 2.10.1 in [422]). The results do not contradict each other, as they rest on different conditions (stationarity versus "richness" and infinite depth), but the exact boundary between them has not been established (problem 19 in Section 10.4).

*Hypergames.* Among the models of MD and operations research, the closest to point-type reflexive games are hypergames [61, 522]: each player sees its own game $G_i$ (composition, action options, preferences) and takes it to be common knowledge; second-level hypergames additionally specify beliefs $G_{ij}$ about which game each other player sees. The shared core is subjective point beliefs without probabilities, which may be inconsistent and false; the game $G_{ij}$ is the analogue of the belief $\theta_{ij}$, and the "image" of player $j$ in the eyes of $i$ is a phantom agent. If the players disagree only in preferences, a level-$k$ hypergame can be assigned a point-type structure of depth at most $k$ (at the last level a player takes its game to be common knowledge), and the hyper Nash equilibrium [522] turns out to be an informational equilibrium in which every agent takes its $\theta_i$ to be common knowledge (an observation by the author of this survey). There are two differences between them. In a hypergame the whole game, including the set of players, can be misperceived – this does not fit a point-type structure, but does fit a GMAS with $|\eta(a)| = 1$ and worlds of different composition. A finite-level hypergame is a tree, each branch of which ends in common knowledge of some game, whereas a graph of finite complexity admits cycles (agent 1231 identical to agent 1 for different $\theta_1, \theta_{12}, \theta_{123}$) that are not reducible to common knowledge of a single game. The representability of any hypergame as a Bayesian game is proved in [476]; both constructions are degenerate cases of Bayesian games without a common prior, covering different parts of this class. Models of unawareness [211, 279] describe a different subject – what an agent is unable to conceive; a false value of a known parameter is not considered in them.

*Dynamic epistemic logic* (DEL) provides a direct analogue of MAS and GMAS transformations [513]. A public announcement [451] restricts the Kripke model to the worlds where the announcement is true – exactly the GMAS transformation under a public message; action models [39] describe private and semi-public messages via the product of the model with an event model, and the splitting of an agent into copies by observation value is structured as such a product. The main difference is the source of the updates: in DEL the messages are given from outside, in AST the observations are generated by the agents' own equilibrium actions, so the transformation of the structure is coupled with solving the game. Term-modal logics [213, 432] carry DEL over to a variable set of agents and are the closest analogue of GMAS semantics, but contain no payoffs or equilibria.

### *6.5.4. An Uncertain Set of Participants: MD Analogues of GMAS*

Composition uncertainty in GMAS has four sides: the objective one (a participant does not know who else participates), the subjectivity and inconsistency of beliefs about the composition at all levels of the hierarchy, phantom participants as full-fledged agents of the equilibrium, and the two criteria and transformation rules. In the standard Harsanyi setting the set of players is fixed and common knowledge, and composition uncertainty is introduced by the device of “inactive” types: every potential player has a type “I do not participate” with a single action. The price of the device is twofold: the set of potential players must be common knowledge (the motive of unawareness models), and with many of them the model is unwieldy (the motive of Poisson games). The MD models with an uncertain composition are scattered across several schools (Table 7, Fig. 8).

Table 7. MD models with an uncertain set of participants and their relation to GMAS

| Strand | How the composition is specified | Solution | Relation to GMAS |
|---|---|---|---|
| Auctions with a random number of participants [362] | The number of participants is a random variable with a commonly known distribution | Bayesian equilibrium; the bid is a weighted average of the bids under a known number | A regular probabilistic GMAS with a common prior |
| Poisson games and games with population uncertainty [368, 386, 385] | A random subset of potential players; in Myerson the number of players of each type is Poisson-distributed | Bayesian equilibrium; “environmental equivalence” | The same for large populations |
| Ambiguity about the number of participants [350] | A set of distributions of the number of participants | Maxmin expected utility | Between the interval and the probabilistic GMAS |
| Prebayesian games with an unknown number of players [26] | A state is a set of active players; no probabilities; each knows only that it is active | Safety-level (maximin) equilibrium | A regular interval GMAS |
| Unawareness [211, 280, 279] | A player may be unaware of another player's existence or actions | Generalized Nash equilibrium, rationalizability | A subjective composition, but unawareness is not the same as certainty of absence |

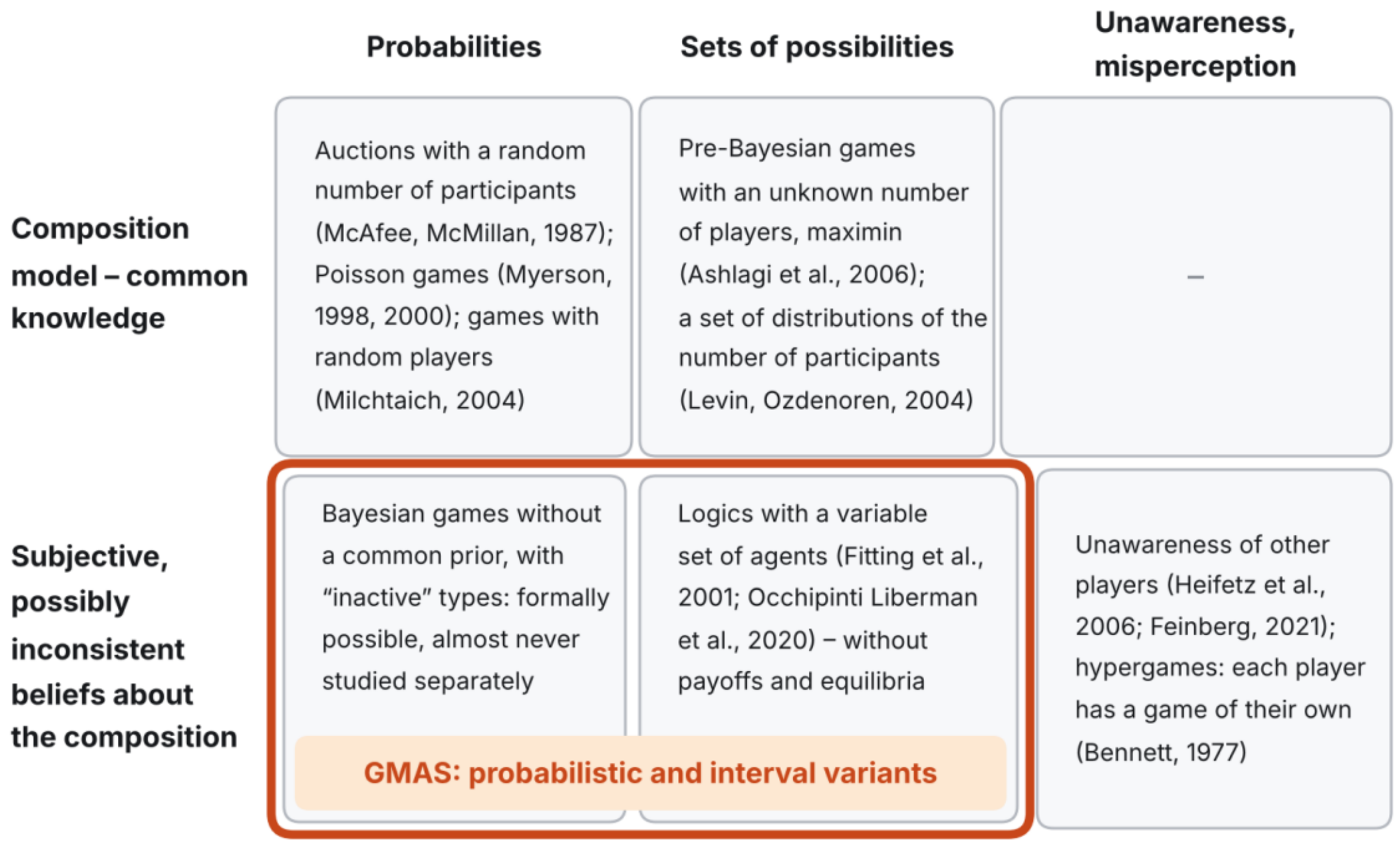

Fig. 8. MD models with an uncertain composition, by two attributes: how the uncertainty is specified (columns) and whether the model of the composition is common to all players (rows). The frame marks the cells where GMAS give a unified model with an equilibrium and transformation rules while the MD analogues are scattered or unconnected to game theory; GMAS contain the top row as a special case

*Formal comparison.* A prebayesian game is a Bayesian game without probabilities on states [26]: a state specifies the set of active players, and a safety-level equilibrium requires every active player to maximize the minimum payoff over the states it considers possible. Assigning to each GMAS agent, real or phantom, a separate potential player, and to each world $\omega$ a state with active players $\xi(\omega)$, we find that the pure-strategy safety-level equilibrium condition coincides with the first condition of (44) for $\eta(a) = H(a)$, i.e. in the regular case; the player's only information is “I participate” (this is how the model of [26] is built, where the existence of a mixed-strategy equilibrium for finite games is proved). Similarly, if the distributions $p_a$ are the conditional distributions of a single prior $\pi$ on $\Omega$ given “the agent participates”,

$$p_a(\omega) = \frac{\pi(\omega)}{\pi\big(H(a)\big)}, \qquad \omega \in H(a), \tag{45}$$

then a regular probabilistic GMAS is a game with population uncertainty [368] with a finite set of potential players, and Poisson games [386, 385] are its limiting case for large populations. Without a common prior one obtains a class of games with subjective and inconsistent beliefs about the composition which, as far as could be established, has not been studied in MD as a class of its own. The translation is formal: all phantom agents must be treated as players, i.e. one must build a set of potential players of which the agents themselves are unaware – the same price as in reducing hierarchies to Harsanyi types. Non-regular GMAS, where an agent is present in a world it considers impossible, do not fit prebayesian games: they require models of fallible belief (KD45 instead of S5).

*The upshot of the comparison.* Every ingredient of GMAS has an MD analogue, but in different schools and unsystematically; there is no model that would simultaneously admit subjective and inconsistent beliefs about the composition at all levels (including false ones), make phantom agents full-fledged participants of the equilibrium, contain both the interval and the probabilistic criterion, specify transformation rules for observations and messages, and pose the problem of controlling beliefs about the composition. In the other direction, GMAS can borrow growing awareness [280, 315], large-population asymptotics [386] and existence theorems [26]. The question of the value of information about the composition was raised in MD long ago: with risk-neutral participants and independent values, disclosing the number of participants does not change expected revenue, while under risk aversion it does [362]; with maximin preferences of the participants the seller may benefit from concealing the number [350]; in resource selection games, common ignorance of the number of players benefits every player, so a revenue-maximizing organizer should disclose the number while a welfare-maximizing one should not [26]; the organizer's shill bids [250] are the literal analogue of a phantom agent created by the principal. In all these works the composition or its distribution is common knowledge, and control reduces to the single decision "disclose or not"; GMAS allow the problem to be posed more broadly – whom to tell what about the composition, taking into account higher-order beliefs and subsequent observations.

### *6.5.5. Stability, Learning, and Bounded Depth of Reasoning*

The stability condition for an informational equilibrium (the observed result coincides with the expected one for every real and phantom agent) is logically analogous to the requirement of self-confirming equilibrium [216] and conjectural equilibrium [52], in which beliefs may be wrong but are not refuted by what is observed; the agent's observation function plays the role of the feedback function. There are three differences. First, in those concepts beliefs concern the opponents' strategies, whereas in an informational equilibrium they concern a hierarchy of point beliefs from which the others' actions are derived as best responses of phantom and real agents; hypergames, in which players perceive the game differently [61], come closest, but they too lack the tree of phantom agents. Second, consistency with observations is part of the definitions of self-confirming equilibrium and Berk–Nash equilibrium [205] (where beliefs are moreover selected by closeness to the true outcome distribution), whereas an informational equilibrium may be unstable, and the awareness structure is given from outside or chosen by the principal. Third, the named concepts admit an interpretation as a stationary state of a "learning" process, while an informational equilibrium is static. Models of unawareness [279] were considered above (Section 6.5.3) and concern a different subject. The kindred cursed equilibrium [307] and the theory of deception [206] show that belief manipulation is studied in MD as well, but there the manipulator is a participant of the game, not a principal choosing the awareness structure.

The closest analogue of stability that accounts for phantom agents is *rationalizable conjectural equilibrium* [204, 472], where beliefs are consistent both with observations and with the others' rationality. Berk–Nash equilibrium [205] explains *why* a false belief is persistent: it explains the data better than the other admissible ones (minimum Kullback–Leibler distance), and the dynamic foundations of misspecified learning show how an overconfident agent attributes failures to external causes and drifts ever further from the optimum [278]. In a reflexive game a false equilibrium is stable if the data simply do not contradict it, and the mechanism of structure formation is not modeled; in multiple structures the link is tighter: a structure that the observation-driven transformation leaves unchanged is the natural candidate for a stable MAS, coinciding in meaning with the set of observation-consistent beliefs in a self-confirming equilibrium.

*Bounded depth of reasoning.* The level-$k$ models [389] and the cognitive hierarchy [134] are the analogue of AST's strategic reflexion. In the level-$k$ model a level-0 player acts nonstrategically and a level-$k$ player best-responds to level $k-1$; in the cognitive hierarchy a level-$k$ player believes the others are distributed over all lower levels in proportion to their shares:

$$f(k) = \frac{e^{-\tau}\tau^k}{k!}, \qquad g_k(h) = \frac{f(h)}{\sum_{l=0}^{k-1} f(l)}, \qquad h = 0, \dots, k-1, \tag{46}$$

and by the authors' data the average depth $\tau$ in most games is close to 1.5, while the survey of experiments finds most subjects at levels 1 and 2 [156]; in the "beauty contest" (the winner is the one closest to 2/3 of the average), level 1 names 33, level 2 names 22, and these are the values most often observed in experiments [389]. The method of reflexive partitions [404] contains both models as special cases: an agent of rank $k$ who takes all others to be of rank $k-1$ yields the level-$k$ model, while one who takes them to be distributed by a truncated Poisson law on $\{0, \dots, k-1\}$ yields the cognitive hierarchy. Both research traditions reach the same conclusion – the real depth of reasoning rarely exceeds two levels – but the status of the conclusion differs: the bounds on the maximal rational rank in AST are statements about when deeper reflexion is useless, whereas the level distributions in MD are statistical estimates of behavior. For control purposes, bounded depth is used in MD along three lines: strategic deception of partners of bounded depth, including the disinformation before the Normandy landings [155] – almost exactly the subject of Lefebvre's reflexive control; mechanism design for level-$k$ agents [169]; and deception under coarse thinking [206]. In AST the synthesis of incentive and planning mechanisms for agents of different ranks has hardly been considered; the first work of this kind is the incentive problem in a reflexive game with a point-type structure and linear best responses [449], where the starting point is a network game with equilibrium $x^* = (I - \beta G)^{-1}\theta$ (Bonacich centrality [38]), the principal communicates incentives to the players individually, and the equilibria under common knowledge and under incomplete awareness are compared.

### *6.5.6. Informational Control and Information Design*

The comparison with information design (Section 4.5) shows different answers to the same question. In problem (26) the principal is bound by commitment and by the consistency of beliefs with the prior, and the gain from controlling information is bounded by the concave envelope. In problem (42) the principal may shape inconsistent and false beliefs, and the disciplining condition is stability. The first model is normatively stricter and admits general machinery (linear programming over the set of Bayes correlated equilibria); the second describes manipulation and reflexive control, which the first excludes by construction. Between these poles the MD literature contains intermediate settings. The principal's commitment can be relaxed: in Min's model [371] the commitment holds with a given probability, and the sender's payoff is nondecreasing in this probability, occupying the space between cheap talk [157] and full persuasion; in the model of Lipnowski, Ravid and Shishkin [354] the sender can alter the report with some probability, and its payoff is described by the concave envelope of a truncated value function. Belief consistency can be relaxed: in Alonso and Câmara [12] the participants hold different priors, and in de Clippel and Zhang [170] the receiver updates beliefs by a non-Bayesian rule, the concavification method surviving for a wide class of rules; in subject matter (belief divergence as the source of gains from controlling information) these models are closer to reflexive control than the common-prior model is. None of these settings admits the false belief hierarchies characteristic of problem (42). There are no publications on Bayesian persuasion by the AST school in 2020–2026, so this direction remains promising for comparative research.

The two research traditions (MD and AST) diverge along two axes – whether the principal is bound by commitment, and how disciplined the agents' beliefs are; classical information design and reflexive control occupy opposite corners of this space, and the MD literature of 2015–2025 is shifting toward the AST corner (Fig. 9). A public signal corresponds to uniform informational regulation, private signals [14] to non-uniform; reflexive regulation (the principal tells an agent what it told the others) has no exact analogue, since the signal structure in information design is common knowledge. Design evaluated at the worst equilibrium [360] is the analogue of AST's guaranteed criterion, and competition among senders [219] of informational confrontation. The main difference between reflexive control and cheap talk is that the agents accept the principal's message without strategic decoding, and the closest analogues are the models with credulous receivers and lying costs [316], deception [155, 206] and persuasion with models [478], where the requirement "do not contradict the data" is the direct analogue of stability. The messages by which MAS and GMAS are transformed are, by contrast, truthful: the message $g(\omega^*) = 0$ is true in the real world and merely excludes the worlds where it is false – in this respect control in multiple structures is closer to verifiable disclosure than to deception, and in Fig. 9 it lies above the point-type models.

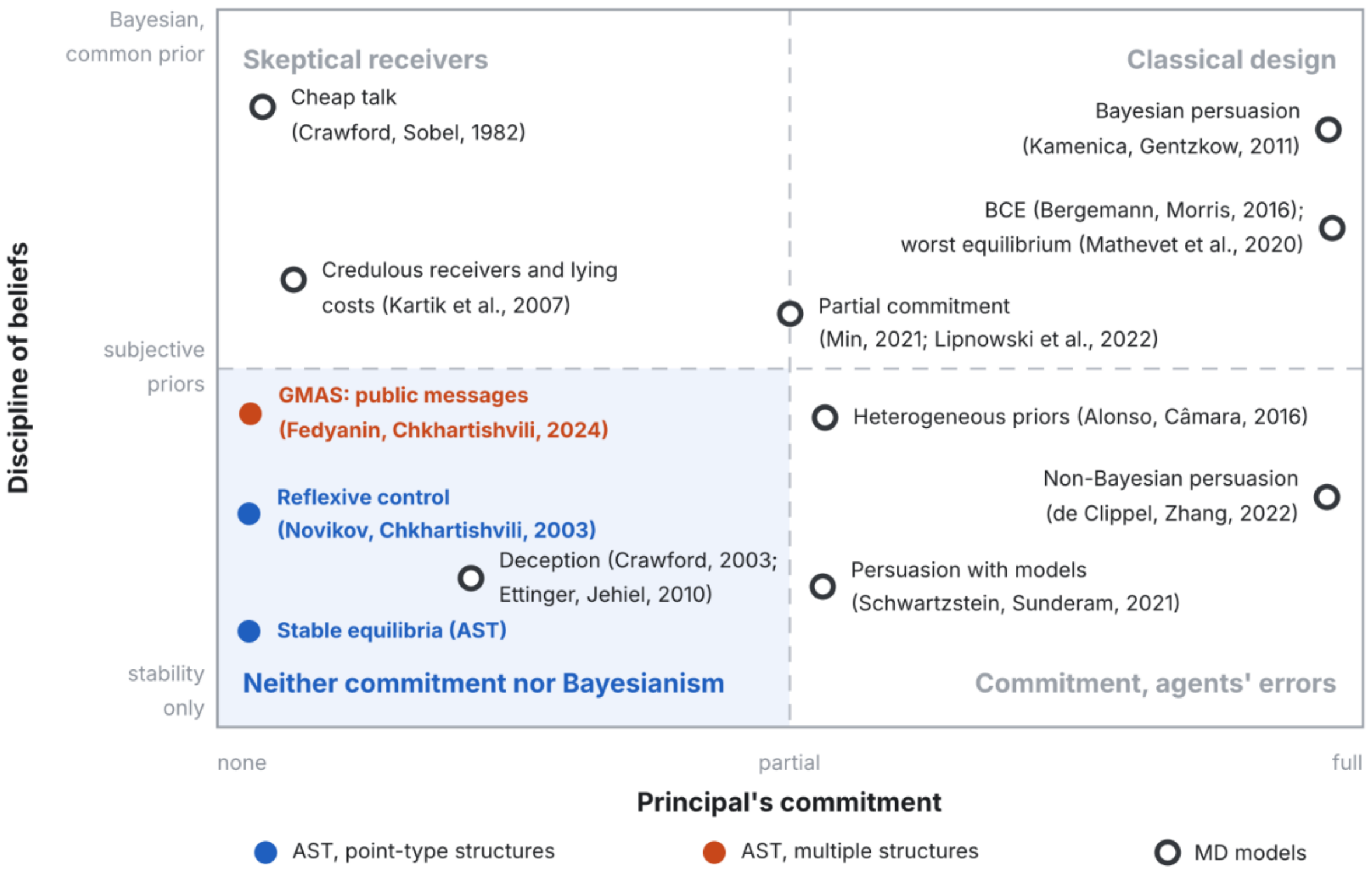

Fig. 9. Models of belief control along two axes: the principal's commitment and the discipline of the agents' beliefs. The positions of the models are the survey author's assessment, accurate up to a zone

### *6.5.7. Reflexion in Oligopoly Games*

*Reflexion in oligopoly games.* M.I. Geraskin's cycle of papers in Automation and Remote Control [224, 222, 223, 221, 220] carries the machinery of reflexive games [422, 403] over to the classical oligopoly model. Agents $i \in N = \{1, \dots, n\}$ choose outputs $Q_i \geq 0$ and receive profits $\Pi_i = P(Q)Q_i - C_i(Q_i)$, where $Q = \sum_{j \in N} Q_j$, the inverse demand function $P$ is decreasing (linear in most of the papers), and the costs $C_i$ are linear, power-law or (in [223]) quadratic. The profit functions are common knowledge; the incompleteness of awareness concerns the rivals' reactions, and the author calls such reflexion strategic [221]. Agent $i$'s beliefs are specified by *conjectural variations* $\rho_{ij} = dQ_j/dQ_i$, i.e. the expected change in rival $j$'s output in response to a unit change in the agent's own output. The equilibrium is defined by first-order conditions in which the variations enter as parameters:

$$P(Q) + (1 + S_i^r)Q_i P'(Q) - C_i'(Q_i) = 0, \quad S_i^r = \sum_{j \in N \setminus \{i\}} \rho_{ij}^r, \quad i \in N, \tag{47}$$

where $S_i^r$ is the sum of the conjectural variations of agent $i$, whose reflexion rank is $r$. The ranks are defined recursively. An agent of rank 0 takes the rivals' outputs as fixed ($\rho_{ij}^0 = 0$, the Cournot hypothesis); an agent of rank 1 takes the rivals to be rank-0 agents and computes the variations by differentiating their first-order conditions, i.e. behaves as a Stackelberg leader; an agent of rank $r$ takes the others to be agents of rank $r - 1$, or leaders of level $r - 1$. For general demand and cost functions the recursion over ranks has the form [222, 223]

$$S_i^r = \left( \frac{1}{\sum_{j \in N \setminus \{i\}} \frac{1}{u_j - S_j^{r-1} + 1}} - 1 \right)^{-1}, \quad S_j^0 = 0, \tag{48}$$

where the coefficient $u_j$ reflects the nonlinearity of demand and of agent $j$'s costs; in the linear case $u_j = -2$. The solution of system (47) under the variations matching the agents' ranks and beliefs is called by the author an informational equilibrium; in effect it is a Nash equilibrium under the agents' given beliefs (a given profile of variations).

The main results concern the properties of the variations and the equilibria. In the linear duopoly $\rho^r = -r/(r + 1)$, i.e. $-1/2$, $-2/3$, $-3/4$ for ranks 1, 2, 3, and growing ranks – and outranking one's rival – increase total output and shift the distribution of output and profit in favor of the "reflexive leader" [221]. With power-law costs with positive economies of scale and with nonlinear demand, the sum of variations can become positive and unbounded: the right-hand side of (48) has a discontinuity, which the author calls a bifurcation of beliefs, and there may be two equilibria, one, or none [223]. The survey [222] names the passage to reflexive control a promising direction for further work.

The control problem is posed in the 2025 paper [224]. The principal is a coalition of the other market participants, and the controlled variable is the sum of variations $S_i$ of the controlled player, i.e. its belief about the environment's reaction. To make the player adopt the value the principal needs, the environment acts for one period according to a "phantom" reaction function; in the linear duopoly, as follows from (48), the target value $S_i$ requires the phantom variation $S^f = -1/S_i - 2$ [224]. In this setting there are no agent messages, payments or incentive compatibility conditions: the agent's belief is changed while the rules of the game stay fixed, which corresponds to informational control in the TCO classification (problem (42)), the signal being the principal's observable action. A close line of research is formed by the works of G.I. Algazin and D.G. Algazina on processes of reflexive collective behavior in oligopoly models and the conditions of their convergence [9, 8].

Geraskin relates his results to oligopoly theory through the notion of conjectural variations, and his papers contain no comparison with level-$k$ models (work on depth of reasoning is mentioned in the introduction of [221]). The rule "an agent of rank $r$ takes the others to be agents of rank $r - 1$" coincides with the rule of level-$k$ models [169, 404], but what is iterated is not the action but the slope of the expected reaction function, and the solution is a joint system of first-order conditions rather than a profile of best responses to reasoning; there is no population distribution of levels – the ranks are assigned to firms by name and estimated from market shares.

Mechanisms robust to bounded depth of reasoning [169] and the identification of ranks from market data solve mutually complementary problems; their combination is formulated below as problem 16 (Section 10.4).

### *6.5.8. Network Models and Mob Models*

*Network models* form a branch of their own. In the informational influence models of Gubanov, Novikov and Chkhartishvili [261, 147], the agents' opinions are averaged through a trust matrix $A$, and the principal acts on the initial opinions:

$$X(u) = A^{\infty}(x^0 + u), \qquad \Phi(u) = H\big(X(u)\big) - c(u) \to \max_{u \in U}, \tag{49}$$

where $A^{\infty} = \lim_{t\to\infty} A^t$; control of trust and reputation and the informational confrontation game of several principals are also considered. The first edition of the monograph on social networks [261] (2010; reissued in 2018, English edition 2019 [147]) consists of three parts – influence models (a survey of threshold, cascade, epidemic and Markov models and influence indices), informational control (the principal's influence on initial opinions with a known trust matrix; with linear costs and a limited resource the problem reduces to linear programming, and the entire resource goes to the agent with the maximal resulting influence) and informational confrontation (a game of two principals controlling the opinions of disjoint groups: in the antagonistic case there is a dominant-strategy equilibrium that is not Pareto efficient; in the nonantagonistic case a Nash equilibrium is found in explicit form; hierarchical and reflexive settings are considered). In the threshold mob behavior models [81], the share of acting agents follows the dynamics $x_{k+1} = F(x_k)$, where $F$ is the distribution function of thresholds, and control of thresholds, of the composition and of reputation is considered, along with informational confrontation. A general model of the joint dynamics of opinions and actions under informational control is proposed in [401] and identified on data from an online social network [260].

The MD analogues of the network models are well known. Opinion averaging through a trust matrix coincides with the DeGroot learning model [171], which the authors of the informational influence models cite; the threshold dynamics [81] coincides in statement with Granovetter's model [251]; choosing the initial set of active agents to maximize coverage is studied as the network influence problem [317]; strategic influence on opinion formation is considered in games [249], and principals' competition for coverage in the work on competitive contagion [248]; targeted influence on nodes with a game-theoretic response of the agents is analyzed by Ballester et al. [38] (the key player) and by Galeotti, Golub and Goyal [217] (optimal interventions). The specificity of the AST school is the formulation of the network control problem for the principal and the inclusion of the awareness structure.

A convenient junction point of the two traditions' network models is the Friedkin–Johnsen model [215], in which an agent with weight $1 - \lambda_i$ retains its initial opinion: $x^{t+1} = \Lambda A x^t + (I - \Lambda)x^0$, $x^{\infty} = (I - \Lambda A)^{-1}(I - \Lambda)x^0$. In form, this is the same inverse operator as in the network game equilibrium of [449]; through it the principal's influence on initial opinions and on the trust matrix is transmitted, which is what makes the control problems of both research traditions computable. The basic models (DeGroot averaging, threshold models) are shared by the two traditions, so comparing results here requires no translation of statements from one language into the other.

## 6.6. Structures, Complexity, Complex Activity

*Hierarchies.* The structure control problem is formulated as discrete optimization: find a hierarchy over a given set of workers minimizing the managers' total costs [375, 246]. The notions of an organization performing a given set of functions and of a sectional cost function were proposed by A.A. Voronin and S.P. Mishin in the late 1990s and are set out in their monograph [519]. For sectional cost functions (a manager's costs depend only on the groups managed by its immediate subordinates) general results have been obtained: under monotonicity in groups an optimal tree exists; for "narrowing" functions the optimal hierarchy gives every manager at most two subordinates, for "widening" ones the two-level (fan) hierarchy is optimal. In the model of a symmetric production line with manager costs $\varphi(x) = x^{\alpha}$ depending on the flow $x$ it controls, for $\alpha \leq 1$ the two-level hierarchy is optimal, while for $\alpha > 1$ the optimal span of control $r^{*}$ is the same for all managers:

$$r_0 = \frac{\alpha + 1}{\alpha - 1}; \qquad (50)$$

the number of managers then equals $(n-1)/(r^{*}-1)$ if $n-1$ is divisible by $r^{*}-1$; otherwise the corresponding costs serve as a lower bound exceeded by at most a factor of $1 + (r^{*}-1)/(n-1)$. The span of control does not depend on the number of workers and decreases as $\alpha$ grows, which the author interprets as instability of the environment [375]. For homogeneous cost functions, homogeneous trees with a constant span of control are optimal or near-optimal [246]. Mishin relates his results to the MD models of hierarchies: in one special case the optimal span of control coincides with one known from the MD literature [375].

The monograph [245] gives a general theory for sectional and homogeneous cost functions, including lower bounds on the costs of the optimal hierarchy and algorithms for constructing suboptimal hierarchies, and the survey [262] relates it to the MD models of hierarchies (Williamson, Calvo and Wellisz, Keren and Levhari, Qian, knowledge hierarchies, multilevel information processing); this TCO strand, unlike most others, has been built in explicit dialogue with organizational economics.

The comparison with Mookherjee's survey [377] shows the following. When the revelation principle applies, centralization cannot be worse than delegation, and delegation matches it under observability of the subcontract, top-down contracting and risk neutrality; the gain from delegation arises with communication costs and contract complexity. The AST decentralization theorems [408] answer a different question – which classes of planning mechanisms admit an arbitrary partition of agents into subsystems without changing the equilibrium – and in this sense complement the MD setting without coinciding with it.

*Multilevel systems and structure as a consequence of the game.* A game-theoretic model of a multilevel organizational system is built in [408]: the efficiency of a hierarchy is determined by five factors – aggregation, economic, uncertainty, organizational and informational – and a structure is called rational if any admissible change of centralization reduces efficiency (the principle of rational centralization). The formal results: the maximum efficiency of incentives in a three-level system is attained under complete awareness of the principal and equals that of the two-level system, so aggregation without the informational factor does not raise efficiency (Theorems 1.3.1 and 1.3.2); introducing an intermediate level with its own nonnegative income raises efficiency, with a negative one lowers it; the planning mechanisms that admit arbitrary decentralization are listed in Section 3.2.

In the monograph [410] structure is tied to the type of hierarchical game: the order of moves and the awareness generate the games $\Gamma_0$, $\Gamma_1$, $\Gamma_2$, and the structural synthesis problem is posed as choosing the number of levels, the distribution of roles and the type of game. It is shown that when a dominant-strategy equilibrium exists, the structure does not affect the outcome of the game $\Gamma_1$ (Proposition 1 ibid.), that in homogeneous systems the fan structure is optimal (Proposition 5), and that with side payments and an external principal the two-level structure is optimal (Proposition 13), so that multilevel structures are explained by constraints on the capabilities of the governing bodies. The MD analogue in subject matter is the theory of hierarchical agency and the theory of the firm, including the conditions under which delegating the contracting to an intermediate tier does not reduce the principal's payoff [363, 377]; there is no MD analogue in method (structure as a consequence of the game type and the distribution of roles).

*Staff.* In the staff control problem with additive income of the principal and convex costs of the agents (zero at zero actions), the maximal staff is optimal; if, however, the income per participant decreases with their number, $H = g(n) \sum_i y_i$, and the costs are of Cobb–Douglas form, there is a critical rate of decrease separating the cases where the minimal and the maximal staff are optimal (Theorems 10.1 and 10.2 in [428]).

Karavaev [313] proposed a classification of staff control problems (formation "from scratch", expansion, reduction, replacement; the criterion $\Phi(I, I_0)$ depends on the initial and final staff) and showed the nesting of the problems of controlling a fixed staff, the staff, and the staff plus structure. For unified incentive schemes with ordered agent types, differential conditions on the optimal actions of agents of adjacent types are obtained (Theorem 2.3.5); in dynamics it is proved that improving any agent's type does not increase the principal's optimal costs, so replacing the worst agent by a better one pays (Proposition 3.1.1), and under staff control through training it is most profitable to improve the type of the agent with the maximum ratio of the benefit of the improvement to the cost of training (Theorem 3.2.1); the formation of a governing staff with one agent and several principals is considered separately (Ch. 4, Section 6.4). The author contrasts this setting with contract theory, where the number of workers is determined without regard to their strategic behavior, and with labor economics, where the optimum is set by equating the marginal product and marginal costs.

Coalitional models in staff control are considered by Goubko [244] (Section 2.3). If the principal does not optimize the staff, the agents can covertly exclude an inefficient participant by redistributing its plan and incentives; the core of this game is nonempty if there is an agent whom it pays to exclude from every coalition containing it, and then the rest distribute the plan the way the principal would with the total plan fixed (Theorem 4). If the agents can bring in an outside contractor, their game is balanced (Theorem 5; the proof reduces it to the distributed control model, with the agents in the role of principals). It is also shown there that, when a new agent is admitted, the principal's right to recompute the others' plans adds only a second-order gain compared to admission without recomputation, and the initial staff formation problem reduces to an enumeration of staffs which, for linear costs, shrinks to single-agent systems but admits no a priori ranking of candidates by cost functions.

The staff control problem has been continued in the direction of personnel development. The brochure by Galinskaya, Ivashchenko and Novikov [218] (2005) considers models of managing personnel development through professional training and the content of development programs (courses, certification, trainings); the material entered Novikov and Ivashchenko's monograph on managing a firm's innovative development [423]. This line is linked to staff optimization models by Karavaev's result that under staff control through training it is most profitable to improve the type of the agent with the maximum benefit-to-cost ratio (Theorem 3.2.1 in [313]), and by the third part of the book [48], where the staff is formed from the identified cost functions (Section 3.5). In economics the related questions belong to human capital theory and the train-or-hire choice; in MD they correspond to problems of information acquisition by agents and of incentives for type investments, but the setting with a principal choosing whom to train and in what, over a given set of candidates, does not appear in the MD literature.

*Complexity and bounded rationality.* The works [417, 399] show that the analytic complexity of solving control problems decreases with the number of hierarchy levels when problems are decomposed, and that real-time requirements rule out finding the optimum; complexity and error limit both the growth of hierarchies and the use of sophisticated mechanisms, and push toward standard solutions. This conclusion coincides with the motivation of algorithmic MD and the simple mechanisms program, but was obtained in control theory terms.

*Integration of mechanisms.* The works [114, 121] pose the problem of assembling an integrated mechanism from individual mechanisms with known properties so that these properties are preserved; a general theory of such assembly apparently exists in neither research tradition. An example of an integrated production-cycle control mechanism (counter plans, incentive-compatible planning, incentives for truthfulness, predictive self-control) is given in [107]. The same engineering line includes the identification of integrated rating mechanisms (criteria-tree convolutions of indicators) from data [122]. In the West, the closest result in structure is the theorem on the composition of smooth mechanisms preserving equilibrium quality guarantees [489] (truthfulness, by contrast, does not survive composition). Adjacent lines: the complementarity of a firm's instruments [291, 369], which justifies the joint use of incentives, asset ownership and control restrictions, and decision bundling [305], where copies of one problem are united by a common message budget; in the format “properties of the components – property of the complex” these works prove no theorems.

*Complex activity and the enterprise.* Since 2017 Belov and Novikov have been developing the methodology of complex activity [54] and, on its basis, a unified optimization model of the enterprise incorporating models of contract theory, technology management, human capital and resources [59]. The Russian editions of this cycle are [54, 58, 56], the English ones [57, 59]. The book [56] deals with coordinating the life cycles of the subjects, subject matters, technologies and resources of activity; in [59] the enterprise control problem is posed as a single optimization problem comprising static and dynamic multi-agent contract models (Ch. 4), models of technologies and human capital (Chs. 5, 6) and algorithmic planning models in hierarchical dynamic multi-agent active systems (Ch. 7); models of adapting dynamic contracts under stochastic uncertainty are given in [53]. The extensions of the basic model along uncertainty, the number of agents and dynamics, named promising in [120], are thereby joined in a single model in this cycle, albeit at the price of passing from theorems to algorithmic formulations.

A more general context is set by Novikov's work on *cybernetics* [415, 398, 397], which names four challenges for control theory: interdisciplinary fragmentation; decentralization (networks, systems of systems, multi-agent systems); strategic behavior of the controlled subjects; and complexity and dimensionality.

*The link with psychology* is analyzed in Novikov's article “Control, Activity, Personality” [400] and the brochure of the same name [416]. The goal is to build an “interface” between personality psychology, the methodology of activity and control theory, so that control problems for systems with people can be posed in shared terms; a five-stage scheme of joint research is proposed (shared concepts; the structures of personality and activity as objects of control; joint problem statements as chains with measurable variables; mathematical modeling; joint interpretation), of which the article covers the first two. The structure of personality is taken after K.K. Platonov (orientation, experience, mental processes, biopsychic properties, abilities, character) with a review of theories from Rubinstein to Bandura; the structure of activity as the process chain “need – motive – goal – tasks – technology – action – result – reflexion” after S.L. Rubinstein, A.N. Leontiev, A.M. Novikov and Engeström and after the methodology of complex activity [54, 57]. The main result is a correspondence between the components of personality and the components of activity (the motivational forms of orientation set the need and the motive; the worldview forms the goal, norms and criteria; cognitive processes the tasks; experience the technology; biopsychic properties the action) and a typology of control by the component acted upon: institutional (restricting the sets of conditions, norms, technologies and actions), motivational (incentives changing the criteria), informational (shifting beliefs, including through consensus and threshold models of social influence) and “personality control” (changing the subject's type – effectiveness, learning rate). In the basic model the subject chooses an action maximizing a function of the result and its own type, and control acts on the admissible set, the technology, the criterion, the state of nature or the type; the multi-agent extension reduces to Nash equilibrium. The article contains no behavioral economics terms, and bounded rationality is developed by Novikov separately, as real-time computation and communication constraints [417].

The comparison with behavioral contract theory [336] shows two routes for incorporating psychology: there, the form of the goal function changes (reference dependence, loss aversion, present bias, overconfidence) while the principal–agent setting is preserved; here, personality enters through the agent's type and through the source of norms, criteria and technology while the standard AST model is preserved, and the dependencies between the components are declared a subject for future empirical research; the article contains no formal behavioral utility functions. Hence two lines of possible convergence: transferring empirically grounded behavioral utility functions into TCO models, and using the typology of control by activity component to classify behavioral interventions (nudging as informational control, choice restriction as institutional).

The book [419] formulates a program of moving “from systems to active milieus”: an *active milieu* is defined as a collection of autonomous subjects, objects and technologies that has no fixed composition, boundary or single goal; it is controlled not by directives but by setting mechanisms of functioning (the mediation principle); the game-theoretic layer – game theory, mechanism theory, TCO – forms one of four levels of formalization; and the open problems named are the computability of two-level optimization problems on hypergraphs, scaling, and the formalization of the generation of the new. The conception is explicitly declared a development of the theory of active systems, with the emphasis shifted from hierarchical structures to polysubject self-developing formations.

## 6.7. Schools Adjacent to AST

*Hierarchical games.* The school of Yu.B. Germeier and N.N. Moiseev studies the principal's maximum guaranteed result (MGR) under various degrees of its awareness of the subordinates' choice. In the notation of Section 3.5, in the game $\Gamma_1$ the principal announces its choice $x_1$ first, and

$$G_1 = \sup_{x_1 \in X_1} \min_{x_2 \in B(x_1)} f_1(x_1, x_2), \tag{51}$$

where $B(x_1)$ is the second player's best-response set; in the game $\Gamma_2$ the principal announces a function $\tilde{x}_1(x_2)$ of the expected information about the subordinate's choice, and the MGR is given by Germeier's theorem (formula (17)) [226, 225]. Games with information exchange, the role of information about the partner's goal function and the structure of the optimal "program plus punishment" strategy were studied by A.F. Kononenko [321, 320], a textbook exposition is given in [334], the English translation of Germeier's monograph appeared in 1986 [227]; dynamic models of conflicts and hierarchical games are systematized in [243].

*Origins and surroundings.* The formulation of the problems of hierarchical systems theory belongs to the joint 1971 paper of Germeier and Moiseev [229]; the methodological frame (controlled systems, simulation, hierarchical systems, expertise) is given in Moiseev's book [376]. The computational side of the guaranteed result principle is formed by Fedorov's numerical maximin methods [208]; a textbook exposition of informational extensions of games and hierarchical games is given in the manual of Kukushkin and Morozov [334]; alternative equilibrium principles for games with non-opposing interests were developed by Smolyakov [484]. The history of the operations research sector and department of the Computing Center and Germeier's bibliography are given in Kostyuk's survey [333]; the ORM2018 conference was dedicated to Germeier's centennial [302].

*Monographs and preprints of the Computing Center school, 1980s–2000s.* The school's results after 1976 are collected in several books and preprints of the Computing Center of the USSR Academy of Sciences (from 1991, of RAS). The book by Gorelik and Kononenko [239] treats four classes of models from a unified standpoint: static models of hierarchical systems, decision making under uncertainty, regulated equilibrium, and dynamic models of hierarchical systems; the AST literature and the Rostov school cite it as the basic work on the information theory of hierarchical systems. The monograph by Gorelik, Gorelov and Kononenko [238] consists of two parts. The first covers models of optimal control of hierarchical systems: static and dynamic, with myopic and farsighted behavior of the lower level, program control under uncertainty (dynamic maximin problems), and game models of interacting systems (stable compromises, negotiation procedures). The second covers the choice of rational information exchange procedures: a formalization of the economic mechanism and of incentives, dynamic models in the class of feedback controls under a priori and stepwise information transmission, conflict under uncertainty and risk, and numerical methods. The second part is closest to the problems of MD: the information protocol (what, when and in what aggregated form the subordinate reports to the principal) is treated as a design variable, which corresponds to choosing the mechanism's message space (Section 3.1); the book itself makes no comparison with contract theory. The dynamic branch goes back to the chapter by Danilchenko and Kononenko in the 1979 volume [162] and to Kononenko's 1980 result on the structure of the optimal strategy in dynamic controlled systems [322]: the "program plus punishment" structure carries over to dynamics.

The book by Kononenko, Khalezov and Chumakov [325] is devoted to decision making in hierarchical games under uncertain factors and risk; adjacent to it are the paper by Kononenko and Chumakov on a two-level system under uncontrolled external factors [323] and the 2000 paper by Kononenko and Khalezov [324], which gives a general solution of the principal–agent problem with asymmetric information under uncertainty and risk. The Computing Center preprints by Aliev and Kononenko on exact aggregation in game-theoretic models [11] and by Gorelov and Kononenko on informational aspects of decision making under conflict [242] (both cited by Gorelov in his work on information exchange and aggregation [241]) are the direct predecessors of the 2000s–2020s results on the dimension of the message space. Later, Kononenko and Shevchenko proposed operational games – scenario-game models of strategic interaction under uncertainty with applications to corporate governance [326].

*The work of F.I. Ereshko.* The preprint [196] deals with hierarchical compromises under general couplings and parallel threats; the same subject matter – models with dictate, counter and parallel threats, side payments, the hierarchical compromise, coordination in a general equilibrium model – is collected in the 1998 dissertation [197], which shows, in particular, that side payments do not reduce the principal's guaranteed payoff (cf. formula (52)). The two 1984 IIASA papers, written with V.V. Fedorov and S.K. Zavriev [200] and with A.S. Zlobin [203], are rare English-language publications of the Computing Center school of that time (apart from translations of papers in the USSR Computational Mathematics and Mathematical Physics journal). The first poses the problem of distributing authority between the principal and the subsystems; the games $\Gamma_1$, $\Gamma_2$, $\Gamma_3$ are interpreted as control by prices, fixed payments and resources; the game $\Gamma_1$ reduces to a maximin with coupled variables, theorems on ε-optimal strategies are proved for $\Gamma_2$ and $\Gamma_3$, and stochastic algorithms are constructed. The second gives numerical methods for a lower level of several subsystems: allocation of scarce resources, a price algorithm, construction of the Pareto set, and equilibrium search. In 1999–2001 Ereshko (jointly with Yu.V. Lokhnygina) introduced the term "reflexive strategies in controlled systems" [202, 198, 199]; the case of single-valued response functions of the active elements is considered along with conditions on the payoff functions ensuring such single-valuedness; this line is independent of the TCO reflexive games (Section 6.5) and precedes the monograph [421] by two years, but it contains no explicit game formulation (an equilibrium depending on mutual awareness).

*Side payments.* In the paper by Germeier and Ereshko [228], the game $\Gamma_2$ is augmented with side payments: the first player, besides a strategy-function, promises the second a payment $z(x_2)$ bounded from above by $z_0$. Following the exposition in the encyclopedic article of Vatel and Ereshko [516], if $L_2 = \max_{x_2}\min_{x_1} f_2(x_1, x_2)$ is the second player's guaranteed result under punishment, then the first player's maximum guaranteed result equals

$$G_2^z = \max_{x_1, x_2:\, f_2(x_1,x_2) \geq L_2 - z_0} \min\{f_1(x_1, x_2),\ f_1(x_1, x_2) + f_2(x_1, x_2) - L_2\}, \tag{52}$$

where the second term under the minimum is the principal's payoff net of the compensation $L_2 - f_2$ that must be paid to the second player so that the pair the principal needs is no worse for it than the guaranteed $L_2$, and the optimal strategy has the form "reward (the target pair plus the payment) – punishment". This is the direct predecessor of the compensatory incentive schemes of AST (formula (16)) and of the theorem on the sufficiency of quasi-compensatory contracts: the principal compensates the agent exactly for the loss relative to its guaranteed alternative. In the same school, Ereshko and Kononenko solved the game with the right of the first move under incomplete information about the partner's intention [201], and Germeier and Vatel the games with a hierarchical vector of interests [230]; a popular exposition of the guaranteed result principle and of games with non-opposing interests is given in the brochure [515].

M.A. Gorelov's recent results formalize arbitrary information exchange as an extension of the game by a mapping $P$ from the second player's choice set into a message space $W$ and link the attainability of a given MGR to the dimension of the message space [241]; games with deliberate distortion of information by the subordinate and selective auditing have been studied [240]. In Gorelov's assessment, feedback models were studied practically simultaneously and independently in the theory of hierarchical games, the theory of active systems and contract theory [241]. The results on the dimension of the message space continue the line begun by Hurwicz (the informational efficiency of mechanisms) and developed in communication complexity theory.

*The theory of sustainable development control for active systems.* The strand took shape at Southern Federal University; its results are summed up in the monographs [236, 441] and the survey [512], which defines it as one of the directions of AST's development. The object is an extended active system: a hierarchy of control subjects together with a controlled dynamical system. The control methods are classified as compulsion (acting on the set of admissible strategies), inducement (acting on the payoff functions) and persuasion (a transition to cooperation); the first two correspond to the institutional and motivational control of TCO.

The main emphasis is on dynamic formulations of control problems (differential and difference games), their solution by simulation, and applications to various subject areas – ecological-economic systems [440], sustainable development [438], and others.

The sustainable development requirement joins two conditions of different nature: the homeostasis (viability) condition $x(t) \in X^*$, external to the agents, and the coordination of interests expressed by the equilibrium of the game. In the dynamic Stackelberg leader–follower game with compulsion [332], the state $x$ obeys the equation $\dot{x} = f(x,u)$, $x(0) = x_0$, and the players' payoffs equal

$$J_L = \int_0^T e^{-\alpha t}\,[g_L(u,x) - c_L(q) - M\rho(x,X^*)]dt \to \max_{q(\cdot)},$$
$$J_F = \int_0^T e^{-\alpha t}\, g_F(u,x)\,dt \to \max_{u(\cdot)}, \quad u(t) \in U\big(q(t)\big), \tag{53}$$

where $q$ is the leader's control, which specifies the follower's set of admissible controls $U(q)$, $c_L$ is the cost of control, $\rho(x,X^*) = 0$ if $x \in X^*$ and $\rho(x,X^*) = 1$ otherwise, and $M$ is the penalty for violating homeostasis (for $M \to \infty$ the requirement is mandatory). Under inducement the leader acts not on the set of admissible controls but on the follower's payoff function, including with feedback $p = p(u)$; persuasion is formalized as a passage to maximizing the sum $J_L + J_F$. For several agents, the sustainable development control problem is solvable if $\mathrm{NE} \cap U^* \neq \varnothing$, where NE is the set of Nash equilibria of the agents' game and $U^*$ the set of controls under which the homeostasis condition holds; in the authors' exposition, the principal's mission is to select economic and administrative instruments making this intersection nonempty [512].

*Protocols and solution methods.* Germeier's information protocols (Section 3.5) are carried over to dynamics: the game $\Gamma_1$ corresponds to a Stackelberg game, the game $\Gamma_2$ to a game in which the leader announces a strategy with feedback on the follower's control (in the school's English-language work, Germeier games or incentive Stackelberg games [437]; in recent years also inverse Stackelberg games, following control theory terminology [434, 462]). In the second case the solution follows the pattern of Germeier's theorem (formula (17)): the punishment strategy and the follower's guaranteed payoff are found, then the leader's optimal control problem is solved subject to the follower receiving no less than the guaranteed payoff (strictly more in [142]), and the leader presents the strategy "reward for following the desired control, punishment for deviation" [142, 511]. Definitions of equilibria incorporating sustainable development requirements and algorithms for constructing them for two- and three-level systems are given in [509, 510]. Analytic solutions (Pontryagin's maximum principle, the Hamilton–Jacobi–Bellman equation) are obtainable in models of special form; in the general case discretization and simulation are used. A separate result concerns dynamic stability: according to [436], compulsion and inducement as optimality principles lack time consistency, while persuasion (the transition to cooperation) admits a regularization that ensures it. On this point the Rostov school touches the work of L.A. Petrosyan's school on dynamic stability [446, 531].

*Incentives in dynamics.* The school's most rigorous result was obtained by D.B. Rokhlin and G.A. Ougolnitsky [462]. In a discounted stochastic game, $m$ followers control a Markov process with state $x_t$ by choosing actions $a_t = (a_{1t}, \dots, a_{mt})$, and the leader chooses incentive functions $c_i(x, a) \geq 0$. Follower $i$ maximizes $\mathbf{E} \sum_t \beta^t (c_i - g_i)$, where $g_i(x, a)$ is its costs, and the leader maximizes $\mathbf{E} \sum_t \beta^t (f - \sum_i c_i)$ at the followers' stationary equilibrium worst for itself. Under conditions guaranteeing the existence of a stationary Markov equilibrium, it is proved that the leader's value of the game equals the value of the centralized Markov problem with criterion $f - \sum_i g_i$, and the incentive scheme

$$c_i^{\varepsilon}(x, a) = g_i(x, a) + \frac{\varepsilon}{m}(1 - \beta)(1 - |a_i - u_i^*(x)|)^+ \qquad (54)$$

is $\varepsilon$-optimal, the followers' equilibrium being unique and consisting in following the optimal control $u^*$ of the centralized problem [462]. This is the dynamic stochastic analogue of the cost compensation principle (15) and the decomposition theorem (16): the principal compensates the costs and adds a small bonus for following the plan, which makes the desired equilibrium unique. Like its static prototypes, the result is obtained under the principal's complete awareness of the cost functions.

*Multilevel models.* Another distinctive feature of the Rostov school is the systematic study of hierarchies with more than two levels. In the general three-level model "regional administration – municipalities – local agents" [437], only the lower-level agents act directly on the system's state, while the upper levels set framework conditions for them:

$$\begin{gathered} J_0 = \sum_{t=1}^{T} \delta^t g_0^t \to \max_{r^t, s^t}, \quad J_i = \sum_{t=1}^{T} \delta^t g_i^t \to \max_{p_i^t \in P_i(r^t),\, q_i^t \in Q_i(s^t)}, \\ J_{ij} = \sum_{t=1}^{T} \delta^t g_{ij}^t \to \max_{u_{ij}^t \in U_{ij}(s^t,\, q_i^t)}, \\ x^{t+1} = f(x^t, u^t), \quad x^t \in X^*, \quad t = 1, \dots, T. \end{gathered} \qquad (55)$$

Here $s^t$ and $q_i^t$ are the administrative instruments of the top and middle levels (acting on the admissible control sets of those below), $r^t$ and $p_i^t$ the economic instruments (acting on their payoff functions), $u_{ij}^t$ the agents' controls. As the author remarks [437], in the general case such models can be solved practically only by simulation. Particular structures were studied earlier: fan structures in three-level systems for managing ecological-economic objects [508]. The same class includes the "principal – supervisor – agent – object" hierarchy, in which corruption is described as bribe feedback: the supervisor relaxes the administrative or economic requirements on the agent in exchange for a bribe, and the criterion for success in combating corruption is the fulfillment of the sustainable development conditions rather than the ratio of the damage to the cost of auditing [142, 511]. The type of game at each pair of levels is determined by the control method; under bribe feedback the supervisor–agent game has protocol $\Gamma_2$ [142].

*Coordinating public and private interests.* In the models of coordination of public and private interests (SPICE models in the later English-language work), agent $i \in N$ divides its resource $r_i$ between private activity and a contribution $u_i$ to a public good [237]:

$$g_i(u) = p_i(r_i - u_i) + s_i c(u) \to \max_{0 \le u_i \le r_i}, \quad i \in N,$$
$$g_0(u) = \sum_{j \in N} p_j\left(r_j - u_j\right) + c(u), \quad \sum_{j \in N} s_j = 1, \tag{56}$$

where $p_i$ and $c$ are concave private and public income functions, $s_i$ is the agent's share in the public income, and $g_0$ is social welfare. The model is called system-compatible if every Nash equilibrium maximizes $g_0$; this is equivalent to the price of anarchy – the ratio of the smallest equilibrium value of $g_0$ to the maximum – being equal to one. It is proved that for a nondegenerate public income function system compatibility requires every agent to be an "individualist" ($u_i = 0$) or a "collectivist" ($u_i = r_i$), and that the proportional allocation mechanism $s_i(u) = u_i / \sum_{j \in N} u_j$ is system-compatible for a linear public income function [237]. The same models are studied in dynamics, including with corruption [511].

*Comparative analysis of the efficiency of modes of interaction of active agents.* Three basic ways of organizing active agents are distinguished: competition (simultaneous and independent decision making), hierarchy (one of the agents becomes the leader), and cooperation (all agents unite into the grand coalition and jointly maximize the total payoff). Mathematically, these yield respectively Nash and Stackelberg equilibria (solutions of Germeier games) in the normal-form game in the first two cases, and the reduction of the game to an optimization problem in the last, after which the grand coalition's payoff is divided among the agents according to some optimality principle for cooperative games. The total payoff is maximal under cooperation, and one can only bound the losses from selfish behavior (the price of anarchy); for individual agents cooperation is not necessarily more profitable, and here a comparison of efficiency is needed. Such work has been done, in particular, on the example of the Cournot oligopoly with various effects, dynamics among them [439].

*Comparison with MD.* The authors of the SPICE work relate their theory to AST and TCO, to the theory of hierarchical games and to contract theory and MD, and themselves state the difference from the latter: the goal of control is maximum social welfare in equilibrium (system compatibility), not the elicitation of private information [237]. First, almost all the school's models are built under complete awareness; they contain no strategy-proofness conditions or information rent, and in their setting they are closer to hierarchical games and to full-information regulation theory than to MD. Second, the school's own contribution is the phase constraint as a separate requirement on the mechanism: in dynamic MD (Section 4.3) and in continuous-time contract models [475] constraints on the state trajectory are usually not imposed. Third, the system compatibility index in the static models [237] is the price of anarchy viewed as a function of the principal's controls (in the dynamic models its normalization varies from paper to paper); this solves the problem of choosing a mechanism by the price-of-anarchy criterion, which algorithmic MD poses for auctions [469] (cf. problem 1 of Section 10.4). Fourth, the three-level scheme with a supervisor corresponds to the MD models of collusion in hierarchies [501], where the supervisor and the agent form a coalition with hidden transfers and the principal must constrain the incentive schemes. The Rostov school's models describe the same link differently: corruption acts as bribe feedback in a dynamical system under complete awareness, and the criterion is sustainable development; the questions of the price of collusion and of when delegating authority to the supervisor pays [37, 339] are not posed in the school's work. The school's authors cite the MD literature but make no detailed comparison. The bulk of the results consists of algorithms and numerical computations for specific organizational systems; general theorems are few, and result (54) stands out among them. Carrying these settings over to the case of agents' private information is formulated as problem 15 (Section 10.4).

*The Samara school.* At Samara State Aerospace University (now Samara University) a strand formed around V.G. Zaskanov and G.M. Grishanov that is called the Samara school of AST (M.I. Geraskin's article in the volume for AST's 55th anniversary). The results are published mainly in the Vestnik of SSAU, in Control Sciences, and in the monograph on material incentives co-authored with D.A. Novikov [514]. The subject is *coordinated incentives* in production systems: the principal assigns plan targets in the system's interests, and the elements must find it profitable to fulfill them. In V.D. Bogatyrev's work [75, 76, 74], coordination is achieved not only by bonuses for plan fulfillment but also by *parametric coordination* – changing the coordinating parameters (prices, norms, profit-sharing rules); the synthesis problem consists in simultaneously choosing the plan and the parameter increments from the intersection of the set of instruments compensating the elements' losses from fulfilling the plan and the set of instruments under which the principal's total effect exceeds the total losses, the non-emptiness of the intersection being the existence condition for a coordinated mechanism; in the supplier–customer system [51] a necessary and sufficient coordination condition is expressed through the sensitivity coefficients of the goal functions to the parameters; the combined scheme [76] uses bonuses until the fund is exhausted. In D.Yu. Ivanov's work [301, 300] the model is specialized to machine-building enterprises: the reward depends on the result and an intensity coefficient, the "domain of interest coordination" is the set of result–coefficient pairs in which the worker's choice coincides with what benefits the principal, and a synthesis algorithm is proposed with a check that the pair falls into this domain under a constraint on the average rate. G.M. Grishanov and coauthors [253] formulate a balance condition for the interaction of the elements of a production system: the total effect is no less than the sum of the additional effects the elements require. The declared applications are AVTOVAZ (assembly and press production, the procurement service), the Samara grain elevator, and special machine-building and aircraft enterprises. Relative to AST, the Samara work develops the coordinated planning of the 1980s [100, 101, 111], with penalties replaced by the parameters of the economic mechanism; relative to MD, the closest setting is incentives through prices and transfer prices within a firm, but the Samara school formulates the coordination conditions as requirements on the ratios of the participants' payoffs under complete awareness, not as incentive compatibility constraints under private information.

*The Kazakh school.* The collaboration of A.A. Ashimov (Alma-Ata) with V.N. Burkov began in 1972 and in the 1980s produced a cycle of work on coordinated control of active production systems: the four-author monograph [22] and three papers in Automation and Remote Control [21, 20, 19]. The subject is operational (calendar) coordinated planning of a two-level production system: for continuous production, the properties of incentive functions allowing constructive construction of sets of coordinated calendar plans were studied [19]; for discrete (batch) production, on the example of lead refining, a model of the operational control system was built in AST terms [20], along with a branch-and-bound algorithm for coordinated planning and mechanisms incentivizing data truthfulness [21]. By Burkov's testimony [88], the concept of optimal coordinated planning, in which the optimum is sought over the set of plans profitable for the executors, was worked out in detail in the 1980s and deployed at industrial enterprises of Kazakhstan; the school's dissertations relate to nonferrous metallurgy, chemical production, machine building and production-transport systems. In the 2000s–2010s Ashimov's school moved on to the *theory of parametric regulation* of macroeconomic systems, set out in three Springer books co-authored with D.A. Novikov [18, 23, 24] and in a journal series [17]: economic policy parameters (tax rates, the share of government spending, the interest rate) are treated as controls in a mathematical model of the economy (systems of nonlinear differential equations, static equilibrium models, computable general equilibrium models, dynamic stochastic models); the theory's components are a library of models, assessment of structural stability (weak structural stability by Robinson's theorem), the choice of regulation laws as solutions of variational or mathematical programming problems, bifurcation analysis of extremals with respect to uncontrolled parameters, and the production of recommendations; the applications are the economy of Kazakhstan, then the Eurasian Economic Union and a global model. Neither AST nor MD is mentioned in these works: parametric regulation is optimal control of macro-model parameters without strategic behavior of agents, and lies closer to computational economics and the theory of economic policy than to mechanism synthesis; the continuity with AST is personal (Novikov's co-authorship, the continued Alma-Ata–ICS collaboration) and lies in the general formulation "the principal chooses the parameters of the mechanism of functioning". For the comparison with MD it is significant that a school that grew out of coordinated planning moved to structural stability and robustness of macro models, bypassing Bayesian mechanisms; the interest in robustness here parallels MD's (Section 4.1) but is realized by different means.

*The Volgograd school.* At Volgograd State University, A.A. Voronin and S.P. Mishin began the study of structure control problems for organizations (Section 6.6) [519]. Voronin with his students and co-authors then developed game-theoretic models and control mechanisms for the socio-natural-economic system "Volzhskaya HPP – Volga-Akhtuba floodplain"; the basis is formed by GIS and hydrodynamic models of flood-water dynamics. The paper [518] builds an optimization model of the flood regime and a game-theoretic model of controlling an economic agent with a pay-for-damage mechanism (hierarchical game is among the keywords of the paper). The paper [521] adds optimization of the hydrograph of the Volzhskaya HPP release, spatial planning of a channel restoration project and mechanisms for managing the economic agents of the territory. The paper [520] considers a fine for exceeding the permitted damage, a payment per unit of damage above the limit, a tax rate reduced according to the damage level and co-financing of resource saving; with several agents the outcome is evaluated at a Nash equilibrium. In [299] the strategic development of the floodplain system is posed as long-term conditional optimization of socio-natural-economic potentials under risk constraints, and the control mechanism is, according to the authors, described by the corresponding hierarchical game, which the paper does not solve. In their formulation these models are close to regulation problems with damage payments and fines in a "regulator – economic agents" system; a direct comparison with MD models is not made in this survey.

The development of the AST school's strands is shown in Fig. 10, and a summary of the results - in Table 8.

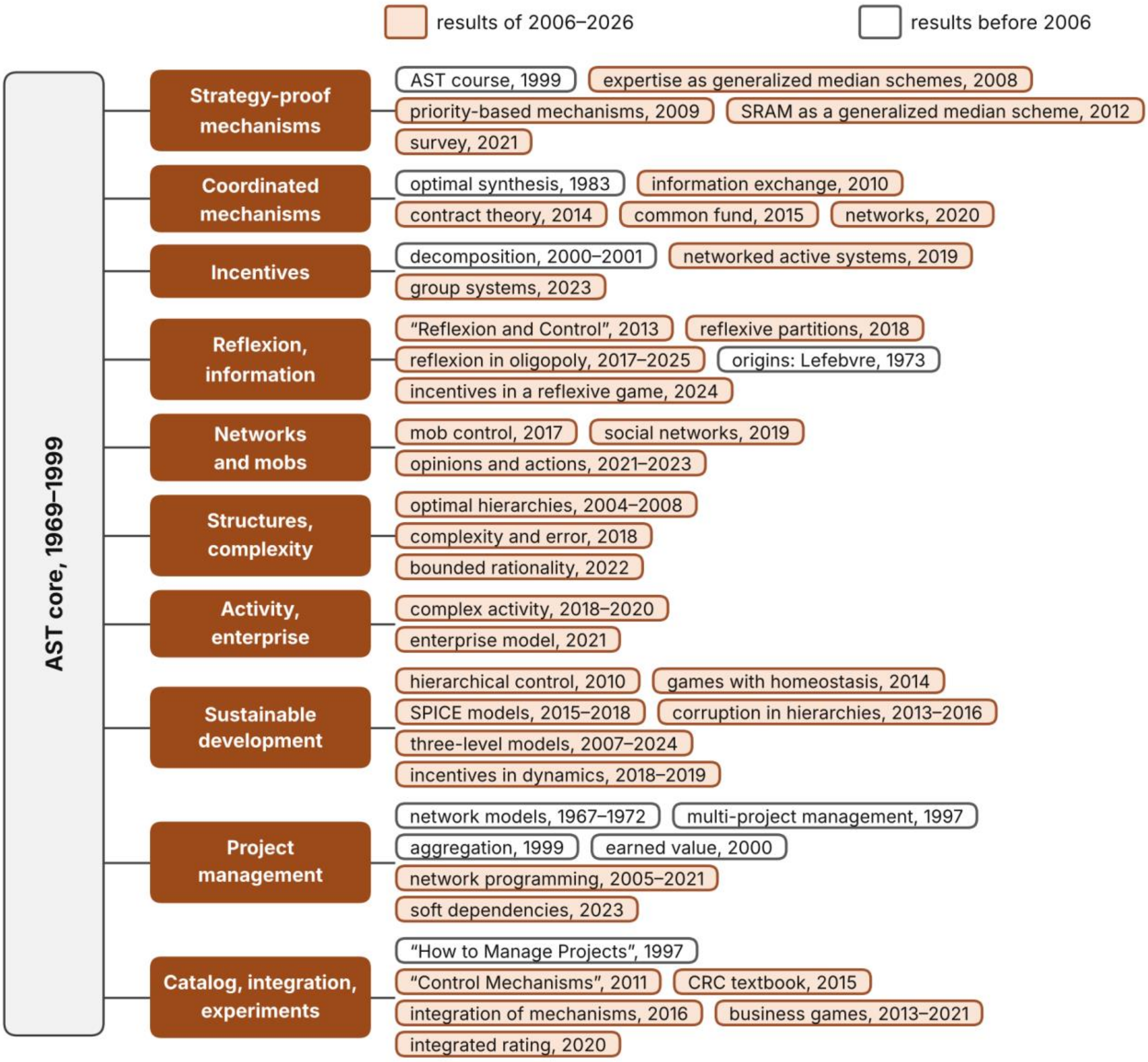

Fig. 10. The development of AST and TCO: the 1969–1999 core and the 2000–2026 branches (by the publications included in the survey)

Table 8. The main results of AST and TCO in 2006–2026 and their MD analogues

| Strand | Result | Sources | MD analogue |
|---|---|---|---|
| Expertise | Mechanism (7) is a median scheme; separability as the criterion in the multidimensional case | [110] | Median schemes [41, 78, 381] |
| Resource allocation | Representation (10); the SRAM as a generalized median scheme; equivalence of priority-based mechanisms | [327, 328] | The uniform and sequential rules [15, 42, 486] |
| Survey and program | Delegation as the basis of strategy-proofness; NTU and TU; active planning | [102] | The surveys [40, 139] |
| Experiments | Business games on mechanism (13); a simple rule no worse than mechanisms with transfers | [329, 330, 331] | Experimental implementation [141] |
| Coordinated mechanisms | Optimality of proper mechanisms; inequality (39); the link with adverse selection | [189, 187, 190, 191] | Screening [50]; verification and limited penalties [60, 388] |

| Strand | Result | Sources | MD analogue |
|---|---|---|---|
| Incentives | Networked active systems; group schemes; multiple principals | [93, 313, 428; 430, 55; 73] | Contracts for teams and networks [161, 528] |
| Reflexion | Informational equilibrium (41); problem (42); reflexive partitions | [422, 449, 404] | Information design [67, 311]; level-$k$ models [169] |
| Reflexion in oligopoly | Reflexion rank as the level of Stackelberg leadership; recursion (48); bifurcation of beliefs; estimating ranks from market shares; informational control of a player's beliefs | [224, 222, 223, 221] | Conjectural variations in oligopoly theory; level-$k$ models [169] |
| Networks and mobs | Problem (49); confrontation; threshold models; identification on data | [260, 81, 147, 401] | Network games |
| Hierarchies | Optimal trees, the span of control (50) | [245, 262, 375, 246] | Organizational economics |
| Complexity | Analytic complexity and error; the principle of bounded rationality | [417, 399] | Algorithmic MD [470] |
| Integration of mechanisms | The mechanism assembly problem; identification of integrated rating | [114, 122, 107] | Composition of mechanisms [489] |
| Activity and the enterprise | The methodology of complex activity; the life cycles of organizational and technical systems; a unified enterprise model with multi-agent and dynamic contracts | [54, 53, 58, 56, 57, 59] | The systemic view of incentives [288] |
| Hierarchical games | Informational extensions, the dimension of messages; information distortion | [241, 243, 240] | Communication complexity; verification of messages |
| Sustainable development | Viability plus equilibrium; compulsion, inducement, persuasion | [441] | Regulation, public goods |
| Dynamics with phase constraints | Stackelberg and Germeier games with the homeostasis condition (53); time inconsistency of compulsion and inducement; $\varepsilon$-optimal incentives (54) | [332, 436, 462, 510] | Dynamic contracts [475]; inverse Stackelberg games |
| Multilevel systems and corruption | Three-level models (55); fan structures; the bribe as feedback in the principal–supervisor–agent hierarchy | [142, 437, 508] | Hierarchical agency, collusion in hierarchies |
| Coordination of interests | SPICE models (56); system compatibility; the proportional mechanism; comparison of modes of organization | [237] | The price of anarchy [469]; public goods |
| Project management: optimization | Resource allocation on networks, perfect aggregation (61), the dual network model (results of 1972–2002); soft dependencies (2023) | [43, 44, 95, 115, 86] | Scheduling theory, project scheduling. |
| Network programming | Bounds (62); convexity of the generalized dual problem (63); exact solution for trees | [96, 94, 128, 90, 127] | Dynamic programming, Lagrange multipliers, branch and bound |
| Project management: mechanisms | Mechanisms by project phase and incentives for crashing durations (64) (1997–2000); the path–cut duality; groups with a unified incentive scheme; comparison of priorities in joint financing | [93, 119, 123, 319, 483] | Incentives in procurement and regulation [338] |
| Active milieus | The program “from systems to active milieus”; the mediation principle; four-level formalization; open problems of computability and scaling | [419] | Multi-agent systems; mechanisms for AI agents |

## 6.8. Dynamic Active Systems: Farsightedness, Horizons, Commitment

*Origins.* Dynamic models in AST go back to adaptive mechanisms [124] and to models with constraint dynamics, in which an agent's choice in a given period changes its admissible sets in later ones. A.V. Shchepkin [481, 482] introduced *farsighted elements* into such models: in every period the agent chooses a state and forecasts its states several periods ahead, and the future enters the criteria of both the agent and the principal. Conditions on the distributions of farsightedness were obtained under which the chosen states coincide both with the forecast and with the plan, and it was shown that in a number of cases the dynamic problem reduces to a static one in an extended parameter space [427, 482]. For coordination in dynamics, the triangle inequality for weighted sums of penalties – the dynamic analogue of condition (39) – suffices; if the penalties of individual periods are not coordinated, sufficiently strong penalties in future periods secure coordination [427].

*V.V. Tsyganov's adaptive mechanisms.* The cycle of papers by Burkov and Tsyganov in Automation and Remote Control [124, 502, 503] introduced adaptive mechanisms of functioning: the principal does not know the active elements' potential and refines their model from observed inputs and outputs (*active identification*), and a mechanism is called *progressive* if a farsighted element finds it profitable to reveal its potential fully rather than understate its results for the sake of future plans; the second part gives necessary and sufficient optimality conditions, the third learning and control. An industry-level exposition is given in the monograph [504]. In the 2000s the machinery was carried over to the level of the organization and its development: in the book "The Intelligent Enterprise" [507] the mechanism for managing an organization's evolution is a three-level self-organizing system (an evolution regulator – a governing body – active elements) built from progressive adaptive mechanisms and simplified models ("archetypes") of the acquisition of capital and power; the monograph [505] transfers the same mechanisms to "humanitarian systems" (the individual, the collective, society). Among recent papers we note [506], where, for a two-level system with simultaneous tuning of the element's constraint model and incentives, sufficient conditions of *correctness* are obtained (the element finds it profitable to use its potential fully), with an example from the energy efficiency of railway transport. Relative to MD, adaptive mechanisms with active identification correspond to dynamic contracts with the ratchet effect [214, 525]: progressivity is the requirement that revising plans from observations leave the agent no incentive to understate its result, and the AST school obtained it in terms of conditions on the adaptation procedure rather than in terms of an optimal dynamic contract.

*The general statement.* Incentive problems in dynamic and multi-agent systems are reviewed in [406], the general theory of deterministic DAS is built in the monograph by D.A. Novikov, I.M. Smirnov and T.E. Shokhina [427], and the multi-agent incentive problem and the coordination of the plans assigned to the agents are treated in the paper [426]. Regard for the future is described by the distribution of farsightedness $\delta^t(\tau)$ – the weight with which a participant in period $\tau$ accounts for the payoff $w^t$ of period $t \geq \tau$; the participant's criterion in period $\tau$ is

$$W(\tau) = \sum_{t=\tau}^{T} \delta^t(\tau)\, w^t, \tag{57}$$

Classes of distributions are identified in which the assessment of a period depends only on the period itself, on its remoteness and the moment of assessment, or only on its remoteness; the last class includes ordinary discounting $(\delta)^{t-\tau}$. The derived characteristics are the degree of farsightedness $\xi(\tau)$, i.e. the number of future periods taken into account, and the decision horizon $L(\tau)$ – the number of periods for which the participant in period $\tau$ chooses strategies:

$$\xi(\tau) = \max\{t \geq \tau\colon \delta^t(\tau) > 0\} - \tau, \quad L(\tau) \leq 1 + \xi(\tau), \tag{58}$$

By the principal's horizon, three control regimes are distinguished: current (decisions only for the current period), program-based (for all periods at once, before the first), and rolling (in each period, several periods ahead with new information taken into account). Hence four basic models: DAS1 – the current regime; DAS2 and DAS3 – the rolling regime without and with the principal's commitment to honor the announced plans; DAS4 – the program regime. By default the agent's farsightedness and horizon are assumed not to exceed the principal's [427].

*Incentive synthesis.* The cost compensation principle and the principle of decomposition of the agents' game (Section 3.5) carry over to DAS. If the costs, income and admissible sets of period $t$ depend on the action history $y^{1,t}$, the optimal scheme is the history-dependent compensatory one, which pays the period's costs only if the plan has been fulfilled in all periods from the first through the current:

$$\sigma^t(x^{1,T}, y^{1,t}) = \begin{cases} c^t(x^{1,t}), & \text{if } y^\tau = x^\tau, \quad \tau = 1, \dots, t; \\ 0 & \text{otherwise,} \end{cases} \tag{59}$$

and the optimal planned trajectory solves the incentive-compatible planning problem

$$x^{1,T} \in \operatorname*{Argmax}_{y^{1,T} \in A_0^{1,T}} \sum_{t=1}^{T} [H^t(y^{1,t}) - c^t(y^{1,t})],$$

where $A_0^{1,T}$ is the set of admissible trajectories. The agent's actions then coincide with the plans, and the efficiency is maximal (Theorem 1 in [427]); in a multi-agent system the same holds with compensation independent of the other agents' actions (Theorem 3). If, however, the period's reward depends only on that period's action, only coordinated trajectories are implementable – those on which the agent's losses in the periods of deviation from the plan are no smaller than its gain from the change in costs in the periods of fulfillment (Theorem 2) – the dynamic analogue of the set (38). The same efficiency as (59) is delivered by a scheme compensating the period's costs when the current plan is fulfilled given the actually realized history (Corollary 1). The entire complexity of the dynamics is thus shifted into the planning problem, solved by dynamic programming. In its setting, scheme (59) corresponds to the history-dependent long-term contracts of contract theory.

*Farsightedness and commitment.* Let $K_1, \dots, K_4$ be the control efficiencies in the basic models. In the general case only the bound $K_4 \geq \max\{K_2, K_3\} \geq K_1$ holds, and the other relations can be arbitrary. In the examples of [427], growing farsightedness lowers efficiency ($K_1 > K_2$: a participant who accounts only for the next period sees the gain from a violation in the second period but not the punishment in the third), while the principal's commitments raise it ($K_3 > K_2$). The adequacy principle makes it possible to order the regimes. Let $J$ be the system's memory, i.e. the number of past periods affecting the current payoff, $L_0$ the number of periods for which the principal commits, and $\xi_0$ the principal's farsightedness. If

$$J + (L_0 - 1) \leq \xi_0, \tag{60}$$

then under history-dependent incentives the regimes with and without commitment are equivalent, $K_2 = K_3$ (Theorem 5a in [427]). Hence commitments for $\max\{1, \ \xi_0 - J + 1\}$ periods do not reduce efficiency, and the principal should revise its decisions no less often than every that many periods. If growing the principal's farsightedness does not reduce its awareness of the future, commitments do not raise efficiency (Theorem 6). In two-period systems $K_1 \leq K_2 = K_3 = K_4$ (Theorem 7), and in three-period ones, under the hypothesis of the principal's conservatism, the rolling regime without commitments is no worse than the regime with them (Theorem 8).

Criterion (57) with weights depending on the moment of assessment is a special case of Strotz's general model of dynamically inconsistent preferences [488]: for $\delta\hat{}t(\tau) = \delta(t - \tau)$ it includes exponential discounting (consistent), the quasi-hyperbolic discounting of Phelps–Pollak [450] and Laibson [341], and truncated exponential discounting with a window $\xi(\tau)$ (a rolling horizon). The difference in interpretation is essential: in those models, and in the work on contracts with such agents [172, 277], the source of inconsistency is present bias and possibly the agent's mistaken expectations about its own future preferences, whereas in DAS the weights reflect incomplete regard for consequences within the horizon, and the divergence of assessments at moments $\tau$ and $\tau + 1$ is due to a new period entering the horizon while the relative weights of the others are preserved. The DAS control regimes have a structural parallel in the classical distinction between precommitment and sequential plan revision, and the ordering $K_4 \geq \max\{K_2, K_3\} \geq K_1$ is consistent with this distinction. The effect "growing farsightedness lowers efficiency" does not, in its setting (efficiency for the principal under a given incentive scheme), coincide with O'Donoghue and Rabin's result [431] that an agent's awareness of its own bias can worsen the outcome, but it is analogous to it in the logic of nonmonotonicity in awareness.

*The role-exchange effect.* If the agent is more farsighted than the principal, it can deviate from the plan in an early period in order to change the plans that the myopic principal will assign later; in the example of [427] such preemptive behavior benefits both the agent and the principal. The effect does not arise under history-dependent incentives or under history-independent incentives with coordinated plans (Theorem 4). In its channel of influence (a current deviation changes the future decisions of a principal not bound by commitment) the effect resembles the ratchet effect [214, 337], but the mechanism is different: in DAS there is no private information, the principal is limited in farsightedness and does not foresee the agent's manipulation, and the outcome can benefit both sides; it therefore cannot be called a deterministic analogue of the ratchet. Accumulation effects, where income or costs depend on the sum of past actions, are also considered, with an application to a new product's market launch [427].

*Comparison with MD.* First, dynamic MD (Section 4.3) is built around private information changing over time, while DAS theory is built around limited farsightedness under complete awareness of the current parameters; formulas (24) and (60) therefore answer different questions. Second, the control regimes correspond to the MD settings with and without commitment [337] and with limited commitment [176], but the value of commitment in DAS is determined not by information disclosure, as in the ratchet effect, but by the relation among the principal's farsightedness, the system's memory and the length of the commitments; the MD literature has no constructive analogue of condition (60). Third, the distribution of farsightedness and the horizon (58) are close in setting to the later limited forecast horizon of Jehiel [306] and the planning horizon of the theory of simplicity [455]. Shchepkin's condition on the coincidence of choice with forecast and plan [482] is close to the requirement of correct forecasts in the limited-horizon equilibrium [306], obtained later. The horizon is used differently: in [455] it sets a simplicity standard for the mechanism with worst-case evaluation beyond the horizon, in DAS it is a constraint to which the principal fits plans and commitments.

*Adjacent dynamic settings of the AST school.* DAS theory does not exhaust the school's dynamic models. In the sustainable development control models (Section 6.7) the dynamics is created not by the participants' farsightedness and the commitment regime but by a controlled system with phase constraints; the participants' horizons there are as a rule identical, and time preferences are given by discounting. Result (54) carries cost compensation over to controlled Markov processes with an infinite horizon; the authors themselves present it as a generalization of the incentive problem in dynamic active systems [426, 462]. The third group of dynamic settings is formed by the project scheduling problems (Section 6.9), in which time enters through the network schedule and the participants are, as a rule, passive.

## 6.9. Project Management and Network Programming

*Project management* is one of the most extensive applied areas of the AST school and at the same time its oldest topic: V.N. Burkov worked on network planning and scheduling from the mid-1960s, before the notion of an active system appeared [116, 121], and in the 2000s project management became, by the authors' own assessment, the theory's most intensively developing application, with the independent schools of A.V. Tsvetkov and S.A. Barkalov [121]. The work forms two corpora that overlap only weakly.

In the optimization corpus (network models, resource allocation, aggregation, network programming) the project participants are passive, and problems of mathematical programming and scheduling theory are solved.

In the game-theoretic corpus (the book [119], the line of D.A. Novikov and A.V. Tsvetkov [319, 361], joint financing, incentives for crashing durations) the participants are active, and it is comparable with MD. The book [319] shows the transition between the corpora: its first chapter is built for passive participants under complete awareness, the second drops this assumption and considers expertise, incentive and planning mechanisms.

*Network models and resource allocation.* Operation $i$ of the project has volume $W_i$ and is performed at a rate $v_i(u_i)$ depending on the amount of resource $u_i$; the total amount of the resource is $R$. For independent operations with concave $v_i$ there is a duration-optimal solution in which every operation runs at constant intensity and all operations finish simultaneously, the minimum duration $T$ being found from the equation $\sum_i v_i^{-1}(W_i/T) = R$ [115]. For power-law dependences $v_i(u) = u^\alpha$, $\alpha \le 1$,

$$T_{\min} = \frac{W_e}{R^\alpha}, \quad W_e = \left(\sum_{i=1}^{n} W_i^{1/\alpha}\right)^\alpha, \tag{61}$$

where $W_e$ is the equivalent volume of the complex of operations: the complex is replaced by a single operation without loss of accuracy (perfect aggregation). For $\alpha < 1$, a complex of operations with any network structure aggregates perfectly, and the equivalent volume is a convex homogeneous function of the operations' volumes [44, 115]; the authors trace the original result to the paper [86]. Aggregation is interpreted as a reflection of the management hierarchy: the upper level allocates resources among projects by aggregated descriptions, the project managers among operations [44, 115]. The same corpus contains the priority rule for operations by degree of criticality, optimal for independent operations and for independent chains under an additional condition but not in general [44]; the project sequencing rules by the forgone-benefit criterion [115]; time-cost optimization and the "dual" network model, which accounts for crews moving between sites and already in simple cases leads to traveling-salesman-type problems [43, 47]; and models with "soft" dependencies between the projects of a program [95]. A systematic textbook exposition of this subject matter is given in [49].

*The network programming method.* For the discrete optimization problems to which a large part of this corpus reduces, V.N. Burkov and I.V. Burkova proposed the method of dichotomous [89] and then of network programming [94, 128, 90]. A function of many variables is represented as a superposition of simpler functions, and the structure of the superposition is depicted by a network whose inputs correspond to the variables and whose terminal node to the function itself. Suppose $f(x)$ is to be maximized subject to the constraints $\varphi_j(x) \le b_j$, $j = 1, \dots, m$, and $x \in X_{m+1}$. The objective function is split into summands $h_j$, and for each constraint its own estimation problem is solved:

$$F_j(h) = \max_{x:\, \varphi_j(x) \le b_j} h_j(x), \quad j = 1, \dots, m,$$
$$F_{m+1}(h) = \max_{x \in X_{m+1}} \left[ f(x) - \sum_{j=1}^{m} h_j\,(x) \right]. \tag{62}$$

**Theorem (Burkova, 2009; Burkov, Burkova, 2010).** *For any feasible solution $x$ and any partition $h = (h_1, \dots, h_m)$, the quantity $F(h) = \sum_{j=1}^{m+1} F_j\,(h)$ is an upper bound: $F(h) \ge f(x)$. The function $F(h)$ is convex, so the generalized dual problem*

$$F(h) \to \min_{h_1, \dots, h_m} \tag{63}$$

*is a convex programming problem* [128, 90].

If the network representation of the problem is a tree (in particular, for additive functions), the solution at the network's terminal node gives the exact optimum, and the successive solution of the estimation problems from the network's inputs to its output is analogous to the dynamic programming scheme; the authors regard Bellman's method as a special case [96, 90, 127]. For $h_j = \lambda_j \varphi_j$, $\lambda_j \ge 0$, one obtains a bound no worse than that of the Lagrange multiplier method, which the authors also regard as a special case, and for linear programming without integrality the generalized dual coincides with the ordinary dual [90]. In the general case the bounds (62) are used in branch and bound. The applications cover the formation of target programs and project portfolios [91, 92, 127] and scheduling with a time–cost dependence [96]. By construction (splitting the objective among the constraints and minimizing the sum of the bounds), the method is akin to the Lagrangian decomposition used in the integer programming literature; the authors themselves compare it with dynamic programming and Lagrange multipliers, and a comparison with decomposition methods would deserve a separate study.

*Project control mechanisms.* The book [119] is organized by project phase: selection of options and expertise; formation of the contractor pool and resource allocation; financing (joint financing, insurance, self-financing, counter-expensive mechanisms); incentives for contractors; operational control (contract renegotiation, predictive self-control). These are in effect the standard AST mechanisms (Sections 3 and 6) tied to the project life cycle. A characteristic "project" problem is incentivizing a reduction of duration [319]. Let $T_0$ be the planned and $T \ge T_0$ the forecast completion time, $y \in [0, T - T_0]$ the reduction of duration the contractor chooses at costs $c(y)$, and $\chi(\cdot)$ the penalties the principal (the project manager) bears for the delay. The principal chooses an incentive function $\sigma$, then the contractor chooses $y$:

$$\Phi(\sigma, y) = \sigma(y) + \chi(T - T_0 - y) \to \min_{\sigma}, \quad f(\sigma, y) = \sigma(y) - c(y) \to \max_{y},$$
$$y^* \in \operatorname{Arg} \min_{y \in [0,\, T - T_0]} [c(y) + \chi(T - T_0 - y)]. \tag{64}$$

The optimal scheme is the quasi-compensatory one: $\sigma^*(y^*) = c(y^*)$ and $\sigma^*(y) = 0$ for $y \neq y^*$ (Theorem 2.3 in [319]). This is the cost compensation principle (15) with the principal's criterion given by the penalties for delay. Under linear penalties and a sufficiently large delay, the optimal reduction does not depend on the forecast $T$. The same book extends the result to sequential and parallel operations and to interval, probabilistic and fuzzy uncertainty (Theorems 2.4–2.7), growing uncertainty never raising the efficiency of incentives. For a network schedule with a unified linear incentive scheme, a duality theorem is proved: the maximum capacity of a path equals the minimum capacity of a cut [123]; the problem of partitioning jobs into groups, each with a single incentive scheme, is reduced to a shortest path search [93]. In the joint financing mechanism the contractors submit bids, the funds are allocated by straight or reverse priorities, and the Nash equilibria are compared; it is shown, in particular, that under both principles an agent with a lower priority invests more of its own funds [483]. Project portfolio formation problems are reduced to knapsack problems and to resource allocation on networks [361].

*The Voronezh school.* The Voronezh school (the control department of Voronezh State Technical University; S.A. Barkalov, P.N. Kurochka and their colleagues) traces its interest in AST to the 1994–1997 seminars of SOVNET, the Russian project management association, which began many years of contact with V.N. Burkov [47]. With his participation, the aggregation of large-scale network models was developed [44]; the school then concentrated on resource and calendar planning in construction, models with resource relocation, and incentive schemes for contractors [43, 47]; recent joint work includes the problem of forming generations of new technology, posed as a set covering problem [45]. Jointly with the Moscow group, the three-volume work [46] and the textbook [49] were published. The main content of the Voronezh work is optimization-based; active participants appear in it mainly in incentive models.

*Comparison.* The optimization corpus corresponds to scheduling theory, resource-constrained project scheduling and combinatorial optimization, but not to MD: it has no active participants. The game-theoretic corpus corresponds to the theory of incentives in procurement and regulation [338] and to contract models. Problem (64) is a "duration–reward" contract problem under complete awareness, and joint financing and contractor tournaments are allocation rules with submitted bids. The AST school's distinctive feature is tying the set of mechanisms to project phases and placing them next to the optimization models of the schedule; the MD literature's distinctive feature is the analysis of the contractor's private information about costs and durations. Settings in which the project's network model, with its critical paths and aggregation, is combined with the private information of the individual operations' contractors and with a strategy-proofness requirement are few in the AST work: [319] gives the fair play principle for a single contractor and, referring to general AST results, points to a strategy-proof mechanism for shortening the production cycle under Cobb–Douglas-type cost functions, but there is as yet no general result for an arbitrary network of active contractors (problem 17 of Section 10.4).

# 7. Comparative Analysis

## 7.1. The Shared Core and Areas of Divergence

The comparison in Sections 3–6 allows the results to be divided into three groups (Fig. 11).

*The shared core.* The sufficiency of direct truthful mechanisms and the characterization of strategy-proofness; strategy-proof mechanisms under single-peaked preferences (expertise and re-source allocation); the contract under an observable action (incentives for the joint result are similar only in the setting, Section 3.5); the basic hidden-type model; the hierarchical principal–agent game with a strategy-function; evaluation of a mechanism by the best or the worst equilibrium. In these problems the results coincide up to notation and the criterion (Table 3); what differs is the order in which they were obtained. The priorities are assigned by the rules of Section 2.2 (the detailed arguments are in Sections 2.2, 3.2, 3.5 and 4.4): the fair play principle (1969–1971) precedes Gibbard's and Myerson's formulations of the revelation principle but follows the work of Hurwicz and Vickrey; the theorem on the value of the game $\Gamma_2$ and the guaranteed-result criterion belong to Germeier's school, while the setting "the principal announces a function of an observable variable" arose in economics independently and in the same years; the construction of the equivalent direct mechanism for priority-based resource allocation mechanisms (1989) precedes Sprumont's characterization of the uniform rule (1991) but follows Benassy's introduction of the rule (1982); the price of unification for incentive schemes (1999) precedes the multidimensional loss bounds of 2013–2020; the median schemes, the VCG mechanisms and the Bayesian theory of contracts were first obtained in MD.

*MD only.* Bayesian optimization with virtual valuations and impossibility theorems of the Myerson–Satterthwaite type; implementation theory with equilibrium refinements and its robust version; dynamic design with changing private information; multidimensional screening; information design with commitment; matching and market design; approximation, complexity and learnability; relational and incomplete contracts; mechanisms for algorithmic agents. In these areas the AST school either has no work or has only individual comparative papers.

*AST and TCO only.* The unified typology of controls covering staff, structure, institutions, incentives and information; the theory of coordinated planning with explicit plans and penalties of bounded growth, in which the planning procedure and the incentive scheme are synthesized jointly (Section 6.3); the decomposition of the agents' game in dominant strategies; the catalog of standard mechanisms and the formulation of the problem of their integration (a general theory of integration has been created in neither research tradition); reflexive games with an awareness structure and an informational equilibrium (Section 6.5); the formulation of mob control problems and of informational confrontation of principals in networks (the MD models of network influence, Section 6.5, cast the principal's strategic role differently); the consistent use of the result guaranteed over unknown parameters from the very beginning (a close criterion was used in MD in the 1978 paper of Hurwicz and Shapiro but became widespread only in robust design); the taxonomy of uncertainty with fuzzy models [407]; the theorems on the invariance of equilibrium to the decentralization of mechanisms [408] (MD analogues are studied for other settings, Sections 3.2 and 6.6); structure as a consequence of the type of hierarchical game [410]; and the distributions of farsightedness, the control regimes and the adequacy principle in dynamic systems [427]. The list also includes the cooperative formulation of the control problem, in which the set of implementable actions is determined by the solution of the agents' cooperative game and the stability criterion is the balancedness of the game [244]; in MD coalitional stability is studied for individual classes of mechanisms (coalitional strategy-proofness, coalition-proof equilibria, core-selecting mechanisms) but not as a general synthesis criterion.

For reflexive games and informational control, the correspondence of the two traditions' notions is collected in Table 9; the detailed justification is given in Sections 6.5.2–6.5.6. Multiple structures with a variable composition (GMAS) add to the "AST only" list a unified model of a subjective and inconsistent set of participants with an equilibrium and transformation rules.

Table 9. Correspondence between the notions of TCO reflexive games and informational control and the MD models

| AST and TCO | MD analogue | AST's main difference |
|---|---|---|
| Point-type awareness structure [421] | Belief hierarchy, type space [366]; hyper-game [61] | Point beliefs, no common prior |
| Informational equilibrium (41) | Bayesian equilibrium on a finite type space without a common prior [144] | Degenerate beliefs |
| MAS [145] | Kripke models [207]; hierarchies without probabilities [195, 357] | The criterion is the guaranteed result |
| Interval and probabilistic GMAS [210] | Prebayesian games with an unknown number of players [26]; games with population uncertainty and Poisson games [368, 385] | Non-regular structures, phantoms, a subjective and inconsistent composition |
| Transformation of the structure by observations and messages | Dynamic epistemic logic [39, 451, 513] | In AST the observations are generated by the equilibrium of the game |
| Stability | Conjectural and self-confirming equilibrium [52, 216, 472] | In AST it is required of phantom agents as well; there is no learning model |
| False stable equilibrium | Berk–Nash equilibrium [205]; analogy-based equilibrium [307] | In MD the false belief is also the one that best explains the data |
| Strategic reflexion, reflexive partitions [404] | Level-$k$ and cognitive hierarchy models [134, 389] | Theoretical rank bounds versus experimental estimates |
| Reflexive control | Deception [155, 206]; hypergames | TCO agents accept the message without strategic decoding |
| Control of beliefs about the composition | Disclosure of the number of auction participants [350, 362]; the value of ignorance [26]; shill bids [250] | In MD, a single "disclose or not" decision under a commonly known model of the composition |
| Informational confrontation; trust networks, mobs | Competition among senders [219]; the DeGroot and Friedkin–Johnsen models, threshold models [171, 215, 251] | TCO agents average opinions rather than being Bayesian; the basic models are shared |

*The number of dimensions.* The comparison shows that TCO covers more dimensions of the problem than MD in the proper sense. In MD the objects of choice are the game form and the payments; in TCO they also include the set of participants, the structure, the admissible sets (institutional control), the awareness (informational control) and the agents' beliefs about one another (reflexion) [245, 408, 410, 412, 422, 403, 414]. MD analogues of these dimensions exist, but they belong to the neighboring theories – information design, the theory of delegation, organizational economics – and have not been brought into a single statement; models of reflexion with inconsistent beliefs, or of optimizing staff and structure as kinds of control, are not found, and institutional control in its general form (the choice of admissible sets) is represented only by the theory of delegation (Section 5.4). Under the broad reading of MD (including contract theory and the adjacent theories) adopted in Fig. 12, this difference shows up in the AST school's higher expert scores in the rows for reflexion, staff, structure, standard solutions and integration; under the narrow reading it would be even more pronounced.

| MD, contracts, implementation only | The shared core | AST and TCO only |
|---|---|---|
| • Bayesian optimization, virtual valuations | • sufficiency of direct truthful mechanisms | • the unified typology of controls |
| • impossibility theorems | • characterization of strategy-proofness | • coordinated planning: plans and penalties |
| • implementation theory and its robust version | • expertise: median schemes | • decomposition of the agents' game |
| • dynamic design | • resource allocation: the uniform rule, the SRAM | • the catalog of standard mechanisms |
| • multidimensional screening | • the contract under an observable action | • the mechanism integration problem |
| • information design with commitment | • the basic hidden-type model | • reflexive games, informational equilibrium |
| • matching, market design | • the hierarchical principal–agent game | • formulations: mob control, confrontation |
| • approximation, complexity, learnability | • evaluation by the best or the worst equilibrium | • the guaranteed result |
| • relational and incomplete contracts | | • theorems on invariance to decentralization |
| • (outside AST: dynamic stability – Petrosyan's school) | | • fuzzy uncertainty; the price of unification |
| • mechanisms for algorithmic agents | | • cooperative models |

Fig. 11. The shared core of the two research traditions and the areas developed in only one of them

The divergence is explained by three circumstances unrelated to the mathematical machinery. The first is subject matter and audience. In the realm of markets and regulation, MD answered questions of economic policy (privatization, regulation, spectrum allocation, the organization of markets) and therefore developed models with prices, revenue and voluntary participation. But for planning and intra-firm incentives, MD authors of the 1970s–1980s worked on closely related problems on the material of the Soviet planning system and corporate practice [214, 286, 355, 524] without citing Burkov's school; and the school's 2021 survey [102] contains no references to this MD work either. The cause of the divergence is thus the difference of scientific communities (economics versus control theory) and the partial cessation of the AST school's English-language publications after 1971, rather than a difference in the problems solved; the terminological barrier is secondary.

The second is the one-sidedness of citation. After 1971 the AST school had no original English-language publications for a long time, apart from the cover-to-cover translation of Avtomatika i Telemekhanika and a few papers at IFAC congresses [102]. Yet this cannot be called isolation in the strict sense: Germeier's monograph was translated into English in 1986 [227], two papers by Ereshko and coauthors were published by IIASA in 1984 [200, 203], the survey [99] appeared in 1994 in an international economics journal (Mathematical Social Sciences), and in 2013–2021 translations and original English-language monographs of the school were published [57, 59, 81, 85, 106, 147, 402, 403, 398, 414]. Nevertheless, the reference lists of the classical papers on strategy-proof rules under single-peaked preferences [42, 381] contain no references to Burkov or Germeier, and in the subsequent core MD literature (Econometrica, JET, GEB, Theoretical Economics, AER) no references to the editions listed have been found; the citations that do exist belong to Russian authors and engineering applications. AST results published in journals and in core MD terms are cited – Danilov's paper [163], for example. The influence in the other direction was stronger: since the 1990s the AST school has systematically related its results to those of MD [102, 104, 125]. Thus the results of the Burkov and Germeier schools (Table 3) are not reflected in the MD literature, and the absence of translations cannot explain this.

The third is the preferred type of result: existence, uniqueness and impossibility theorems in MD, versus optimal mechanisms in parametric classes and standard solutions in the AST school.

## 7.2. A Coverage Map of Problem Classes

Fig. 12 gives the survey author's expert assessment of how developed each class of problems is in each research tradition as of 2026. The scale has four levels with pre-set criteria: 0 – no work; 1 – one or two works, or comparative papers only; 2 – several works by different authors, with surveys or monograph chapters, but with basic questions still open; 3 – a completed theory: the main theorems formulated and proved, with monographs and textbook expositions. The assessment is based only on the sources included in the survey and reflects the depth of the theory, not the number or volume of publications.

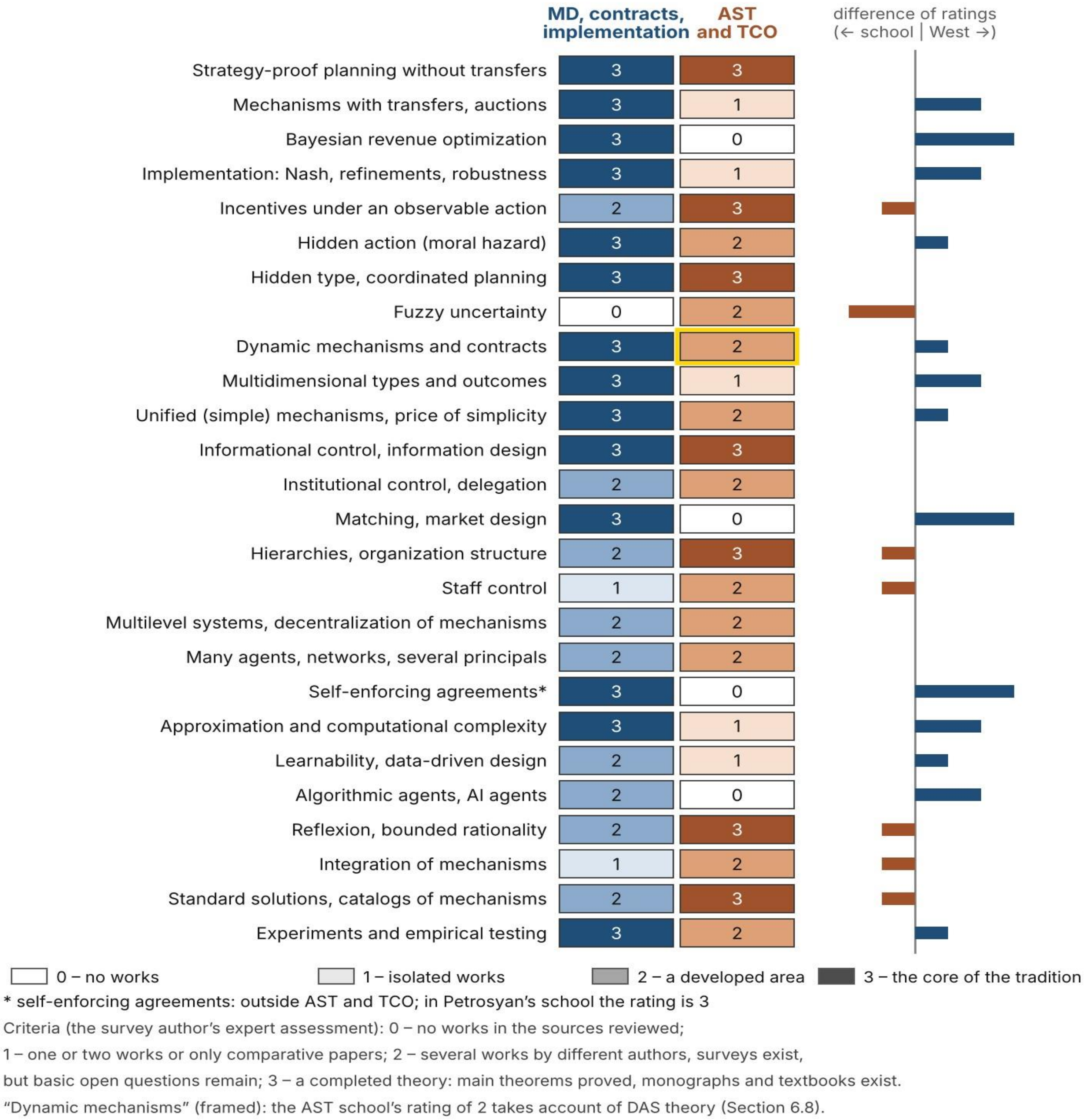

Fig. 12. Coverage of problem classes by the two research traditions as of 2026 (the survey author's expert assessment on a 0–3 scale)

The map shows three zones. In the parity zone (strategy-proof planning without transfers, hidden type and coordinated planning, multilevel systems and decentralization, networks and systems with several principals) the scores coincide. In the MD advantage zone the gap is largest for Bayesian revenue optimization and matching, where the AST school has no work, and substantial for mechanisms with transfers, implementation theory, dynamic and multidimensional design, approximation and complexity, and self-enforcing agreements (studied outside AST, in Petrosyan's school). The AST advantage zone contains incentives under an observable action, reflexive models, organizational structure, staff control, fuzzy uncertainty, the integration of mechanisms and standard solutions; the MD results here either exist but are scattered across disciplines, or are absent. The AST school's high scores in the rows “staff control”, “fuzzy uncertainty”, “unified mechanisms and the price of simplicity”, as well as in hidden action, informational control, institutional control and organizational structure, are raised in view of the monographs [245, 375, 408, 410, 407, 412, 422, 403, 414]. The AST school's score in the row “dynamic mechanisms and contracts” is 2 in view of DAS theory [406, 427, 426, 481, 482] (Section 6.8); it is not raised above 2 because this theory does not cover private information changing over time.

The four-level scale of Fig. 12 is convenient for a map but coarsens the differences within the zones. Fig. 13 gives the same assessment on a 0–10 scale in the axes of the TCO classification of problems [412]: the basic model of motivational control, kinds of uncertainty, information revelation, multilevel and multi-agent systems, distributed control, cooperative models, dynamics, activity constraints, staff, structure, and informational control; uncertainty is split into interval, probabilistic and fuzzy, since it is here that the research traditions differ most. The level criteria: 0 – no work; 1–3 – individual works or comparative papers only; 4–6 – a developed strand with several authors and surveys but with open basic questions; 7–9 – a completed theory with the main theorems, monographs and textbook expositions; 10 – the core of the research tradition. MD is taken in the broad reading, the AST school without the adjacent schools. The AST contour is higher along the axes of the basic model, interval and fuzzy uncertainty, multilevel and multi-agent systems, cooperative models, activity constraints, staff, structure and informational control; the MD contour along probabilistic uncertainty, information revelation and dynamics; for distributed control the scores are close (in MD – common agency [73], in AST – [244, 313, 430]). The AST score for cooperative models rests on the monograph [244] and the coalitional strategy-proofness conditions [444], the MD score on [28, 72, 167] and the collusion models [339]. The MD score for informational control is below AST's, since information design covers only the case of the principal's full commitment under consistent beliefs of the agents and contains no analogues of informational equilibrium, informational control or control of beliefs about the composition (Section 6.5); the MD scores for multilevel systems and activity constraints are low, since hierarchical contracts and the theory of delegation describe three-link chains and interval delegation to a single agent, not the structure and admissible sets of a multi-agent system.

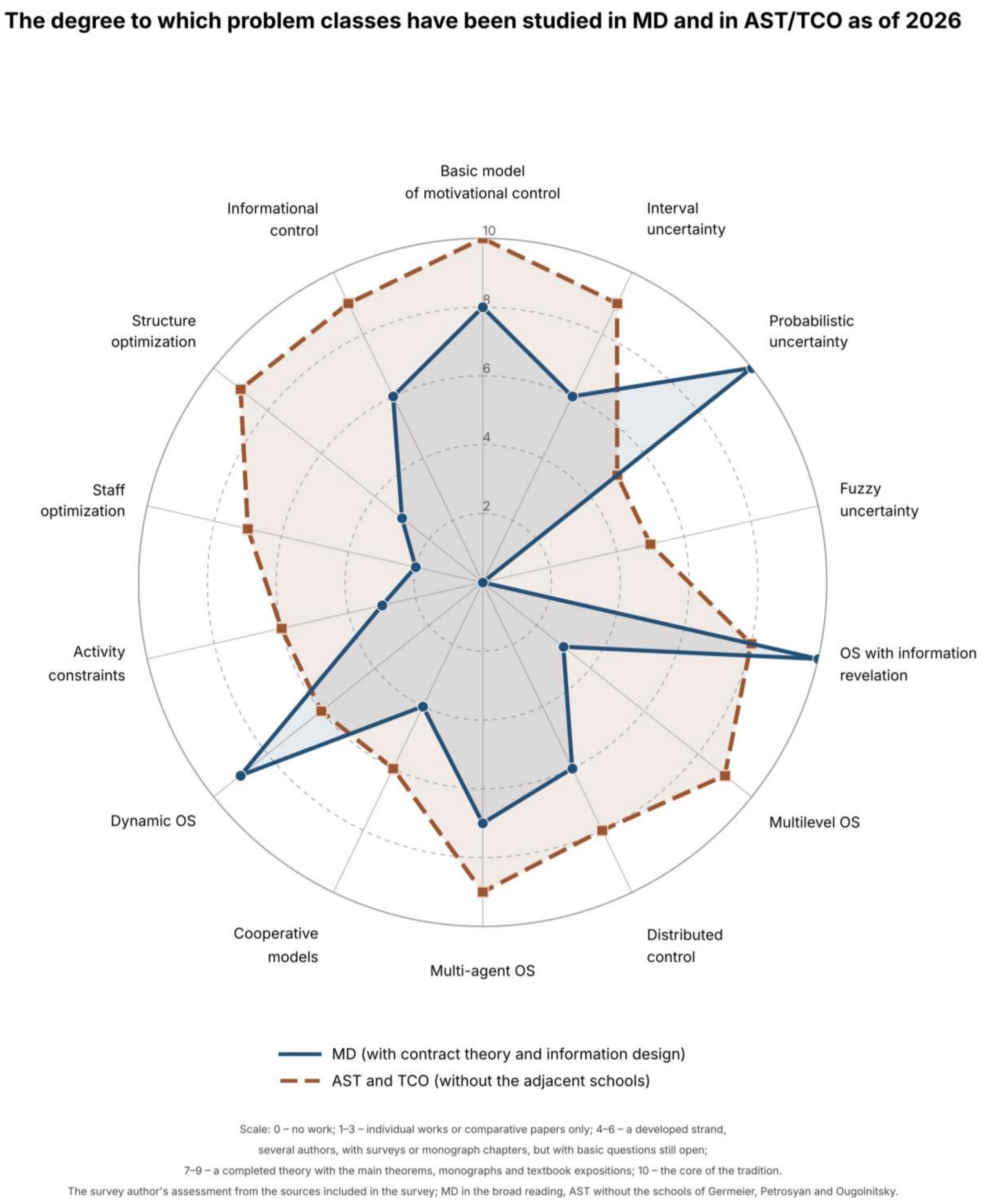

Fig. 13. The degree to which problem classes have been studied in MD and in AST/TCO as of 2026, on a 0–10 scale (the axes are the TCO classification of problems)

## 7.3. Strengths and Weaknesses

The summary assessment is given in Table 10. It rests on the material of Sections 3–6 and on the assessments the research traditions themselves give of their state [102, 139, 179, 288].

Table 10. Strengths and weaknesses of the research traditions (the survey author's assessment)

| | Mechanism design, contract theory, implementation theory | AST and TCO |
|---|---|---|
| Strengths | Generality of the statements and exact characterizations; a developed impossibility theory marking the boundaries of the attainable; the Bayesian apparatus and its robust alternatives; computational and statistical theory; experimental and empirical validation; large-scale documented deployments | A single scheme uniting incentives, plans, institutions, information, staff and structure – more dimensions than in MD; the result guaranteed over unknown parameters as the primary criterion; a taxonomy of uncertainty including interval and fuzzy; the theory of multilevel systems and of mechanism decentralization; the orientation toward the optimal mechanism in a tractable class and toward standard solutions; the explicit modeling of actions after the plan is received; the tradition of business games; readiness for integrated mechanisms |
| Weaknesses | Fragmentation into subcommunities with different models; the dependence of Bayesian mechanisms on unobservable details; optimal mechanisms often too complex to apply; a weak theory of combining several mechanisms in one organization | Few results in international journals and no citations from the MD core despite the translated monographs; a narrow class of preferences and solution concepts; approximation-type bounds obtained only for tournament mechanisms and the price of unification; no theory of sample complexity or of continuous-time dynamics for incentive problems with private information (the Rostov school does have full-information dynamic hierarchical games); weak empirical validation and no independent assessment of the deployments |
| Opportunities | Transfer to the economics of AI agents; combination with machine learning; distributional and public goals | Connecting the catalog of mechanisms to digital management systems; translating the core into the language of robust and algorithmic design; organizations with a mixed staff of humans and AI agents |
| Risks | A gap between theory and platform practice, where mechanisms change faster than they are analyzed; collusion of learning algorithms | Withdrawal into its own terminology; loss of comparative advantages as MD masters robustness and simplicity |

# 8. A System of Classifications of Problems and Results

## 8.1. A Faceted Classification of Problems

The existing classifications are tied to the traditions: in TCO problems are divided by the object of control and by the extensions of the basic model [412], in the MD literature by the historically established branches. The comparison requires a classification neutral to terminology. A faceted scheme is proposed: a mechanism synthesis problem is described by a set of values of ten independent attributes,

$$z = (\varphi_1, \varphi_2, \dots, \varphi_{10}), \qquad \varphi_k \in \Phi_k, \tag{65}$$

where the value sets $\Phi_k$ are listed in Table 11. The first six facets describe the model (what the principal chooses, what it knows and observes, what instruments it has, how the system is organized), the next two the criterion and the solution concept, and the last two the restrictions on the mechanism and the nature of the agents. The scheme is depicted in Fig. 14. Facet F2 describes the principal's awareness and, through the subvalue "the agent's own", the agent's own uncertainty; the agent's attitude to uncertainty (ambiguity aversion) belongs to F10. The scheme is proposed by the author of this survey; its purpose is to make comparable the statements written in the traditions of the different research schools and to point out unfilled combinations of attributes.

Table 11. Facets of the classification of mechanism synthesis problems

| Facet | Values | Note |
|---|---|---|
| F1. Object of control | The decision rule (the plan); payments (incentives); the sets of admissible actions; the agents' awareness; the staff; the structure | Coincides with the TCO types of control; MD studies only the first two values |
| F2. The principal's awareness | Complete; a set of parameter values; a fuzzy set (a membership function); a distribution; a set of distributions or beliefs; a data sample | From AST (a set, a fuzzy set [407]) to Bayesian MD (a distribution), robust MD (a set of distributions) and learning-based MD (a sample). The agent's uncertainty and its perception (internal and external uncertainty in AST terms, ambiguity in MD [79, 173, 529]) are coded by the subvalue "the agent's own" and facet F10 |
| F3. Observability | Actions; the result, the result with noise; messages only; messages with selective verification | Distinguishes the observable action, moral hazard, the hidden type, verification |
| F4. Utilities and instruments | Nontransferable utility (NTU); transferable (TU); transfers with limited liability; penalties of bounded growth | Determines the class of strategy-proof mechanisms |
| F5. System structure | One agent; many agents; several principals; a multi-level hierarchy; a network | |
| F6. Time and commitment | Statics; dynamics with full commitment; dynamics without commitment (renegotiation, relational contracts); adaptation; limited farsightedness of the participants and the rolling regime with commitments (Section 6.8) | |
| F7. The principal's criterion | Implementation of a given rule; the full-information value; the expected value; the guaranteed value; a share of the optimum; regret under learning | Formulas (1), (12), (3), (25) |
| F8. Solution concept | The single agent's optimum; dominant strategies; ex post; Nash and refinements; Bayesian equilibrium; rationalizability; obvious dominance; informational (reflexive) equilibrium; learning dynamics | In the absence of a game – an optimization problem |
| F9. Restrictions on the mechanism | None; a parametric class (simplicity); computational complexity; message volume; incentive compatibility for the agent and/or for the principal itself | |
| F10. Nature of the agents | Rational; maximin (ambiguity-averse); boundedly rational and reflexing; coalitions; learning algorithms; language models and AI agents | |

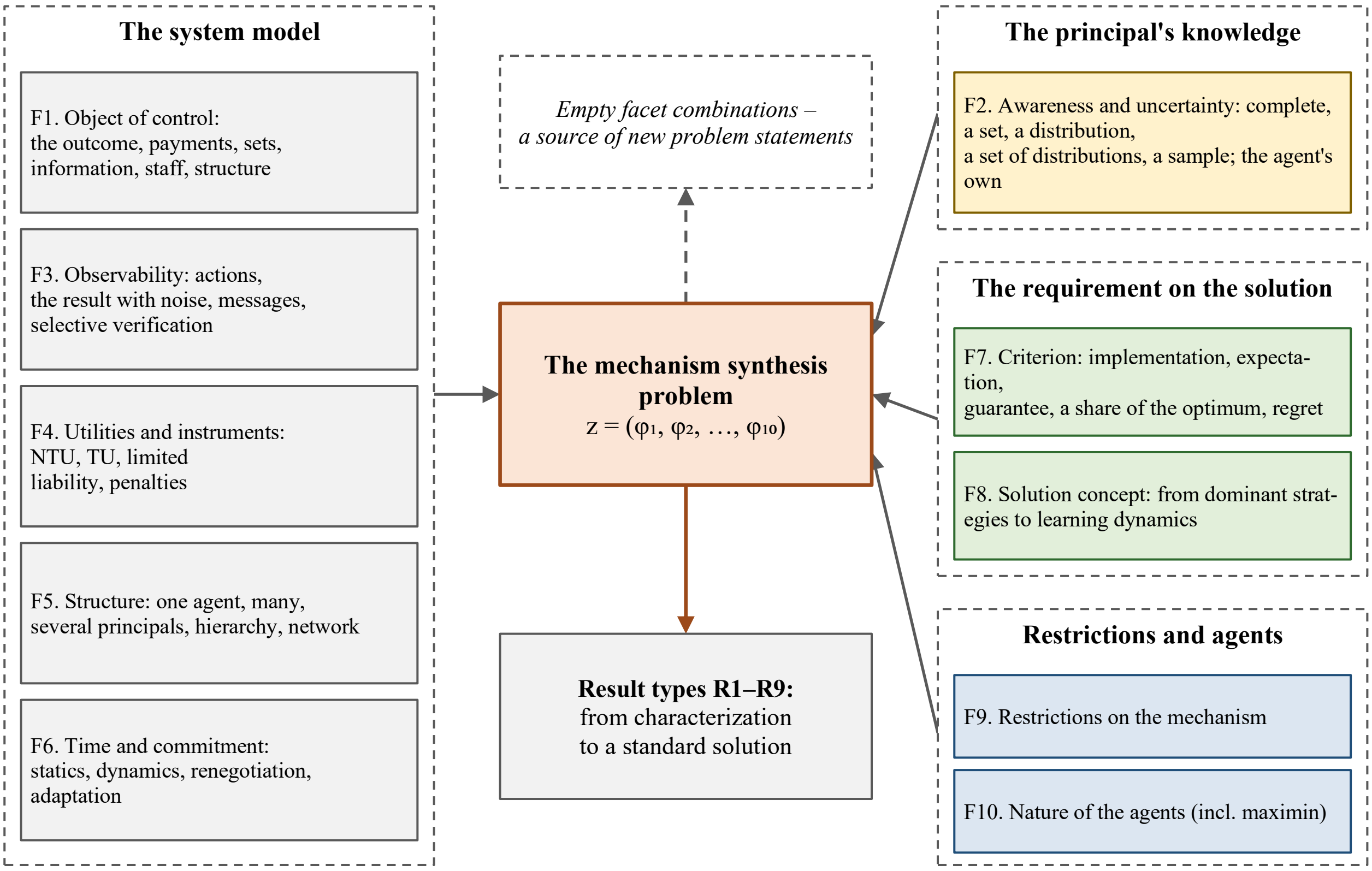

Fig. 14. The faceted classification of mechanism synthesis problems: the ten facets and the groups they form

The projection of the scheme onto two facets – the principal's awareness (F2) and the solution concept (F8) – is shown in Fig. 15. It shows that classical AST and classical Bayesian MD occupy different cells of the same table, while the post-2005 development fills the cells between them: robust MD moves toward the guaranteed criterion while partly keeping the Bayesian solution concept, algorithmic MD combines dominant strategies with the worst case, and learning-based design and mechanisms for algorithmic agents open a new row and a new column.

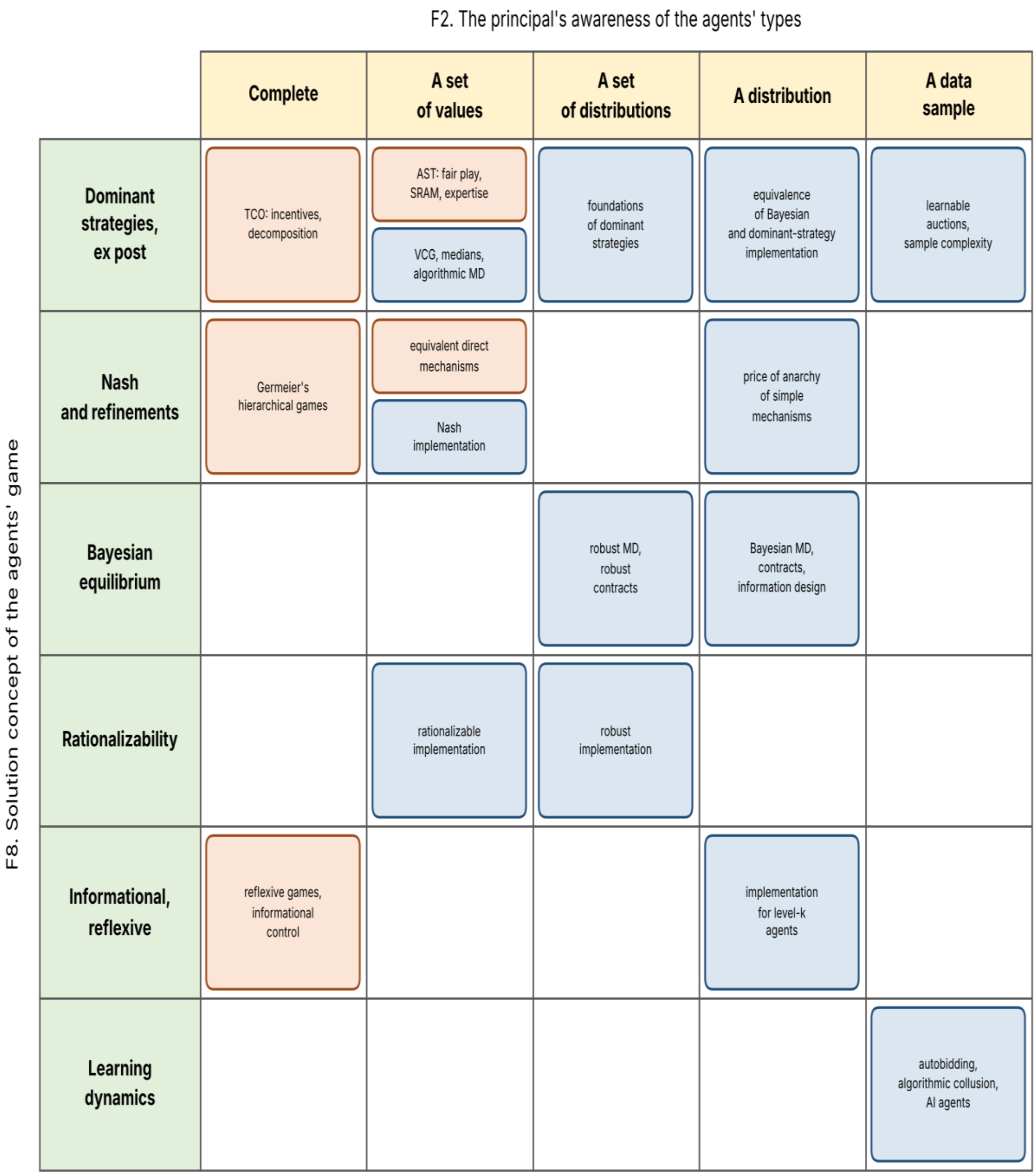

Fig. 15. The projection of the classification onto the facets "the principal's awareness" and "solution concept": the location of the main strands

The second projection – onto the facets "object of control" (F1) and "system structure" (F5) – is shown in Fig. 16. It depicts the difference between the research traditions in the number of dimensions (Section 7.1): the MD results are concentrated in the two top rows, the decision rule and the payments, where they are present in every column except planning problems with several principals; in the rows of admissible actions and of the agents' awareness MD is represented by the neighboring theories – delegation, information design and network influence models – while the staff and structure rows are filled almost exclusively by the AST school's work. The empty cells point to unfilled combinations of attributes: institutional control with several principals, in hierarchies and networks; informational control and staff control in multilevel systems; staff control in MD under any system structure.

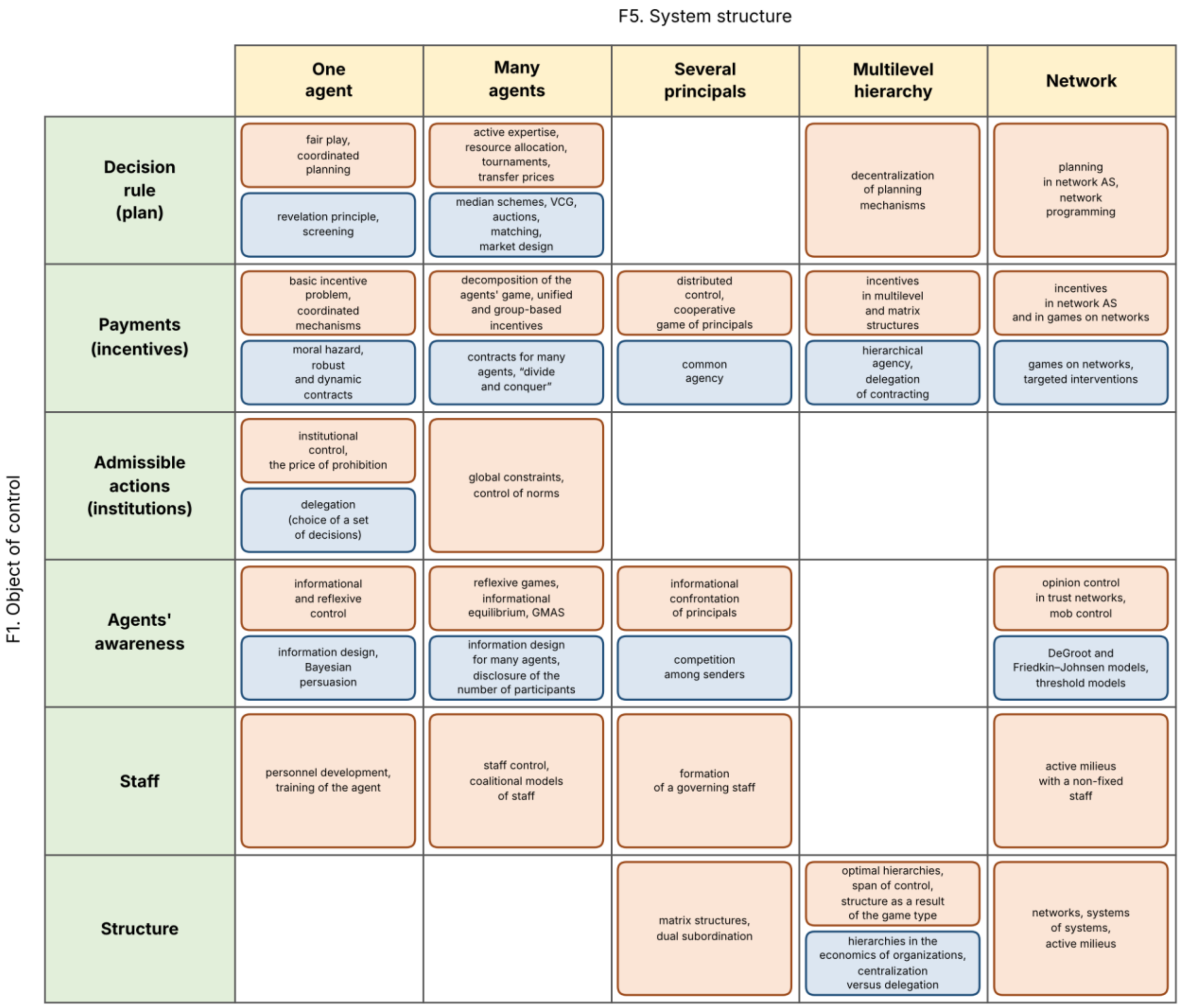

Fig. 16. The projection of the classification onto the facets "object of control" and "system structure": the location of the main strands

The coding of the canonical problems is given in Table 12 (the abbreviation "lim. liab." means transfers with the agent's limited liability). It shows that problems considered distinct within the research traditions often differ in only a few facets. The basic hidden-type model and Enaleev's coordinated mechanism (in the special case of an unlimited fund and an unbounded penalty) differ in the principal's awareness and the criterion; the uniform rule and the SRAM coincide in all facets; Carroll's robust contract differs from the AST incentive problem in the observability of the action, the principal's awareness of the agent's technology, and limited liability.

Table 12. Coding of the canonical problems in the proposed classification (the main facets)

| Problem | F1 | F2 | F3 | F4 | F7 | F8 |
|---|---|---|---|---|---|---|
| Optimal auction [384] | Outcome, payments | A distribution | Messages | TU | Expectation | Bayesian |
| VCG mechanism [257] | Outcome, payments | A set | Messages | TU | Implementati on | Dominant |

| Problem | F1 | F2 | F3 | F4 | F7 | F8 |
|---|---|---|---|---|---|---|
| The uniform rule [486], the SRAM [97] | Outcome | A set | Messages | NTU | Implementati on | Dominant |
| Active expertise [110], median schemes [381] | Outcome | A set | Messages | NTU | Implementati on | Dominant, Nash |
| Nash implementation [358] | Outcome | A set | Messages | NTU or TU | Implementati on | Nash |
| The quasi-compen-satory scheme, de-composition [428] | Payments | Complete | Actions | TU | Value or guarantee | The agent's optimum, dominant |
| Moral hazard [284] | Payments | The noise distribution | The result | TU | Expectation | The agent's optimum |
| The robust contract [136] | Payments | A set of technologies | The result | TU, lim. liab. | Guarantee | The agent's optimum |
| Screening [50] | Outcome, payments | A distribution | Messages | TU | Expectation | The agent's optimum |
| The coordinated mechanism [189] | Plan, incentives, penalties | A set | Messages and actions | Penalties | Guarantee | Dominant |
| Delegation [13] | Sets | A distribution | Actions | NTU | Expectation | The agent's optimum |
| Bayesian persuasion [311] | Information | A distribution | Actions | NTU | Expectation | Bayesian |
| Informational control [422] | Information | Complete | Actions | NTU or TU | Value or guarantee | Informational |
| The optimal hierarchy [375, 246] | Structure | Complete | Actions | TU | Value | No game |
| The neural-network auction [181] | Outcome, payments | A sample | Messages | TU | Expectation | Dominant, approximately |
| The token auction [182] | Outcome, payments | A set | Messages | TU | Implementati on | Dominant |
| The contract with a maximin agent [79] | Payments | A distribution for the principal; a set for the agent | The result | TU | Expectation | The agent's optimum (maximin) |
| The mechanism with maximin agents [529] | Outcome, payments | A distribution for the principal; a set for the agent | Messages | TU | Expectation | Bayesian, maximin |

## 8.2. A Typology of Results

The results obtained for problem (65) fall into nine types (Table 13). Types R1–R4 make up the classical "exact" theory, R5–R7 the theory of approximate and robust solutions that took shape after 2000, R8–R9 the empirical and engineering layer.

Table 13. The types of results and their examples in the two research traditions

| Type of result | Content | Examples: MD | Examples: AST and TCO |
|---|---|---|---|
| R1. Characterization | Necessary and sufficient conditions for implementa-bility or strategy-proofness | Maskin monotonicity [358]; median schemes [381]; the contraction condition [65] | Condition (4) [98]; separability [110] |

| Type of result | Content | Examples: MD | Examples: AST and TCO |
|---|---|---|---|
| R2. Impossibility | Incompatibility of a set of requirements | The Gibbard–Satterthwaite and Myerson–Satterthwaite theorems; the trilemma [6] | MD results are used; few of its own |
| R3. Optimal mechanism | An explicit solution of the optimization problem | Myerson's auction; Sannikov's contract; the linear contract (32) | The fair play law; schemes (15), (16); Enaleev's mechanism |
| R4. Equivalence | Coincidence of the results of mechanisms or concepts | Revenue equivalence; Bayesian and dominant-strategy implementation [231] | Equivalent direct mechanisms [405]; equivalence of priority-based mechanisms [328] |
| R5. Approximation | A guaranteed share of the optimum for a simple mechanism | Bound (25); reserve prices [276]; the price of anarchy [469] | The guaranteed efficiency of tournament mechanisms [120]; the price of unification [408]; the error of active planning [102] (a formulation) |
| R6. Complexity | Computational, communication, sample | NP-hardness, the price of truthfulness [175]; sample complexity [264] | The complexity and error of control problems [399]; the dimension of messages [241] |
| R7. Robustness | Resilience to errors of the model, beliefs, information | Robust implementation [64]; non-robustness [5] | The guaranteed criterion as the starting point; generalized solutions |
| R8. Empirical validation | Experiment, data, assessment of deployment | Laboratory and field studies [141, 453] | Business games [330]; identification on network data [260] |
| R9. Standard solution | A mechanism developed into a regulation or a program | Auction formats, matching algorithms [370, 465] | The catalog of basic mechanisms [153]; integrated mechanisms [107] |

Joining the two classifications yields the morphological matrix "problem class – result type" (Fig. 17). Its occupancy differs: in MD almost all rows are filled, R4 and R9 more weakly than the others; in AST the results are concentrated in rows R3 and R9, noticeable in rows R1 and R4, and almost absent in rows R2, R5 and R6 (the exceptions in row R5 are the guaranteed-efficiency bounds for tournament mechanisms and the price of unification). The empty cells of the matrix serve as a source of problem statements for Section 10: they include, for example, approximation bounds for the standard AST mechanisms, impossibility theorems for integrated mechanisms, and characterizations for informational control.

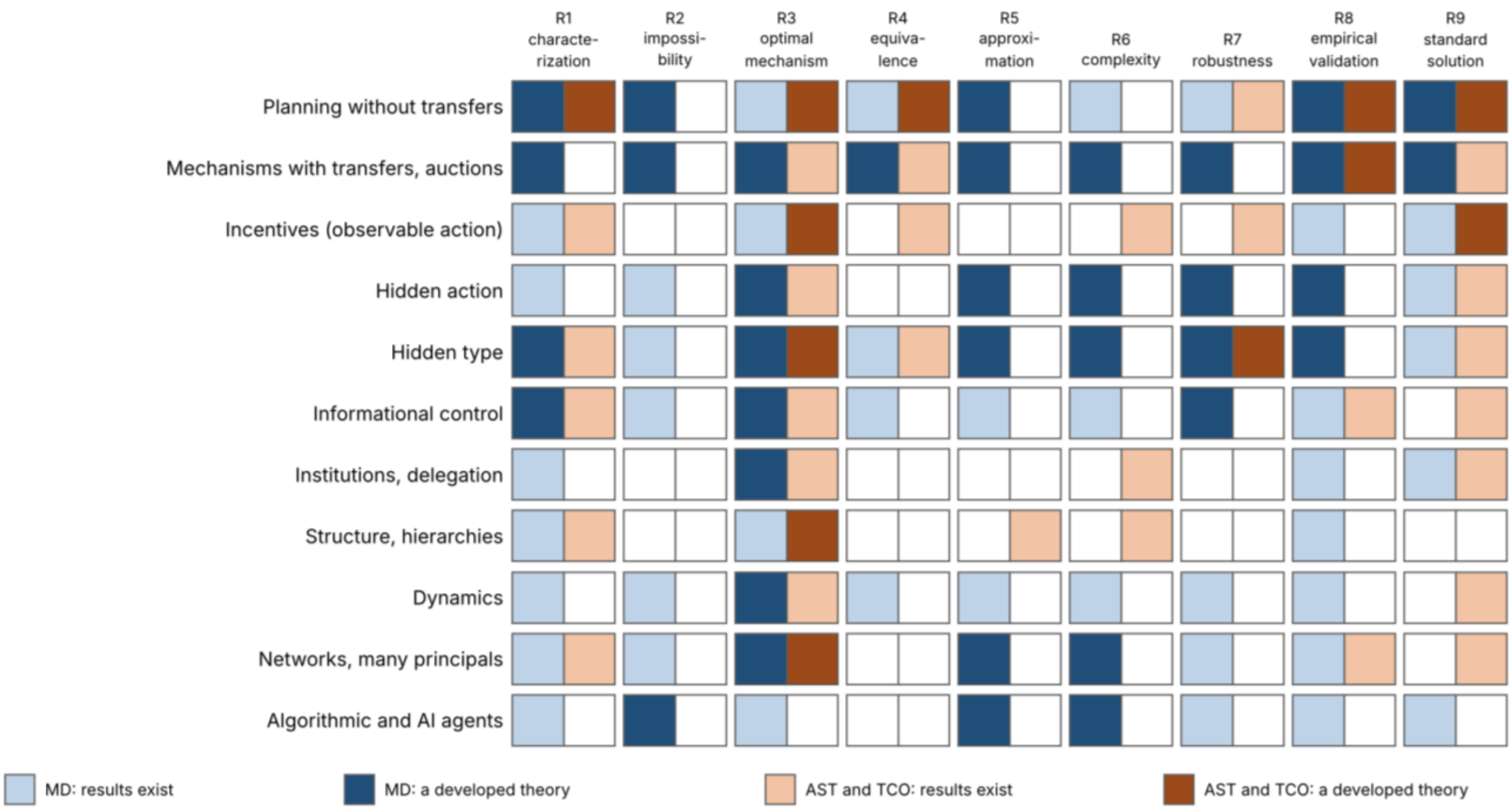

Fig. 17. The morphological matrix "problem class – result type": the presence of results in MD (the left half of a cell) and in AST and TCO (the right half); the survey author's assessment

# 9. Applications of the Theoretical Results

## 9.1. MD: Markets, Platforms, Public Institutions

The applications of MD are distinguished by scale and documentation: the design decisions are published and their results measured. The main areas are collected in Table 17 in the Appendix.

*Auctions.* The simultaneous multi-round auction was first used by the U.S. Federal Communications Commission in 1994; over 1994–2014 auctions of this format raised more than $120 billion in the United States and more than $200 billion worldwide [497]. The incentive auction of 2016–2017 was the first to reallocate spectrum between uses: a reverse auction bought out the rights from television broadcasters, a forward auction sold them to telecom operators; its design required solving an economic and a computational problem jointly, since checking the feasibility of repacking the stations became a constraint of the mechanism itself [370, 497]. In online advertising the generalized second-price auction is studied in [186], and in the 2020s the bids are set by algorithms [3]. Auctions also sell electricity in wholesale markets [154], government bonds and emission quotas [497].

*Matching and allocation without money.* The chain of reforms begins with the assignment of medical school graduates [465] and school students [1, 496] and continues with kidney exchange [466], course allocation [83] and the distribution of food among food banks [453]. The generalization of this experience is Sönmez's "minimalist" market design: identify the institution's purpose and change only what obstructs it; among the reforms he describes are housing allocation, school choice, organ exchange, the assignment of military academy graduates to branches of service, and the allocation of medical resources during the pandemic [490].

*Decentralized systems.* The EIP-1559 fee mechanism has operated in the Ethereum network since August 2021: the first-price auction was replaced by variable-size blocks, a history-dependent reserve price and the burning of the base fee [468]. It is one of the first protocol-level mechanisms whose incentive compatibility has been analyzed systematically [467, 468]; the Bitcoin fee market was analyzed earlier [292, 344], and the proposal itself is described in [129].

*Contracts and regulation.* Contract theory (CT) is applied to executive compensation, insurance, the regulation of natural monopolies and public procurement [185, 338, 492]. The theory's own authors are reserved in their assessments: with several tasks, strong incentives on measurable indicators crowd out effort on unmeasurable ones, and pay for performance is not always appropriate [290, 288].

## 9.2. AST and TCO: Organizations, Projects, Industries

The applications of the AST school are documented mainly by the authors themselves; independent (third-party) assessments of the effect are almost nonexistent, so what follows lists the declared areas of application [109, 121, 120] (see Table 18 in the Appendix).

In the 1970s–80s AST mechanisms were deployed in industry-level management systems: coordinated planning systems at the nonferrous metallurgy enterprises of Kazakhstan (A.A. Ashimov's school, Section 6.7 [22, 88]), a performance assessment system in instrument making, integrated systems in the radio industry, multi-channel mechanisms in the “Operator Advisor” system in ferrous metallurgy; in the late 1980s a two-year experiment with counter-expensive taxation was run in research organizations. State prizes were awarded for part of this work [121]. In the 1990s–2000s the applications shifted to enterprise restructuring, project management, offset schemes, safety programs and integrated rating [119, 121]. The Samara school declares deployments of coordinated incentives at AVTOVAZ and the Samara grain elevator (Section 6.7 [75, 300]). In the 2010s–20s the declared areas include the management of federal and regional projects and programs, ecological-economic systems [402] (including control of the hydrological regime of the Volga-Akhtuba floodplain [299, 518, 520, 521]), production systems and “extended enterprises” [55, 59], and the analysis of online social networks [260]. The proceedings of the 2024 anniversary conference give an example with a quantitative result: after the introduction in 2022 of a system of departmental project offices, the share of control points completed late fell from over 20% to about 13% [498].

Two deployments of the 1980s are documented better than the others. The *multi-channel mechanisms* in ferrous metallurgy were developed jointly with V.P. Avdeev's Novokuznetsk school (the Siberian Metallurgical Institute) [31]. The original problem was a process control system with an “operator advisor”: a passive advisor is of little use, since during training its recommendations are poor, and an experienced operator stops heeding them. The solution is an *active advisor*: a simulation model estimates what the result would have been had the recommendation been followed, and the operator is rewarded if his control proved better and penalized otherwise; this organizes a competition of decision channels, after whose introduction the operator began to follow the model in normal situations; the generalization is several advising channels based on different models (multi-channel mechanisms) [88]. The deployment in the “Operator Advisor” systems at ferrous metallurgy enterprises, with the participation of A.K. Enaleev and T.V. Kiseleva, produced, by the school's testimony, a significant economic effect and was recognized with a State Prize [88]; in the 2000s the line continued as *multi-variant active systems* [32, 318], combining an organizing system, a system for evaluating the efficiency of variants, and variant systems with physical, model and mixed physical-model objects. In MD terms the active advisor is a tournament-type mechanism between a human and a model in which the incentives depend on a counterfactual result; the contract theory literature has no analogues of this setting.

*The AKKORD system* (automated quantitative integrated performance assessment) was created by order of the USSR Ministry of Instrument Making, Automation Equipment and Control Systems; the task of building an integrated system for assessing the performance of the industry's enterprises that would stimulate technological progress was set by Minister M.S. Shkabardnya and ICS director V.A. Trapeznikov, and the work was led by A.M. Cherkashin [121]. The methodology of integrated quantitative assessment for research institutes and design bureaus is set out in [126], and the theoretical basis in the monograph [113], where performance assessment mechanisms and incentives for taut plans are treated as an instrument for improving the economic mechanism. By the school's testimony, the system "became a real management tool in the industry", and the outcome of the 1980s was a flexible integrated performance assessment system experimentally deployed in the Ministry [121]. The MD analogue is the discussion of the success indicators of Soviet enterprises and divisional evaluation [355, 524], which was conducted in the same years without references to the work of the Institute of Control Sciences.

The earliest and longest line of applications is network planning and project management: in the 1960s network planning and management systems were deployed at the gold and diamond mining enterprises of Siberia [121]; since the 1990s this line has been developing jointly with the SOVNET association and the Voronezh scientific school of project management, primarily in construction [47] (Section 6.9). The ecological-economic applications of the Rostov school include three-level systems for managing ecological-economic objects [508].

## 9.3. Two Models of Practical Adoption

The ways of carrying theory into practice differ (Fig. 18). MD has developed a public engineering cycle: the problem of a specific institution – a theoretical model – a laboratory experiment – a mechanism design – deployment – measurement of the result on data – publication and correction. Each project is unique, and what accumulates is design principles [464, 490]. The AST school has developed a cycle of standard solutions: a class of management problems – a basic mechanism from the catalog – parameter tuning and a business game – inclusion in the organization's management system. Here it is the catalog of mechanisms that accumulates, while the results of deployment remain inside the organizations.

The first cycle provides verifiability and self-correction but is expensive and applicable where there is a single client and data. The second is cheaper and closer to management practice, but it has no external verification of results. Convergence is possible from both sides: the catalog of standard mechanisms needs measurable guarantees (Section 10.4), and the MD cycle needs a theory of combining several mechanisms in one organization; Holmström's Nobel lecture also points to the need to treat incentives as a system [288].

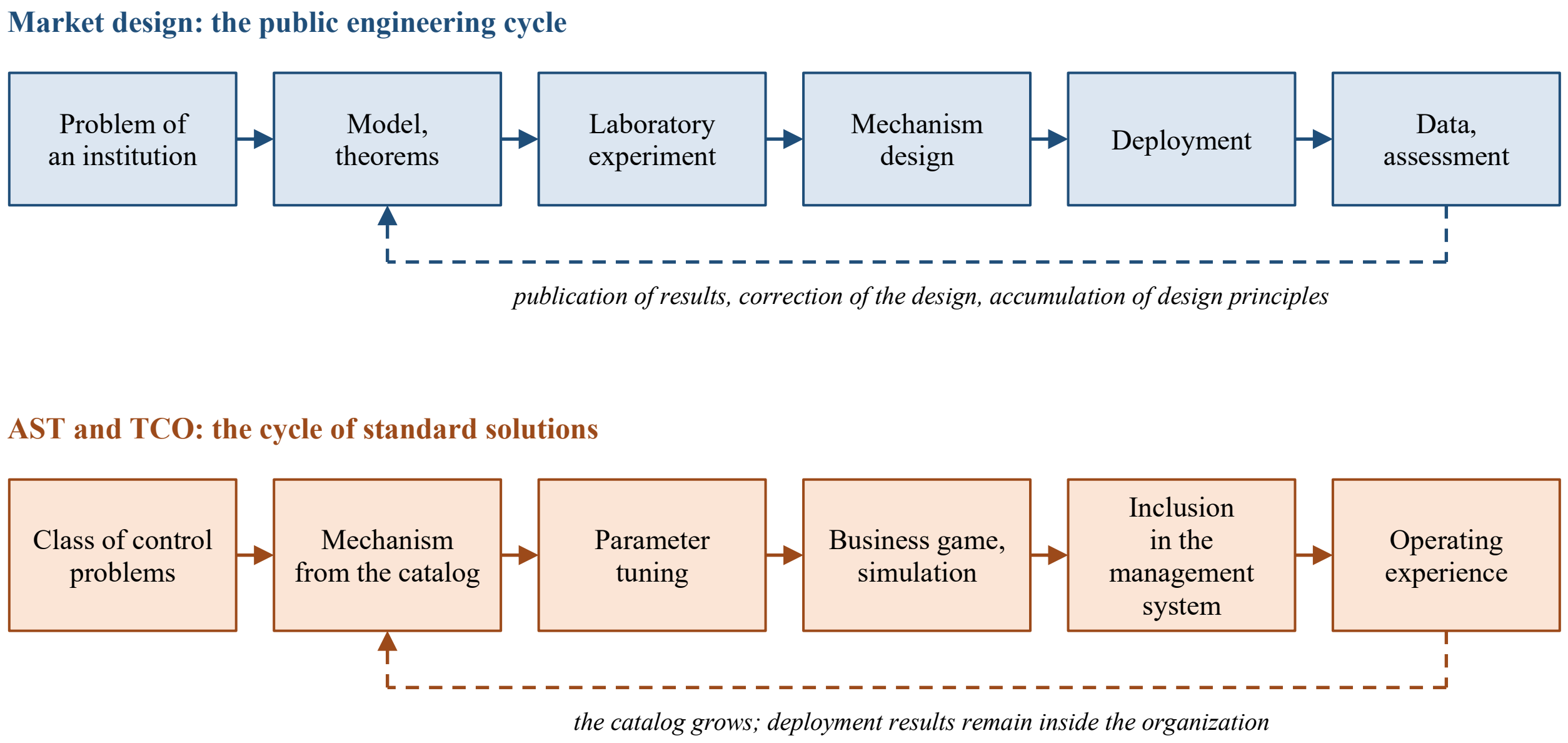

Fig. 18. The two models of carrying theory into practice: the public engineering cycle of market design and the AST and TCO cycle of standard solutions

# 10. Development Forecast

The forecast reflects the survey author's assessment. It rests on the published programmatic texts [102, 114, 139, 179, 397, 470], on the directions manifest in the publications of 2023–2026, and on the unfilled cells of the classification of Section 8. The horizon is up to 2035; confidence is rated on a three-level scale.

## 10.1. Driving Forces

*AI agents as participants.* Economic interactions are increasingly conducted by software agents: autobidders in advertising, pricing algorithms, agents based on language models. The first theoretical results concern autobidding, collusion of pricing algorithms and mechanisms for language models [3, 133, 182, 212]. For the theory this means a change in the value of facet F10: the utility maximizer with given preferences is replaced by a learning system with a known architecture.

*Data and learning.* Ever less often does the principal know the distribution of the uncertain parameters, and ever more often does it have a sample and the ability to experiment. This moves problems from the Bayesian cell of F2 into the “sample” cell and makes questions of sample complexity and regret standard.

*Computation.* The mechanism becomes a program and its design a computational problem; the outcome of twenty years of algorithmic MD is an understanding of where the incentive requirement genuinely makes computation more expensive and where it does not.

*Robustness and simplicity.* After the results on the fragility of finely tuned mechanisms (Section 4.2) and on the price of simplicity (Sections 4.4, 4.8), the requirements of simplicity and robustness enter the problem statements as constraints in their own right.

*Legitimacy and public goals.* Distributional goals, fairness and admissibility enter the criteria, and institutional reforms are judged by how well they preserve the institutions' original purpose [177, 490].

*The self-assessment of economic design.* The volume [494] shows that MD researchers themselves regard robustness under weak structure, the simplicity and transparency of mechanisms, behavioral implementation, the treatment of authority and organizational structure, and learning-based synthesis as the directions of the near future (Section 4); three of these five coincide with the areas of AST's traditional strength, which reinforces the convergence forecast in the problems of Section 10.4.

*The AST school's internal factors.* Generational change; the digitalization of management, creating demand for formalized mechanisms inside information systems; the consolidation of the field in university courses [153].

## 10.2. Forecast for Mechanism Design, Contract Theory, and Implementation Theory

The forecast theses are collected in Table 14 and Fig. 19. In substance they come down to four claims.

First. The central theme of the next five years will be mechanisms whose participants are *learning agents*. This will require new solution concepts (equilibria of learning dynamics instead of Nash equilibria), new notions of robustness to collusion arising without communication [133, 212], and new notions of incentive compatibility for agents whose preferences are given implicitly [182]. The theory of autobidding [3] provides the first completed example of such a branch.

Second. Contract theory will develop as a theory of delegating tasks to AI systems: the hidden action takes the form of the agent's choice of model and computational effort, and contracts must be computed and learned [179]. This also includes treating AI alignment as an incomplete contract and social choice problem [151, 265].

Third. The robust, learning-based and algorithmic approaches will converge into a single theory of mechanisms with guarantees under limited knowledge: the result guaranteed over a set of distributions, sample-based bounds and approximation ratios are different forms of one requirement [139, 470]. Automated design will move from approximate incentive compatibility to verifiable guarantees [159].

Fourth. In market design the shift will continue from optimality to minimal changes of existing institutions [490] and to distributional goals; in implementation theory, to behavioral and experimentally testable implementation [141, 168]; in decentralized systems, to mechanisms that presuppose no trusted organizer [6, 468].

Table 14. The forecast theses (the survey author's assessment)

| Thesis | Grounds | Horizon | Confidence |
|---|---|---|---|
| Mechanisms for learning and AI agents will crystallize into an independent branch with solution concepts of its own | [3, 182] | 2026–2030 | High |
| Algorithmic contract theory will become the main apparatus for analyzing the delegation of tasks to AI systems | [179] | 2026–2030 | High |
| The notions of coalitional stability will be redefined for collusion arising from learning; regulatory standards will appear | [133, 212] | 2026–2032 | High |
| Automated design will deliver mechanisms with verifiable exact strategy-proofness for applied problems | [159, 181] | 2028–2032 | Medium |

| Thesis | Grounds | Horizon | Confidence |
|---|---|---|---|
| Robust and learning-based design will merge into a theory of guarantees under limited knowledge; the Bayesian optimum will remain the comparison benchmark | [139, 264, 470] | 2028–2035 | Medium |
| Information design will join up with the problems of recommender systems and data markets | [62, 67] | 2026–2032 | Medium |
| Social choice and contract theory will enter the practice of AI alignment | [151, 265] | 2026–2030 | Medium |
| Market design will shift toward minimal institutional reforms and distributional criteria | [177, 490] | 2026–2030 | High |
| Mechanisms without a trusted organizer (credible auctions, protocol-level mechanisms) will receive a general theory | [6, 468] | 2028–2035 | Medium |
| Dynamic design will receive general conditions for global incentive compatibility via the link with reinforcement learning | [70, 443] | after 2030 | Low |
| In AST and TCO the main growth will fall on integrated mechanisms in digital management systems and on organizations with a mixed staff | [114, 59] | 2026–2032 | Medium |
| The AST core will be restated in the language of robust and algorithmic design and will receive approximation bounds | [102, 15] | 2028–2035 | Low or medium |

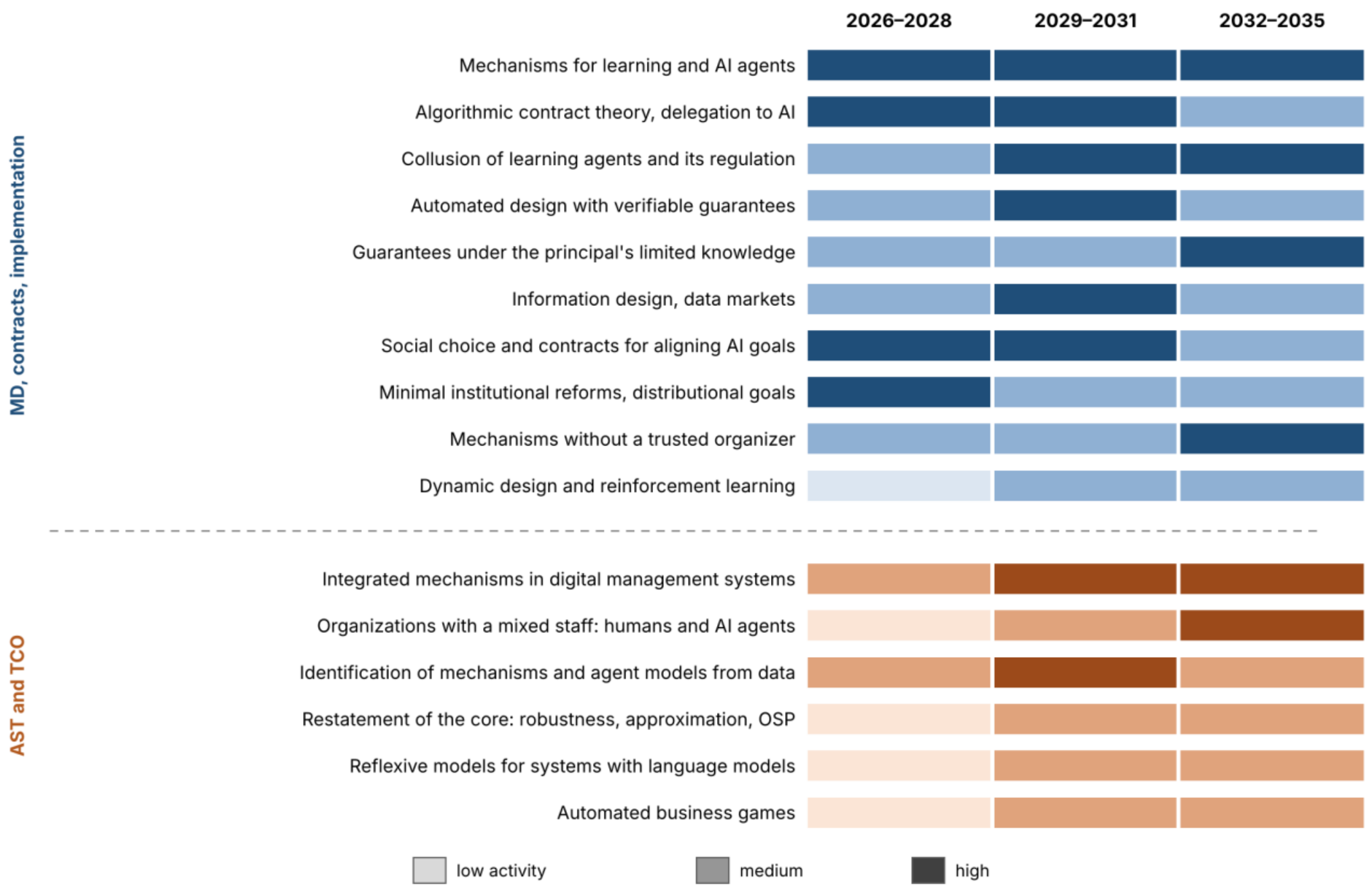

Fig. 19. A roadmap of the expected development of the strands up to 2035 (the survey author's assessment; the saturation of a band corresponds to the expected activity)

## 10.3. Forecast for AST and TCO

The AST school itself names open problems at two levels. The theoretical ones [102]: carrying the positional dictatorship schemes over to systems with transferable utility; abandoning the weak influence hypothesis; the relation between implementability under different equilibrium concepts on “non-rich” domains; accounting for the agents' mutual awareness; computational complexity and the simplicity of mechanisms; robustness. The engineering ones [114, 121]: constructing integrated mechanisms; identifying integrated rating mechanisms from data; multilevel systems; embedding mechanisms into decision support information systems. The general context is set by the challenges of decentralization, strategic behavior and complexity [417, 397].

It is instructive to set this agenda against the forecast the AST school gave a quarter of a century ago. The 1999 book named ten promising problems [120]: incorporating the notions of psychology, economics and sociology into the models; mechanisms when coalitions can form; systems with mixed uncertainty; the synthesis of staff and structure, including multilevel and network structures; systems with strongly coupled elements under different equilibrium concepts; necessary and sufficient strategy-proofness conditions for planning mechanisms of general form; optimality conditions for proper mechanisms; preferences given by relations; the stability of solutions and the adequacy of models; vector control structures. In the survey author's assessment, five of them have been noticeably advanced by 2026: reflexive models and bounded rationality [417, 422], the synthesis of structures and network systems [375, 55, 246], the strategy-proofness of planning mechanisms incorporated into the theory of generalized median schemes [110, 327, 328], the optimality of proper mechanisms [189, 187], and systems with several principals [313, 430]. The items that advanced were those resting on work already begun; coalitional effects, mixed uncertainty and the stability of solutions remained open. This observation is reflected in the confidence ratings of the present forecast.

Given the state of the AST school, three scenarios are possible. Integrationist: the theory keeps developing, including for AI agents; the results on mechanisms without transfers and on coordinated planning are published in international journals as contributions to robust and money-free design, and approximation bounds and work on OSP implementability appear. Engineering: the main product becomes integrated mechanisms and their software implementations in the digital management systems of enterprises and agencies, with the theory developing toward identification, complexity and standard solutions. Inertial: the catalog and the courses persist, the flow of new theorems dwindles, and the field dissolves into general control theory and data analysis. The most likely appears to be the engineering scenario with elements of the integrationist one: it rests on the AST school's real competencies and on demand, while integration will occur at the points where MD itself has arrived at AST's formulations (the guaranteed criterion, simple mechanisms, allocation without money).

The AST school's substantive growth points are as follows. Organizations with a mixed staff of humans and AI agents are a natural generalization of the active system in which some agents have a known architecture. Reflexive games acquire a new object: language models form beliefs about the interlocutor's beliefs, and the machinery of awareness structures [422] is applicable to controlling such systems. Business games can be automated with simulation agents based on language models, which lowers the cost of experimentally testing the catalog. Finally, the enterprise models [59] provide a frame in which individual mechanisms receive a quantitative assessment of their contribution.

The common frame for these growth points is set by the active milieus program [419]: a variable set of participants, control through mechanisms of functioning rather than directives, and a four-level formalization in which mechanism theory and TCO constitute the strategic behavior level; at the same time, the open problems named there – the computability of hierarchical optimization problems, scaling, the generation of the new – coincide with the agenda of algorithmic MD and complexity theory.

## 10.4. Problems Where Convergence Is Expected

Table 15 lists the problems formulated by the survey's author at the intersection of the research traditions. Each fills an empty cell of the matrix of Fig. 17 and rests on results from both sides; the convergence scheme is shown in Fig. 20. Some of the problems deserve comment.

Problem 1 requires assessing the standard AST mechanisms in a relative measure. For a mechanism $\rho$ with equilibria $s^*(r)$ and the principal's criterion $\Phi$, the natural characteristic is

$$\mathrm{PoA}(\rho) = \inf_{r\in\Omega} \frac{\min\limits_{s^*\in P(\rho,r)} \Phi(\pi(s^*),r)}{\max\limits_{x} \Phi(x,r)}, \tag{66}$$

that is, the guaranteed efficiency $K_g$ of (3) normalized by the full-information optimum. The AST school considered bounds of this type for tournament mechanisms, the question of the maximum guaranteed efficiency being named open in the book [120]; for the priority-based mechanisms (8), transfer prices and joint financing there are none, although the equilibria are known in explicit form.

Problem 4 links the weak influence hypothesis with strategy-proofness in the large. The qualitative result (the decay of the gain from manipulation as the economy is replicated) is known for classical exchange economies [456]; for the AST mechanisms with a common control variable it has not been obtained. What is needed is a quantitative estimate of the gain from manipulation in mechanisms with a common control variable $\lambda(s)$ as a function of the number of agents: at what $n$ and at what rate the gain decays, and how it depends on the agents' heterogeneity.

Problem 6 calls for a general model of informational control that includes as special cases Bayesian persuasion (the principal bound by commitment and belief consistency) and reflexive control (the principal unbound). Partial commitment of the principal is modeled in MD by [354, 371], but false belief hierarchies are not allowed there. The natural parameter of such a model is the set of admissible awareness structures $\mathbb{I}$ in (42); one must find out how the problem's value depends on the restrictions on $\mathbb{I}$ and what the "price of commitment" equals.

Problem 13 joins two independent lines: the relational contracts of contract theory (33) and the dynamic stability of cooperative solutions of Petrosyan's school [446, 531]. The question is under what conditions an imputation distribution procedure that makes the cooperative solution dynamically stable is simultaneously supported by an equilibrium of the repeated game with incentives, i.e. is a self-enforcing agreement in the sense of [351]; for TCO this would give a model of a mechanism the principal cannot violate.

*Problem 14* joins the theory of simplicity with DAS theory. In [455] the planning horizon serves as the standard by which a mechanism is judged, in [427] as the parameter to which the principal fits plans and commitments. The natural questions: does the history-dependent compensatory scheme (59) meet the simplicity standards of [455] for a given agent's horizon; how does the price of simplicity depend on the distribution of farsightedness; and can condition (60) be restated as the condition under which the rolling regime with commitments is implemented by a simple mechanism.

Problems 15–17 relate to the strands considered in Sections 6.7, 6.5 and 6.9 respectively. What they share is that the AST school's substantive model (the phase constraint, the reflexion rank, the project's network schedule) is built under complete awareness of the goal functions or with passive participants, whereas MD has the machinery for private information but does not consider these objects in this form: phase constraints and the project network schedule do not occur in MD, and the depth of reasoning is set exogenously [169] and is not estimated from market data. In problem 15 the natural characteristic is the price of viability, i.e. the loss in the principal's criterion from the requirement $x(t) \in X^*$ under the agents' private information compared with the complete-awareness case. In problem 17 the bounds (62) can play the same role as approximation algorithms in algorithmic MD: one must establish for which partitions $h$ the allocation rule built from the bound remains monotone and hence admits a strategy-proof implementation.

Problems 18–20 follow from the comparison of reflexive games with the epistemic models of MD (Sections 6.5.2–6.5.5). Problem 18 concerns a question unresolved in the theory of reflexive games – where the awareness structure comes from and why it persists: under what conditions a stable informational equilibrium is the limit of a learning dynamics, and which false structures survive. Problem 19 concerns the boundary between the result "depth beyond two is unnecessary" and the results on the decisive role of higher orders. Problem 20 is an applied setting for which MD has no ready-made model: when it benefits the principal to disclose, conceal or invent participants, including the existence of mixed-strategy equilibria of the interval and probabilistic GMAS and the exact correspondence of irregular structures to KD45 belief models.

Table 15. Problems at the intersection of the research traditions (formulated by the survey's author)

| No. | Problem | Support in AST and TCO | Support in MD |
|---|---|---|---|
| 1 | Approximation bounds (66) for the standard mechanisms: priority-based, transfer prices, joint financing | Explicit equilibria [97, 328]; bounds for tournaments [120]; the error of active planning [102] | The price of anarchy and smoothness [469, 489] |
| 2 | A maximin justification of fair play: when the law (5) is optimal among all Bayesian mechanisms under agents' beliefs unknown to the principal; extension to interdependent valuations | The Burkov–Enaleev theorem [98] | Justifications of dominant strategies [64, 149] |
| 3 | Obviously strategy-proof sequential procedures for the catalog mechanisms (tournaments, transfer prices, integrated rating); the losses from the obviousness requirement | The catalog of mechanisms [153] | OSP implementation [15, 36, 352] |
| 4 | A quantitative theory of the weak influence hypothesis for a finite number of agents | The weak influence hypothesis [112; 33; 456] | Strategy-proofness in the large [33] |
| 5 | Carrying positional dictatorship over to systems with transfers; strategy-proofness and budget balance | An open problem [102]; mechanism (13) | Affine maximizers; [252, 387] |
| 6 | A unified model of informational control; the price of commitment and the price of false beliefs | Problem (42) [422] | Bayesian persuasion [67, 311]; persuasion under partial commitment [354, 371] |
| 7 | Composition of mechanisms: conditions under which incentive compatibility and strategy-proofness survive in an integrated mechanism | Integration of mechanisms [114, 107] | Composition of smooth mechanisms [489]; incentive systems [288] |
| 8 | Decomposition of the agents' game under unobservable actions and an aggregated result | The decomposition and aggregation theorems [428] | Unique implementation [266, 528]; teams [161, 287] |

| No. | Problem | Support in AST and TCO | Support in MD |
|---|---|---|---|
| 9 | The sample complexity of identifying cost functions and of learning incentive schemes | Identification [122, 260] | Learning contracts from data [179]; sample complexity [264] |
| 10 | Mechanisms for collectives of humans and AI agents: incentives, reflexion, collusion of learning agents | Reflexive games [422, 404] | [182, 212] |
| 11 | The computational complexity of synthesizing incentive-compatible plans, and simple mechanisms with a guarantee | Complexity [399] | The price of truthfulness [175] |
| 12 | A general methodology of experimental testing: business games, laboratory experiments, simulation by language-model agents | Business games [330, 331] | [141] |
| 13 | Self-enforcing agreements: conditions under which a dynamically stable cooperative solution is supported by an equilibrium with incentives; mechanisms the principal cannot violate | Petrosyan's school [446, 447, 531] (outside AST) | Relational contracts [351]; trust in the organizer [6] |
| 14 | Simplicity under bounded farsightedness: which DAS mechanisms meet the simplicity standards for given horizons; the price of simplicity and the price of commitment in the rolling regime | DAS theory: farsightedness, control regimes, condition (60) [427, 426, 482] | The theory of simplicity [455]; a bounded forecasting horizon [306]; bounded commitment [176] |
| 15 | Incentives under phase constraints: mechanisms with agents' private information that guarantee the homeostasis condition $x(t) \in X^*$; the price of viability and the information rent under such a constraint | Models of sustainable development control (53), (54) [441, 332, 462] | Dynamic MD [443]; continuous-time contracts [475] |
| 16 | The depth of reasoning as an object of identification and control: estimating reflexion ranks from market data; mechanisms robust to heterogeneous ranks; the limits of informational control of beliefs | Reflexive games [422]; reflexion in oligopoly (47) [224, 222] | Level-$k$ models [169]; information design [67] |
| 17 | A project with active executors: strategy-proof mechanisms for the time–cost trade-off and portfolio formation problems when the durations and costs of operations are private information; consistency of aggregation with incentives; network programming bounds as the basis of approximate mechanisms | Project management and network programming (61), (63), (64) [119, 319, 90] | Incentives in procurement [338]; the price of truthfulness [175] |
| 18 | Stability as the outcome of learning: conditions under which a stable informational equilibrium is the limit of a learning dynamics; which false structures survive | Stable, true and false equilibria [422]; transformations of MAS and GMAS [210, 145] | Conjectural equilibrium [204, 472]; Berk–Nash equilibrium [205]; misspecified learning [278] |
| 19 | The boundary between “depth beyond two is unnecessary” and “higher orders are decisive”: for which classes of games the first holds, and for which the second | Bounds on the maximum expedient rank [422, 404] | The e-mail game [471]; global games [135]; the structure theorem [523] |
| 20 | Controlling beliefs about the composition: when it benefits the principal to disclose, conceal or invent participants; the existence of GMAS equilibria; the correspondence of irregular GMAS to KD45 logic | GMAS and control of the composition [209, 210] | Disclosure of the number of participants [350, 362]; the value of ignorance and existence [26]; shill bidding [250]; term-modal logics [432] |

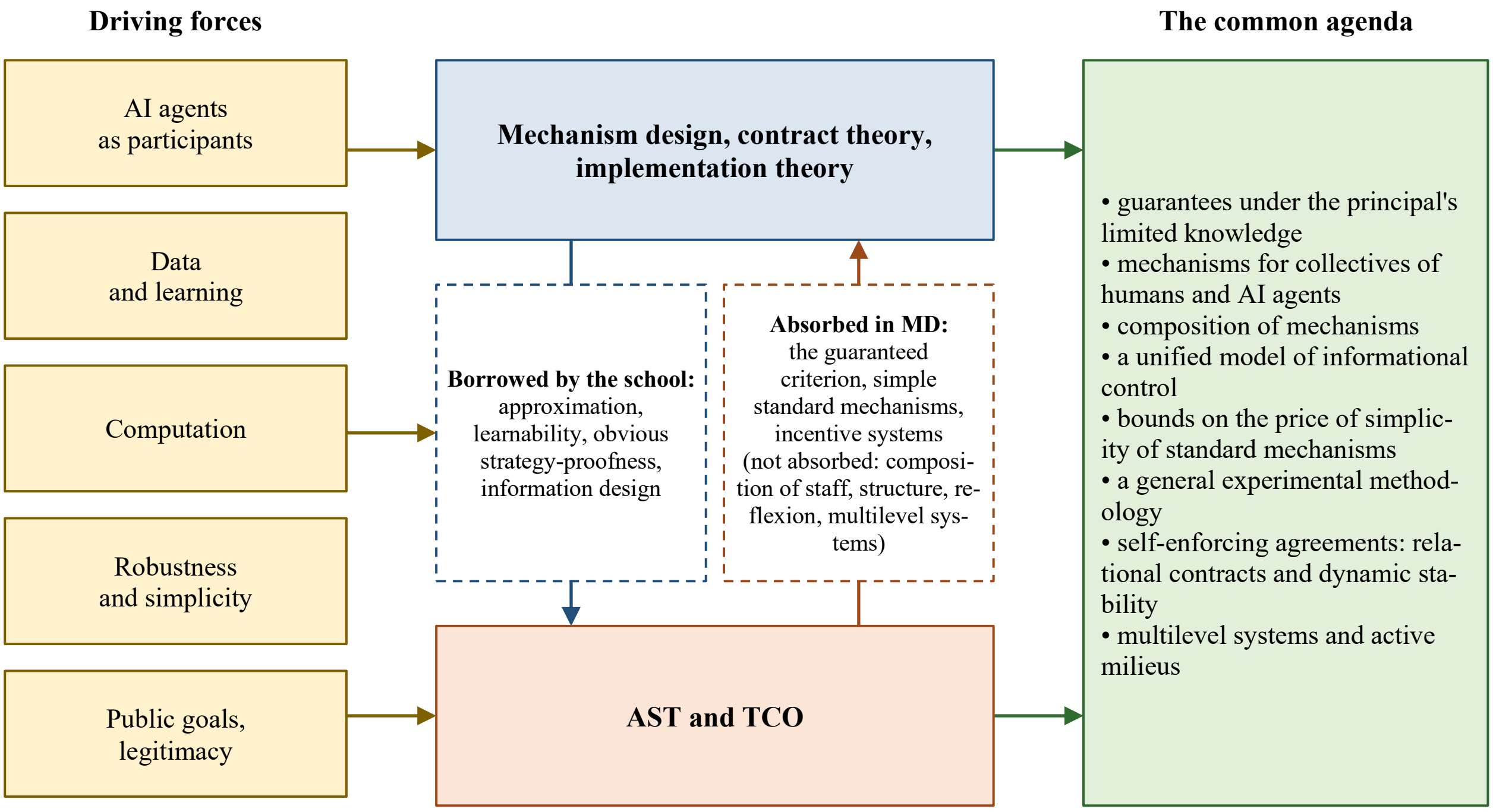

Fig. 20. The scheme of the convergence of the research traditions: the driving forces, the mutual borrowings, and the common agenda

## 11. Conclusion

1. Mechanism design and the theory of active systems are two research traditions that took shape in different scientific communities and realize one idea: the principal chooses the rules, and the result is judged by the solution of the agents' game. In the common core their results coincide up to the criterion: the fair play principle and the perfect coordination condition correspond to the revelation principle and the characterization of strategy-proofness through option sets, the serial resource allocation mechanism to the uniform rule, the active expertise mechanism to median schemes, and the quasi-compensatory incentive schemes to contracts under observable action. Germeier's theorem on the value of the game $\Gamma_2$ and the guaranteed-result criterion (1971–1976) belong to his school, whereas the setting “the principal announces a function of the observable variable” arose in economics independently and in the same years; in the axiomatics of the uniform rule, the AST school's priority covers the direct-mechanism construction for priority-based mechanisms (1989) but neither the rule itself (introduced by Benassy in 1982) nor its characterization. Translations of the basic books have existed since 1986 and 2013 but are not cited in the MD literature (Section 2.2).

2. The main difference between the settings is the description of uncertainty and the criterion. From the very beginning AST describes the unknown parameters by sets and applies the guaranteed result to them (with equilibrium selection subordinated, by default, to the benevolence hypothesis); in MD, sets and guarantees had been applied in individual works even earlier (Hurwicz and Shapiro, 1978), but the approach became widespread only after 2005, in robust, algorithmic and learning-based design. In this respect the development of the last twenty years has brought the research traditions closer together.

3. Over 2006–2026 MD created several new branches (robust, dynamic, information and algorithmic design, algorithmic contract theory, mechanisms for AI agents) that have no analogues in the AST school. Over the same period the AST school incorporated its classical mechanisms into the general theory of strategy-proofness, completed the theory of coordinated mechanisms, and broadened its subject: reflexion, networks, hierarchies, complex activity. Several AST results of the late 1990s – the systematics of uncertainty with fuzzy models, the theorems on the invariance of equilibria under the decentralization of mechanisms, the price of unification – have no direct analogues in MD; MD works close in subject do exist and are set side by side in Sections 3.2, 4.4 and 6.6. Three lines of the AST school have a subject of their own: reflexion in oligopoly games (Section 6.5), sustainable development control with dynamic and multilevel models (Section 6.7), and project management with network programming (Section 6.9). The first two are built on games with complete awareness of the goal functions, the third is for the most part optimizational, and they intersect with MD only partially; the problems at these intersections are formulated in Section 10.4 (problems 15–17).

4. The comparative advantage of TCO is a single scheme covering incentives, plans, institutions, information, composition and structure – more dimensions than MD has (where these dimensions are represented by separate neighboring theories: information design, delegation theory, organizational economics) – together with its orientation toward complexes of standard mechanisms. The comparative advantage of MD is theorems that set the bounds of the achievable, quantitative guarantees, and the public verification of deployments.

5. The proposed facet classification (ten facets) and the typology of results (nine types) make it possible to write the settings of both research traditions in a single coordinate system and point out the unfilled combinations. The twenty problems of Section 10.4 are derived from such combinations.

6. Over the coming decade the subject of both research traditions will change in the same way: the agents will be learning systems, and the principal will know less about them than the classical models assume and more than the theory of games with rational players assumes. For AST, which defined the active element as a controlled object possessing goals of its own, this means a return to the original setting on a new class of objects.

# Appendix. Reference Tables

The appendix collects the reference tables referred to in the main text: the basic monographs and survey chapters of the two research traditions, the documented applications of mechanism design and contract theory, and the declared areas of application of the AST and TCO mechanisms.

Table 16. The basic monographs and survey chapters

| AST publication | Content and main results | MD analogue |
|---|---|---|
| Burkov, 1977 [87] | The active element and the active system, the fair play principle, the degree of centralization, weak influence conditions, the probabilistic incentive problem for a “simple” active element | Hurwicz's works on mechanisms [495] |
| Burkov and Kondratiev, 1981 [112] | Mechanisms of functioning, coordinated planning, the weak influence hypothesis | – |
| Ashimov, Burkov, Dzhaparov and Kondratiev, 1986 [22]; the 1985 papers [21, 20, 19] | Coordinated operational planning of continuous and discrete production; sets of coordinated schedules; branch-and-bound algorithms; deployment in the nonferrous metallurgy of Kazakhstan | Intra-firm planning under private information (Section 2.2) |
| “Large-Scale Systems”, 1989 [97] | Resource allocation and active expertise: equivalent direct mechanisms | The uniform rule, median schemes [381, 486] |
| Germeier, 1976 [226]; the 1971 paper [225]; the 1986 English translation [227] | Games with a fixed sequence of moves, the maximum guaranteed result (17) | The Stackelberg equilibrium; the contract as a strategy-function |
| Burkov and Novikov, 1999 [120] | The outcome of thirty years: the basic model and its extensions, incentive schemes, planning mechanisms, the catalog of mechanisms, ten promising problems | Survey chapters [40, 359] |
| Novikov and Petrakov, 1999 [424] | A course with numbered theorems: incentives (Ch. 2), planning and implementability theory (Ch. 5) | Textbooks [131, 340] |
| Novikov, 1998 [407] | Thirteen basic incentive models by the type and kind of uncertainty and awareness; the optimality of C- and K-type schemes under interval, probabilistic and fuzzy uncertainty; the payment for information | Moral hazard [255, 284]; no fuzzy analogue |
| Novikov, 1999 [408] | Multilevel systems: five factors of hierarchy, perfect aggregation, arbitrary decentralization of anonymous mechanisms, unified schemes and the price of unification, the principle of rational centralization | Hierarchical agency; no analogue of the decentralization theorems |
| Novikov, Smirnov and Shokhina, 2002 [427] | A classification of DAS control problems; distributions of farsightedness and decision horizons; the basic models DAS1–DAS4; cost compensation and decomposition in dynamics (Theorems 1–3); the role-exchange effect (Theorem 4); the adequacy principle and the comparison of control regimes (Theorems 5–8); accumulation effects; a survey of AST, hierarchical-games and contract-theory results on dynamics | Contract dynamics with and without commitment [337]; the planning horizon [455] |
| Novikov and Tsvetkov, 2000 [428] | Multi-agent systems: decomposition of the agents' game, aggregation, control of the composition | Teams, tournaments, unique implementation [287, 345, 528] |
| Novikov, 2003 [410] | Structure as a consequence of the type of hierarchical game; the structural synthesis problem, assignment of the principal, side payments, distributed control, sequential synthesis | The theory of the firm, organizational economics – mathematical models are almost absent |
| Novikov, 2003 [409] | Individual and collective incentives, forms and systems of pay, experimental data on labor supply, incentives under optimization of the staff | Moral hazard [255, 289]; the income–leisure model of labor economics |

| AST publication | Content and main results | MD analogue |
|---|---|---|
| Novikov and Chkhartishvili, 2003 [421] | The first exposition of reflexive games: strategic and informational reflexion, the awareness structure, phantom agents, informational equilibrium, the reflexive game graph, reflexive control | Harsanyi's belief hierarchies [270]; common knowledge; level-k models |
| Novikov, 2004 [411] | Institutional control: constraints (admissible sets) and norms of activity; the price of prohibition and the comparison with motivational control (Proposition 1); coordinated and unified norms; the role of awareness | Delegation [13, 285]; norms as equilibria |
| Novikov, 2005 [412]; 4th ed. [418] | The general theory: the components of the model, the types of control, the general statement and solution scheme, the table of standard mechanisms | No single analogue |
| Novikov, 2013 [414] | The English edition of the monograph [412]: nine chapters – from incentive and planning mechanisms to control of the composition and structure, informational and institutional control | No single analogue |
| Voronin, Mishin, 2003 [519] | Optimal hierarchical structures: an organization as a set of functions to be performed, sectional cost functions | Hierarchy models in organizational economics |
| Mishin, 2004 [375] | Optimal hierarchies: sectional cost functions, the span of control (50) | Hierarchy models in organizational economics |
| Goubko, 2006 [245]; the 2008 survey [262] | Optimal hierarchies under homogeneous and sectional cost functions; lower bounds on costs; suboptimal algorithms; a survey of MD hierarchy models | The hierarchy models of Wiliamson, Calvo and Wellisz, Keren and Levhari, and Qian |
| Novikov and Chkhartishvili, 2013 [422]; the 2014 English edition [403] | Reflexive games: awareness structures, informational equilibrium, the reflexion rank, informational control | Epistemic logic, Bayesian games, information design [67] |
| Belov and Novikov, 2018–2021 [54, 58, 56]; English editions [57, 59] | The methodology of complex activity; coordinating the life cycles of organizational-technical systems; a unified enterprise optimization model with static and dynamic multi-agent contracts and mathematical models of technologies, human capital and planning | Organizational economics; systems engineering – mathematical models are absent |
| Novikov, 2016 [415, 398] | Cybernetics 2.0: history, laws and principles of control, four challenges to control theory, organization theory | No single analogue |
| Novikov, 2026 [419] | Active milieus: definition and criteria, principles of control (including the mediation principle – control through mechanisms), the four-level formalization scheme, open problems | Multi-agent systems; complexity sciences |
| "Control Mechanisms", 2011 [153] | The catalog: 20 basic and 4 integrated mechanisms | No analogue |
| Ougolnitsky, 2016 [441], the survey [512] and the papers [332, 436] | Hierarchical control of sustainable development: homeostasis and the coordination of interests; compulsion, impulsion, persuasion; the extended active system | Regulation theory; dynamic games with state constraints |
| Chernushkin, Ougolnitsky and Usov, 2013 [142] | Corruption in "principal – supervisor – agent – object" hierarchies as feedback on the bribe (after the papers [142, 511]) | Models of hierarchical agency and collusion |
| Burkov et al., 1967 [116] | Network models and control problems: network-based scheduling | Network planning methods (CPM, PERT), scheduling theory |
| Burkov and Novikov, 1997 [119] | Project management mechanisms by phase: expertise, the composition of executors, financing, incentives, operational control | The theory of procurement and regulation [338] |

| AST publication | Content and main results | MD analogue |
|---|---|---|
| Burkov, Kwon and Tsitovich, 1997 [115]; Barkalov, Burkov and Gilyazov, 1999 [44]; Barkalov et al., 2002 [43] | Multi-project management: resource allocation, foregone profit, perfect aggregation (61), the criticality-degree rule, the dual network model | Resource-constrained scheduling |
| Kolosova, Novikov and Tsvetkov, 2000 [319]; Matveev, Novikov and Tsvetkov, 2005 [361] | Earned value; incentives for shortening durations (64) under certainty and uncertainty; project portfolios | The earned value method; contracts with a completion-time bonus. The game-theoretic component is absent |
| “Mathematical Foundations of Project Management”, 2005 [49]; Barkalov, Burkov, Novikov and Shulzhenko, 2003 [46] | The 2005 textbook: an instructional exposition of network models, resource allocation, aggregation and discrete project management problems | Textbooks on project management and operations research |
| Petrakov, 2001 [444] | Dictatorship sets; conditions of strategy-proofness and coalitional strategy-proofness of planning mechanisms; the existence of an equivalent direct mechanism | Generalized median schemes [41] |
| Goubko, 2003 [244] | The cooperative statement of the control problem: the balancedness of the agents' game as a criterion of the stability of the maximal coalition, inessentiality as a criterion of the absence of cooperation; the robustness of incentive schemes to collusion; coalitions of principals under distributed control; coalitional change of the composition; conditions for the non-emptiness of the core in resource allocation; the constant-priority mechanism and the internal market | Aumann's strong equilibrium, coalition-proof equilibria, mechanisms with a core outcome, collusion models |
| Burkov, Kondratiev, Tsyganov and Cherkashin, 1984 [113] | AST as an instrument for improving the economic mechanism; performance assessment and incentives for taut plans (the AKKORD system) | Soviet success indicators [355, 524] |
| Tsyganov, 1991 [504]; Tsyganov, Borodin and Shishkin, 2004 [507]; Tsyganov, 2012 [505] | Adaptive mechanisms: active identification, progressivity, farsightedness; mechanisms of organizational evolution; humanitarian systems | The ratchet effect and dynamic contracts [214, 525] |
| Fedyanin and Chkhartishvili, 2024 [210]; Fedyanin, 2025 [209] | GMAS: a variable set of agents, the interval and probabilistic variants, transformations, the equivalence of regular structures to S5 Kripke models | Pre-Bayesian games [26]; Poisson games [385]; dynamic epistemic logic [513] |
| Vatel and Ereshko, 1973 [515]; Germeier and Ereshko, 1974 [228] | A popular exposition of hierarchical games; side payments in the game $\Gamma_2$ as a precursor of compensatory incentives | Contracts under observable action [485] |
| Gorelik and Kononenko, 1982 [239]; Gorelik, Gorelov and Kononenko, 1991 [238] | Static and dynamic models of hierarchical systems, the regulated equilibrium; the choice of information exchange procedures and feedback control in dynamics | The choice of the message space [293]; dynamic contracts |
| Kononenko, Khalezov and Chumakov, 1991 [325]; the 1990 and 1994 Computing Center preprints [11, 242] | Hierarchical games under uncertain factors and risk; exact aggregation and the informational aspects of decision making | Robust MD; communication complexity |
| Ereshko, 1984 [196]; Ereshko et al., IIASA, 1984 [200, 203]; Ereshko, 2001 [198] | Hierarchical compromises and threats; numerical methods for the games $\Gamma_1$–$\Gamma_3$ and equilibria; reflexive strategies | Delegation and threats; stochastic algorithms; reflexive games (Section 6.5) |

Table 17. Applications of mechanism design and contract theory

| Area | Mechanism | Evidence | Sources |
|---|---|---|---|
| Radio spectrum | The simultaneous multi-round auction; the incentive auction of 2016–2017 | The first auction, 1994: $617 million; the United States, 1994–2014: over $120 billion; worldwide: over $200 billion | [370, 497] |
| Online advertising | The generalized second-price auction; autobidding | Truthful bids do not form an equilibrium; with autobidders, the price of anarchy of the second-price auction equals 2 | [3, 186] |
| Assignment of physicians and school students | Deferred acceptance | In 2025 the NRMP offered 43,237 positions (PGY-1 and PGY-2) for 47,208 active applicants [390], against about 20 thousand in 1999 [465]; New York: 90% fewer enrollments in schools outside the preference list | [1, 465, 496] |
| Kidney exchange | Exchange cycles and chains | The mechanism is efficient and incentive-compatible; computations show a substantial increase in the number of feasible transplants | [466] |
| Course allocation | Approximate competitive equilibrium from equal incomes | Strategy-proofness in the large, bounded envy | [83] |
| Food assistance | An internal market with a scrip currency | A 21% increase in the value of the food distributed | [453] |
| Electricity | Day-ahead and real-time markets | Settlement at locational marginal prices; capacity markets with pay-for-performance | [154] |
| Blockchain | The EIP-1559 fee mechanism | In operation since August 2021; the base fee is burned | [468] |

Table 18. Areas of application of the AST and TCO mechanisms

| Period | Areas and objects | Mechanisms used | Sources |
|---|---|---|---|
| Since the 1960s (a continuous line) | Network-based scheduling and project management: the extractive industries, construction, programs and project portfolios | Network models, resource allocation, aggregation, network programming; since the 1990s, financing and incentive mechanisms | [47, 91, 121] |
| 1970s–80s | Nonferrous and ferrous metallurgy, instrument making, the radio industry, research organizations | Coordinated planning, counter plans, performance assessment, multi-channel mechanisms, counter-expensive mechanisms | [121] |
| 1980s | Ferrous metallurgy (the “Operator Advisor” control system, V.P. Avdeev's school, Novokuznetsk); instrument making (the Ministry of Instrument Making, the AKKORD system) | Two- and multi-channel mechanisms with recalculation models; integrated quantitative performance assessment | [31, 126, 113, 88, 121] |
| 1990s–2000s | Enterprise restructuring, project management, safety programs, offsets | Integrated rating, tournaments, joint financing, insurance, exchange schemes | [119, 121] |
| 2010s–20s | Project offices, regional programs, ecological-economic systems, production systems, online networks | Integrated mechanisms, network planning, group incentives, informational control | [260, 498, 59, 402] |